\documentclass[12pt, letterpaper]{article}
\usepackage[margin=1in]{geometry}

\usepackage[utf8]{inputenc}
\usepackage[english]{babel}

\usepackage{outlines}
\usepackage[hyperfootnotes=false]{hyperref}
\usepackage{csquotes}
\usepackage{amsmath}
\usepackage{amssymb}
\usepackage{graphicx}
\usepackage{xcolor}
\usepackage{chngpage}
\usepackage{float}
\usepackage{amsthm}
\usepackage{mathtools}
\usepackage{bbm}
\usepackage{enumitem}
\usepackage{algorithm}
\usepackage{algpseudocode}
\usepackage{multirow}
\usepackage{setspace}
\usepackage{times}
\usepackage{authblk}
\usepackage{pifont}
\usepackage{subcaption}
\usepackage{tikz}
\usepackage{titling}

\usepackage{caption} 
\newcommand{\cmark}{\ding{51}}%
\newcommand{\xmark}{\ding{55}}%

\DeclareMathOperator{\AEE}{AEE}
\DeclareMathOperator{\E}{E}
\DeclareMathOperator{\PP}{P}
\DeclareMathOperator{\Cov}{Cov}
\DeclareMathOperator{\Var}{Var}
\newcommand{\gbar}{\mathbf{\bar{g}}}
\newcommand{\Gbar}{\mathbf{\bar{G}}}

\newcommand{\Zbar}{\mathbf{\bar{Z}}}
\newcommand{\zbar}{\mathbf{\bar{z}}}
\newcommand{\Ybar}{\bar{Y}}
\newcommand{\wbar}{\mathbf{\bar{w}}}

\newcommand{\boldW}{\mathbf{W}}
\newcommand{\boldX}{\mathbf{X}}
\newcommand{\boldx}{\mathbf{x}}
\newcommand{\boldO}{\mathbf{O}}
\newcommand{\boldo}{\mathbf{o}}

\newcommand{\hOne}{\frac{\mathbbm{1}(\Gbar_t=\gbar_t)}{\E[\mathbbm{1}(\Gbar_t=\gbar_t)]}}

\newcommand{\hZero}{\frac{\mathbbm{1}(\Gbar_t=\gbar_t'){\pi}(\boldx; \gbar_t)}{{\pi}(\boldx; \gbar_t') \mathrm{E}[(\mathbbm{1}(\Gbar_t=\gbar_t'){\pi}(\boldx; \gbar_t)/{\pi}(\boldx; \gbar_t')]}}

\newcommand{\hhone}{h_1(\Gbar_{it})}
\newcommand{\hhzero}{h_0(\Gbar_{it}, \boldX_i; \pi_i)}
\newcommand{\mmu}{\mu_{i,\gbar_t',t^*}(\boldx_i)}
\newcommand{\hhonehat}{\hat{h}_1(\Gbar_{it})}
\newcommand{\hhzerohat}{\hat{h}_0(\Gbar_{it}, \boldX_i; \hat{\pi}_i)}
\newcommand{\mmuhat}{\hat{\mu}_{i,\gbar_t',t^*}(\boldx_i)}
\newcommand{\deltay}{\Delta_{t^*}Y_i}
\newcommand{\pigX}{\pi_i(\boldX_i;\gbar_t)}
\newcommand{\pigprimeX}{\pi_i(\boldX_i;\gbar_t')}
\newcommand{\pighatX}{\hat{\pi}_i(\boldX_i;\gbar_t)}
\newcommand{\pigprimehatX}{\hat{\pi}_i(\boldX_i;\gbar_t')}
\newcommand{\pigx}{\pi_i(\boldx_i;\gbar_t)}

\newcommand{\pig}{\pi_i(\boldX_i;\gbar_t)}
\newcommand{\pigprime}{\pi_i(\boldX_i;\gbar_t')}
\newcommand{\pighat}{\hat{\pi}_i(\boldX_i;\gbar_t)}
\newcommand{\pigprimehat}{\hat{\pi}_i(\boldX_i;\gbar_t')}

\newcommand{\Ig}{\mathbbm{1}(\Gbar_{it}=\gbar_t)}
\newcommand{\Igprime}{\mathbbm{1}(\Gbar_{it}=\gbar_t')}
\newcommand{\pgone}{p_{i1}(\gbar_t)}
\newcommand{\pgonehat}{\hat{p}_{i1}(\gbar_t)}
\newcommand{\pgtwo}{p_{i2}(\pi_i, \gbar_t, \gbar_t')}
\newcommand{\pgtwohat}{\hat{p}_{i2}(\hat{\pi}_i, \gbar_t, \gbar_t')}

\newcommand{\sumin}{\sum_{i=1}^{n}}
\newcommand{\meanin}{n^{-1}\sum_{i=1}^{n}}

\newcommand{\indep}{\protect\mathpalette{\protect\independenT}{\perp}}
\def\independenT#1#2{\mathrel{\rlap{$#1#2$}\mkern2mu{#1#2}}}

\newcommand*\circled[1]{\tikz[baseline=(char.base)]{
            \node[shape=circle,draw,inner sep=.5pt] (char) {#1};}}

\newtheorem{theorem}{Theorem}
\newtheorem{proposition}{Proposition}
\newtheorem{assumption}{Assumption}

\newtheorem{definition}{Definition}

\usepackage[
backend=biber,
style=authoryear,
natbib=true
]{biblatex}

\AtEveryBibitem{\clearfield{url}}
\AtEveryBibitem{\clearfield{urlyear}\clearfield{urlmonth}\clearfield{urlday}}
\AtEveryBibitem{\clearfield{note}}
\AtEveryBibitem{\clearfield{month}}
\AtEveryBibitem{\clearlist{language}}

\allowdisplaybreaks

\title{Efficient nonparametric estimation with difference-in-differences in the presence of network dependence and interference}
\author{Michael Jetsupphasuk\thanks{Corresponding author. Email: jetsupphasuk@unc.edu}, Didong Li, Michael G. Hudgens}
\date{}

\affil{Department of Biostatistics, University of North Carolina at Chapel Hill}

\begin{document}

\maketitle

\vspace{-3em}

\begin{abstract}
    Differences-in-differences (DiD) is a causal inference method for observational longitudinal data that assumes parallel expected potential outcome trajectories between treatment groups under the counterfactual scenario where all units receive a specific treatment. In this paper DiD is extended to allow for: (i) non-identically distributed treatment effects and exposure probabilities; (ii) interference, where treatment of one unit can affect outcomes in neighboring units; and (iii) latent variable dependence, where outcomes, treatments, and covariates may exhibit between-unit correlation. The causal estimand of interest is the network-averaged expected exposure effect if units received a specific exposure level, where a unit's exposure is a function of its own treatment and its neighbors' treatments. Under a conditional parallel trends assumption and suitable network dependency and heterogeneity conditions, a doubly robust estimator allowing for data-adaptive nuisance function estimation is proposed and shown to be consistent, asymptotically normal, and efficient. The proposed methods are evaluated in simulations and applied to study the effects of adopting emission control technologies in coal power plants on county-level mortality due to cardiovascular disease. 
\end{abstract}

\section{Introduction}

\subsection{Background}

Differences-in-differences (DiD) is a method to estimate causal effects in observational studies that relies on a parallel trends assumption. Under the canonical set-up, there is a treated group and an untreated group with the outcome measured at two time periods where treatment only occurs after the first time period \citep{roth_whats_2023}. The parallel trends assumption stipulates that the average outcome in the treated and untreated groups would have changed by the same amount between time periods, under the counterfactual scenario where neither group received the treatment. DiD allows for the identification and estimation of causal effects in the absence of treatment randomization and has been used in many fields, such as in studying the effects of contaminated water on cholera incidence \citep{snow_mode_1855}, minimum wage laws on unemployment \citep{card_minimum_1994}, employment protection on productivity \citep{autor_does_2007}, and Medicare expansion on mortality and medical spending \citep{finkelstein_what_2008}. 

Most DiD methods assume independent and identically distributed (iid) data, which may not be appropriate when data are dependent. In this paper, methods are developed that allow for different types of heterogeneity and dependencies that may occur when units are connected within a network. In such settings there may be treatment effect and exposure probability heterogeneity based on position within the network. For instance, the probability that a state-level policy is enacted may differ in California compared to North Carolina and the effect of that policy may also differ between the two states. This paper adopts a finite population approach where the causal estimand is an empirical average of potentially heterogeneous unit-level treatment effects. 

Dependencies are also present with network data. Dependence may arise due to interference, whereby the treatment status in one unit (e.g., a county or state) may have effects on neighboring units. Interference may be present in settings where DiD methods are often employed, such as in the study of place-based interventions. For example, a tax instituted in a particular county may not only affect consumer behavior in the county receiving the new tax but also affect consumers in neighboring counties \citep{hettinger_estimation_2023}. Methods that accommodate interference often assume a particular form of interference structure. In settings where units form natural, non-overlapping clusters, it is common to assume clustered interference, where interference may exist within clusters but there is no interference between clusters. In other settings, there may be network interference, where treatments in any particular unit may affect outcomes in other units according to a network structure \citep{halloran_dependent_2016}. 

Data of units that are close within a network may exhibit dependence or correlation for reasons beyond interference. For example, there may be latent variable dependence whereby outcomes in one unit are correlated with outcomes from neighboring units through shared unobserved variables \citep{ogburn_causal_2022}. Such latent variable dependence may exist, for example, if health outcomes (e.g., all-cause mortality) measured at the county-level are correlated across counties due to unobserved environmental pollutants that affect neighboring counties similarly. Latent variable dependence may also be present for treatments and covariates. In studies of social networks, correlation between person-level data is often exhibited through homophily, where peers connected in a social network tend to share similar characteristics \citep{morgan_social_2013}. Certain data settings may exhibit any combination of interference, latent variable dependence, and homophily, along with other data dependencies.

In addition to non-iid data, in some settings another challenge is posed when interference takes on a bipartite structure, where outcomes and treatments are measured on different types of units and multiple treatment units may affect the potential outcomes of each outcome unit \citep{zigler_bipartite_2021}. Bipartite interference is particularly relevant in environmental health since outcome data are often defined on the person-level (or some aggregate, such as the census tract or county-level) while interventions are performed on the environment; for example, regulations on air or water quality. Causal estimands of interest under bipartite interference may differ from estimands in the standard interference setting since under the bipartite setting, there may not be a single treatment unit corresponding to a particular outcome unit, complicating definitions of direct and spillover effects \citep{zigler_bipartite_2021}. To distinguish from the bipartite structure, henceforth ``unipartite" is used to refer to the standard setting where outcomes and treatments are defined on the same units. 

This paper builds on recent methodological work regarding DiD with nonparametric, doubly robust estimation and the intersection of DiD with interference. In the iid setting, \citet{santanna_doubly_2020} and  \citet{chang_doubledebiased_2020} proposed a doubly robust estimator of the average treatment effect on the treated (ATT) under a conditional parallel trends assumption. This estimator was shown to be consistent, asymptotically normal, and nonparametric efficient under certain regularity conditions. At the intersection of DiD and interference, several papers assumed two-way fixed effects (TWFE) models where the outcome has a known structural relationship with treatments after adjusting for individual and time fixed effects \citep{clarke_estimating_2017, butts_difference--differences_2021, fiorini_simple_2024}. \citet{hettinger_doubly_2025} considered DiD under interference and spatial correlation, proposing a doubly robust estimator based on a correctly specified exposure mapping; they implemented a multiplier block bootstrap method to conduct inference while allowing for spatial correlation but did not provide a theoretical justification for their proposed method. \citet{shahn_structural_2024} discussed structural nested mean models under parallel trends allowing for clustered or network interference. \citet{xu_difference--differences_2025} considered DiD under similar network dependency conditions as this paper but targeted a different estimand and relied on parametric nuisance function estimators for their doubly robust estimator. 

\subsection{Contribution}

In this paper, a DiD method is developed which can accommodate network dependency including interference and latent variable dependence. The target estimand is defined on a finite population, such that inference is specific to the study sample and observed network. The proposed framework also allows for unit-specific heterogeneity in exposure effects and exposure probabilities (unconditional on covariates). The bipartite interference setting is considered, which includes the unipartite setting as a special case. The proposed doubly robust estimator adapts the estimator introduced by \citet{santanna_doubly_2020} from the iid data setting to the network dependent data setting. The proposed estimator is doubly robust in the sense that if either the outcome regression or propensity score nuisance functions are correctly specified, then the estimator is consistent. When both nuisance functions are correctly specified, the proposed estimator is shown to be asymptotically normal and nonparametric efficient under certain types of network dependencies and a set of sufficient conditions that allow for data-adaptive nuisance function estimators.

The proposed methods are utilized to estimate the effect of emission control technologies in coal power plants on mortality. Coal power plants emit sulfur dioxide (SO$_2$) which interacts with the atmosphere and breaks down to particular matter less than 2.5 microns in diameter (PM2.5). Exposure to PM2.5 may cause increased risk of some cardiovascular diseases (CVDs), where there is evidence based on biological pathways \citep{shkirkova_effects_2020} (e.g., via oxidative stress in heart tissues) and observational epidemiology studies \citep{danesh_yazdi_effect_2022, thurston_ischemic_2016, fann_estimating_2012}. In the motivating data application, the treatment is the implementation of flue-gas desulfurization scrubbers in coal power plants. Scrubbers are an emission control technology that help limit the amount of SO$_2$ emitted. The outcome is annual county-level mortality rate due to CVDs. Bipartite interference may be present since intervention and outcome units differ and atmospheric conditions (e.g., weather patterns) can transport emissions across counties such that the CVD mortality rate for a particular county may depend on scrubber installation in a distant power plant located in a different county. 

The remainder of this paper is organized as follows. Section \ref{sec:methods} introduces notation, defines the causal estimand of interest, provides sufficient conditions for identification, proposes estimators, and derives the large sample properties of the proposed estimators. Section \ref{sec:simulation} evaluates properties of the proposed estimators under simulated finite samples. Section \ref{sec:application} applies the proposed methods to the motivating data application. Section \ref{sec:discussion} concludes and discusses future work.

\section{Methods} \label{sec:methods}

\subsection{Notation and potential outcomes}

Considering the bipartite setting, let $i=1,\dots,n$ index the outcome units and $j=1,\dots,m$ index the intervention (treatment) units. In the data application below, $i$ indexes counties and $j$ indexes power plants. The unipartite setting is a special case where $i=j$ and $n=m$. Time periods are indexed by $t = 0,\dots,T$ where all units are untreated at $t=0$. At time $t$, intervention unit $j$ receives treatment $Z_{jt}$ which may be multi-valued or continuous, and the outcome $Y_{it}$ is measured on outcome unit $i$. Let $z_{jt} \in \mathcal{Z} \subseteq \mathbb{R}$ denote realizations of $Z_{jt}$. Throughout this paper, the notation is adopted that boldface denotes vectors or matrices and overbars denote histories, e.g., $\mathbf{Z}_t = (Z_{1t},\dots,Z_{mt})^{\top}$ is the vector of treatments for all intervention units at time $t$, and $\mathbf{\bar{Z}}_{s:t} = (\mathbf{Z}_s,\dots,\mathbf{Z}_t)$ is the $m \times (t-s+1)$ matrix of treatment histories for all intervention units. For simplicity, also let $\mathbf{\bar{Z}}_{t} = \mathbf{\bar{Z}}_{0:t}$. Realizations of treatment histories $\mathbf{\bar{z}}_t$ are defined similarly. 

Under bipartite interference, the potential outcomes for the outcome units are defined as a function of all intervention units' entire treatment histories for the study period and are denoted by $Y_{it}(\mathbf{\bar{z}}_{T})$ where $\mathbf{\bar{z}}_{T}$ is an $m \times (T+1)$ matrix of treatment histories for all $m$ intervention units up to time $T$. To relate potential outcomes to observed outcomes $Y_{it}$, the following form of causal consistency is assumed.

\begin{assumption}[Causal consistency]
    \label{assump:consistency}
    If $\mathbf{\bar{Z}}_T = \mathbf{\bar{z}}_T$, then $Y_{it} = Y_{it}(\mathbf{\bar{z}}_T)$ for all $i$ and $t$.
\end{assumption}
\noindent

The proposed methods rely on an assumption about the interference structure between outcome and treatment units. Specifically, assume that the interference structure can be described by $\boldW_t$, an $n \times m$ matrix of known interference weights with elements $w_{ijt} \in [0,1]$ that describe the amount of possible interference of the $j$th intervention unit to the $i$th outcome unit at time $t$. When $w_{ijt}=0$, the treatment of intervention unit $j$ is assumed to not impact the potential outcomes of outcome unit $i$, whereas $w_{ijt}>0$ allows for intervention unit $j$ to possibly affect outcome unit $i$ at time $t$. In some settings, it may be reasonable to specify the interference weights as binary, i.e., $w_{ijt} \in \{0,1\}$. For example, if outcome unit $i$ can only be impacted by intervention unit $j$, then $w_{ijt}=1$ and $w_{ikt}=0$ for all $k \neq j$. In other settings, an outcome unit may be impacted by multiple intervention units to varying degrees. In the motivating air pollution study, treatments at power plants in closer proximity to a particular county may have more influence on that county's CVD mortality rate compared to more distal power plants. Thus, it might be assumed that $w_{ijt} > w_{ikt} > 0$ if county $i$ is possibly affected by treatments at both power plants $j$ and $k$ but is closer to power plant $j$ than power plant $k$. When $w_{ijt}$ is continuous larger interference weights reflect greater relative possible influence. 

The interference set for outcome unit $i$ at time $t$ is defined as $\mathcal{I}_{it} = \{j: w_{ijt} \neq 0\}$, i.e., the collection of intervention units that have non-zero interference weights with outcome unit $i$. Define an exposure mapping $g(\mathbf{Z}_t; \mathbf{w}_{it})$ to be a surjective function from the vector of treatments for all intervention units at time $t$ and the vector of interference weights for outcome unit $i$ to a bounded, discrete real scalar, i.e., $g: \mathcal{Z}^m \times [0,1]^m \rightarrow \mathcal{G}$ where $\mathbf{w}_{it} = (w_{i1t}, \dots, w_{ijt}, \dots, w_{imt})^{\top}$ and $\mathcal{G}$ is a discrete set with cardinality $|\mathcal{G}|$ that does not depend on $n$. This broad definition includes many commonly used exposure mappings; for example, the weighted proportion of neighbors that were treated corresponds to $g(\mathbf{Z}_t; \mathbf{w}_{it}) = \sum_{j \in \mathcal{I}_{it}} w_{ijt} Z_{jt} \big/ \sum_{j \in \mathcal{I}_{it}} w_{ijt}$. Ideally, specification of the exposure mapping function and interference matrix would be based on domain-specific knowledge. For instance, in the data example presented in Section \ref{sec:application}, a scrubber installed in power plant $j$ is assumed to potentially affect health in county $i$ only if emissions from power plant $j$ can plausibly reach residents in county $i$. In particular, an atmospheric transport model relating power plant emissions to counties is used to define the interference matrix $\boldW_t$. Though the term ``exposure" is often synonymous with intervention or treatment, here ``exposure" specifically refers to an exposure mapping with $G_{it} := g(\mathbf{Z}_t;\mathbf{w}_{it})$ denoting the random exposure for outcome unit $i$ at time $t$. Also, with a slight abuse of notation let $\gbar(\zbar_t; \wbar_{it}) = (g(\mathbf{z}_1; \mathbf{w}_{i1}), \dots, g(\mathbf{z}_t; \mathbf{w}_{it}))^{\top}$ which may also be written as $\gbar_t$ when the context is clear. Further, let the random exposure histories be denoted $\Gbar_t$.

\begin{assumption}[Interference through exposure mapping]
\label{assump:interference}
    For all $i=1,\dots,n$ and $t=0,\dots,T$, if $\gbar(\zbar_T; \wbar_{iT}) = \gbar(\zbar'_T; \wbar_{iT})$ for any $\zbar_T$ and  $\zbar'_T$, then $Y_{it}(\zbar_T) = Y_{it}(\zbar'_T)$.
\end{assumption}

Assumption \ref{assump:interference} stipulates that potential outcomes depend on treatments only through the exposure mapping and therefore can be expressed in terms of the exposure histories $Y_{it}(\gbar_T)$. This notation is adopted for the remainder of the paper unless otherwise stated. 

Each outcome and intervention unit has a pre-treatment ($t=0$) covariate vector $\boldX_{i}^{\mathrm{out}}, i=1,\dots,n$ and $\boldX_{j}^{\mathrm{int}}, j=1,\dots,m$, respectively. In the unipartite setting, let the collection of covariates associated with outcome unit $i$ be denoted $\boldX_{i} = s_1(\{\boldX_{j}^{\mathrm{int}}\}_{j \in \mathcal{I}_{i0}})$ where $s_{1}(\cdot)$ is a user-specified function that maps intervention unit covariates to a possibly low dimensional space that does not depend on $i$. When there are many intervention units in the interference set of each outcome unit at time $0$, the dimensionality of $\{\boldX_{j}^{\mathrm{int}}\}_{j \in \mathcal{I}_{i0}}$ may be large. In these settings, the function $s_1(\cdot)$ may be useful to reduce dimensionality. For instance, one may consider a weighted average of intervention unit covariates with weights according to the interference matrix, i.e., $s_1(\{\boldX_{j}^{\mathrm{int}}\}_{j \in \mathcal{I}_{i0}}) = \left(\boldX_i^{\mathrm{int}}, \left( \sum_{j \in \mathcal{I}_{i0} \setminus i} w_{ij0} \right)^{-1} \sum_{j \in \mathcal{I}_{i0} \setminus i} w_{ij0} \boldX_{j}^{\mathrm{int}} \right)$. In the bipartite setting, let the collection of covariates associated with outcome unit $i$ be $\boldX_{i} = (s_1(\{\boldX_{j}^{\mathrm{int}}\}_{j \in \mathcal{I}_{i0}}), s_2(\{ \{\boldX_{k}^{\mathrm{out}}\}_{k \in \mathcal{I}^{*}_{j}} \}_{j \in \mathcal{I}_{i0}}))$, where $s_2(\cdot)$ is defined similarly as $s_1(\cdot)$ and $\mathcal{I}^{*}_j = \{i: w_{ij0} \neq 0 \}$.

For each outcome unit, the random data vector $\boldO_i = (\Ybar_{iT},\Gbar_{iT}, \boldX_{i})$ is observed. In the network dependent data setting, $\boldO_1,\dots,\boldO_n$ are not necessarily independent nor identically distributed. Instead, a network model may be assumed to describe dependency between $\boldO_i$ and $\boldO_k$ for $i \neq k$. Consider a size $n$ undirected network $U_n = (\mathcal{N}_n, \mathcal{E})$ where $\mathcal{N}_n = \{1,\dots,n\}$ is the set of nodes and $\mathcal{E}$ denotes the collection of edges between nodes. Each node $i \in \mathcal{N}_n$ is endowed with the corresponding data $\boldO_i$, and an edge connecting nodes $i$ and $k$ denotes possible dependence between $\boldO_i$ and $\boldO_k$. Assume the network is fixed and non-random, though the collection of edges is not necessarily known. In this sense, this paper adopts a finite population framework where the sample of units is considered the population of interest. The data vector $\boldO_i$ is considered a random function of the network $U_n$ for all $i$, and the observed data from all $n$ units are denoted $\boldO_{1:n} = (\boldO_1, \dots, \boldO_n)^{\top} \sim \mathbb{P}$. 

\subsection{Causal estimand}

The causal estimand of interest involves contrasts of potential outcomes under counterfactual exposure histories. Consider an exposure history $\gbar_t = (\gbar_{0:t^*}', \gbar_{(t^*+1):t})$ and a reference exposure history $\gbar_t' = (\gbar_{0:t^*}', \gbar_{(t^*+1):t}')$ for $0 \leq t^* < t$. Let the unit-specific average exposure effect if exposed (AEE) at level $\gbar_t \in \bar{\mathcal{G}}_t$ at time $t \in \mathcal{T}$ be denoted $\AEE_{it}(\gbar_t) = \mathrm{E} \left[ Y_{it}(\gbar_t) - Y_{it}(\gbar_t') \big| \Gbar_{it} = \gbar_t \right]$, where $\bar{\mathcal{G}}_t$ is the set of exposure histories of interest with elements $(\gbar_{0:t^*}', \cdot)$ and $\mathcal{T}$ is the set of time points of interest. All expectations are with respect to the data distribution $\mathbb{P}$ unless otherwise noted. Dependence of the reference exposure $\gbar_t'$ in $\AEE_{it}(\gbar_t)$ is left implicit, since studies typically do not vary $\gbar_t'$. Often, the reference exposure history is defined to be the absence of treatment, e.g., $\gbar_t' = (0,\dots,0)$, though other definitions may be reasonable based on the study context. The exposure effects $\AEE_{it}(\gbar_t)$ compare expected potential outcomes under the same exposure history up to time $t^*$ but differing thereafter. If $t^*$ is set to $t-1$, the estimand isolates the effect of a change in exposure in the time period $t$. 

The proposed causal estimand is the network-averaged exposure effect if exposed, which is defined as the empirical average of all unit-specific exposure effects in the network, 
\begin{align*} \label{estimand:att}
    \AEE_{t}(\gbar_t) &\coloneqq \meanin \mathrm{E} \left[ Y_{it}(\gbar_t) - Y_{it}(\gbar_t') \big| \Gbar_{it} = \gbar_t \right] \\
    &= \meanin \mathrm{E} \left[ \frac{\mathbbm{1}(\Gbar_{it} = \gbar_t)}{\PP(\Gbar_{it} = \gbar_t)} \{ Y_{it}(\gbar_t) - Y_{it}(\gbar_t') \} \right] ,
\end{align*}
where $\mathbbm{1}(\cdot)$ denotes the indicator function, i.e., $\AEE_{t}(\gbar_t)$ is the expected effect at time $t$ of exposure history $\gbar_t$ relative to $\gbar_t'$ if exposed to $\gbar_t$, averaged over units $i \in \mathcal{N}_n$.

Let ``network effect heterogeneity" be defined as heterogeneity of $\AEE_{it}(\gbar_t)$ across units $i \in \mathcal{N}_n$, i.e., network effect heterogeneity is present if $\AEE_{it}(\gbar_t) \neq \AEE_{kt}(\gbar_t)$ for at least one pair $(i,k)$ where $i \neq k$. Network effect heterogeneity is distinct from effect heterogeneity due to observed covariates $\boldX_i$ or exposure groups $\Gbar_{it}$. Similarly, there may be ``network exposure probability heterogeneity" if $\PP(\Gbar_{it} = \gbar_t)$ varies across units $i \in \mathcal{N}_n$. Let ``network effect homogeneity" and ``network exposure probability homogeneity" be defined as the absence of heterogeneity in effects or exposure probabilities, respectively. 

$\AEE_{it}(\gbar_t)$ reduces to the classic ATT in the iid data setting with two time periods, two treatments $z \in \{0,1\}$, and $\gbar_t = (0,1)$ and $\gbar_t' = (0,0)$. The estimand $\AEE_{it}(\gbar_t)$ also reduces to the group-time average treatment effect parameter introduced in \citet{callaway_difference--differences_2021} when there is no network effect heterogeneity, there are two treatments, $\gbar_t' = (0,\dots,0)$, and $\gbar_t = (0,\dots,0,1,\dots,1)$ where treatment groups are specified by the timing of the change from $0$ to $1$ in $\gbar_t$.

\subsection{Identification}

In this section, the AEE is shown to be identifiable under Assumptions \ref{assump:consistency} -- \ref{assump:interference} and the following three assumptions of no anticipation, positivity, and conditional parallel trends. No anticipation in Assumption \ref{assump:no_anticip} states that potential outcomes at time $t$ do not depend on treatments at times $s>t$. In other words, potential outcomes do not vary based on treatments occurring in the future. Accordingly, potential outcomes at time $t$ can be written as depending on treatment history up to time $t$ only, i.e., $Y_{it}(\zbar_t)$ or $Y_{it}(\gbar_t)$ under Assumption \ref{assump:interference}.

\begin{assumption}[No anticipation]
\label{assump:no_anticip}
        $Y_{it}((\mathbf{\bar{z}}_t, \zbar_{(t+1):T})) = Y_{it}((\mathbf{\bar{z}}_t, \zbar_{(t+1):T}'))$ for any $\zbar_{(t+1):T}, \zbar_{(t+1):T}'$. 
\end{assumption}

Under Assumption \ref{assump:positivity}, the two exposure histories being compared in the causal estimand must have a positive probability of occurring. Note that a similar positivity assumption on the intervention unit treatments $\mathbf{Z}$ is not needed. 

\begin{assumption}[Positivity of exposure history]
    \label{assump:positivity}
    There exists $\epsilon > 0$ such that for all $i=1,\dots,n$, $\gbar_t \in \bar{\mathcal{G}}_t$, and $t \in \mathcal{T}$, $\mathrm{P}(\Gbar_{it} = \gbar_t| \boldX_i) > \epsilon \text{ and } \mathrm{P}(\Gbar_{it} = \gbar_t'| \boldX_i) > \epsilon$.
\end{assumption}

The conditional parallel trends assumption in Assumption \ref{assump:parallel} states that the expected trajectories of potential outcomes under the reference exposure $\gbar_t'$ is the same, up to a weighted average, regardless whether the exposure is $\gbar_t \in \bar{\mathcal{G}}_t$ or $\gbar_t'$, conditional on covariates. 
\begin{assumption}[Conditional parallel trends]
\label{assump:parallel}
    For all $\gbar_t \in \bar{\mathcal{G}}_t$, and $t \in \mathcal{T}$,
    \begin{align*}
        &\meanin \frac{\Ig}{\PP(\Gbar_{it}=\gbar_t)} \mathrm{E}[Y_{it}(\gbar'_t) - Y_{it^*}(\gbar'_t) | \boldX_i, \Gbar_{it} = \gbar'_t] \\
        &=  \meanin \frac{\Ig}{\PP(\Gbar_{it}=\gbar_t)} \mathrm{E}[Y_{it}(\gbar'_t) - Y_{it^*}(\gbar'_t) | \boldX_i, \Gbar_{it} = \gbar_t].
    \end{align*}
\end{assumption}
\noindent
A stronger version of Assumption \ref{assump:parallel} could be imposed which assumes that parallel trends holds for every $i$. However, Assumption \ref{assump:parallel} is substantially weaker. A particular unit $i$'s expected potential trajectories need not be the same conditional on receiving different exposure histories. Instead, Assumption \ref{assump:parallel} stipulates that the trajectories, weighted by the inverse probability of receiving the exposure history $\gbar_t$, average out to being equal, only among those that were observed to receive exposure history $\gbar_t$. If the exposure probabilities are homogeneous in $i$, then Assumption \ref{assump:parallel} only requires that conditional parallel trends holds on average, among the units that received exposure $\gbar_t$. 

In the absence of interference, Assumption \ref{assump:parallel} generalizes the parallel trends assumption in \citet{callaway_difference--differences_2021} from the staggered adoption setting (where treatments are binary and irreversible once received) to generic treatment histories. A special case of Assumption \ref{assump:parallel} is the classic conditional parallel trends assumption as in \citet{abadie_semiparametric_2005} where there are two time periods $t \in \{0,1\}$, and the exposures are $\gbar_1'=(0,0)$ and $\gbar_1=(0,1)$. Assumption \ref{assump:parallel}, as the main identifying assumption, shows that the reference exposure history $\gbar_t'$ should be chosen carefully.

Let $\mu_{i,\gbar_t,t^*}(\boldx) := \mathrm{E}[Y_{it} - Y_{it^*} | \boldX_i = \boldx, \Gbar_{it}=\gbar_t]$. Denote the exposure propensity score by $\pi_i(\boldx;\gbar_t) := \mathrm{P}(\Gbar_{it}=\gbar_t|\boldX_i=\boldx)$. The conditional mean outcomes and exposure propensity scores are indexed by $i$ since in the network dependent setting, it is not necessarily the case that $\mu_{i,\gbar_t,t^*}(\boldx) = \mu_{k,\gbar_t,t^*}(\boldx)$ for $i \neq k$, and similarly for the exposure propensity score. 

The AEE is identifiable by Proposition \ref{prop:dr} under Assumptions \ref{assump:consistency} -- \ref{assump:parallel} (all proofs are provided in the Supplementary Material). 

\begin{proposition}
    \label{prop:dr}
    Let $\tau_i(\boldO_{i}) = (h_{i1}(\Gbar_{it}) - h_{i0}(\Gbar_{it},\boldX_i;\pi_i))(\Delta_{t^*} Y_{it} - \mu_{i,\gbar_t',t^*}(\boldX_i))$ where $h_{i1}(\Gbar_{it}) = \mathbbm{1}(\Gbar_{it}=\gbar_t) [\pgone]^{-1}$, $h_{i0}(\Gbar_{it},\boldX_i;\pi_i) = \mathbbm{1}(\Gbar_{it}=\gbar_t')\pi_i(\boldX_i; \gbar_t) [\pgtwo \pi_i(\boldX_i; \gbar_t')]^{-1} $, $\pgone = \mathrm{P}(\Gbar_{it}=\gbar_t)$, $\pgtwo = \E \left[ \mathbbm{1}(\Gbar_{it}=\gbar_t')\pi_i(\boldX_i; \gbar_t) \{\pi_i(\boldX_i; \gbar_t')\}^{-1} \right]$, and $\Delta_{t^*} Y_{it} = Y_{it} - Y_{it^*}$. 
    If Assumptions \ref{assump:consistency} -- \ref{assump:parallel} hold, then
    \begin{align*}
        \tau(\gbar_t, \gbar_t', t, t^*) := \meanin \E[\tau_i(\boldO_i)] = \AEE_{t}(\gbar_t).
    \end{align*}
\end{proposition}  

\noindent
Note that if data were iid, the statistical estimand $\tau(\gbar_t, \gbar_t', t, t^*)$ in Proposition \ref{prop:dr} is equivalent to the estimand in \citet{santanna_doubly_2020}. For notational simplicity, $\tau(\gbar_t, \gbar_t', t, t^*)$ will be denoted by $\tau$, with dependency on $\gbar_t, \gbar_t', t,$ and $ t^*$ left implicit. 

\subsection{Estimation} \label{sec:methods-estimation}

In this section an estimator $\hat{\tau}$ of $\tau$ is proposed. In particular, consider the plug-in estimator $\hat{\tau} := \meanin \hat{\tau}_i(\boldO_i)$, where $\hat{\tau}_i(\boldO_i) = (\hat{h}_{i1}(\Gbar_{it}) - \hat{h}_{i0}(\Gbar_{it},\boldX_i;\hat{\pi}_i)) (\Delta_{t^*} Y_{it} - \hat{\mu}_{i,\gbar_t',t^*}(\boldX_i))$, $\hat{h}_{i1}(\Gbar_{it}) = \mathbbm{1}(\Gbar_{it}=\gbar_t) [\pgonehat]^{-1}$, 
$\hat{h}_{i0}(\Gbar_{it},\boldX_i;\hat{\pi}_i) = \mathbbm{1}(\Gbar_{it}=\gbar_t')\hat{\pi}_i(\boldX_i; \gbar_t) [\pgtwohat \hat{\pi}_i(\boldX_i; \gbar_t')]^{-1}$, and in general $\hat q$ denotes an estimator of $q$. Consider the case when there is either network effect homogeneity or network exposure probability homogeneity. In this setting, $\hat \tau$ is equivalent to an efficient influence function (EIF) based estimator with a one-step bias correction. The EIF of $\tau$, given in Proposition \ref{prop:eif}, characterizes efficiency of nonparametric estimators of $\tau$ in the sense that the variance of the EIF is the nonparametric efficiency bound, i.e., the greatest lower bound for regular and asymptotically linear (RAL) estimators of $\tau$ under nonparametric models \citep{kennedy_semiparametric_2023}. In the following section, $\hat \tau$ is shown to be nonparametric efficient under suitable conditions.

\begin{proposition}
    \label{prop:eif}
    Suppose a nonparametric model $\mathcal{P}$ such that $\mathbb{P} \in \mathcal{P}$. If there is either network effect homogeneity or exposure probability homogeneity, then the statistical functional $\tau$ is pathwise differentiable with efficient influence function $\phi^*(\boldO_{1:n}; \mathbb{P}) = \meanin \phi_i(\boldO_i, \mathbb{P})$, where $\phi_i(\boldO_i, \mathbb{P}) = \tau_i(\boldO_i) - h_{i1}(\Gbar_{it}) \tau$. 
\end{proposition}

The estimator $\hat \tau$ involves estimation of several nuisance functions, such as the exposure probabilities $\pgone$. In the iid setting \citet{santanna_doubly_2020} proposed that $\pgone$ be estimated nonparametrically using an empirical average, i.e., $\hat{p}_{i1}^{\mathrm{emp}}(\gbar_t) \coloneqq \meanin \mathbbm{1}(\Gbar_{it}=\gbar_t)$, since the empirical average is consistent for the true exposure probability by the law of large numbers. If $\Gbar_{it}$ is weakly dependent (as defined in the next section) across units, then the consistency result $\hat{p}_{i1}^{\mathrm{emp}}(\gbar_t) - \meanin \pgone \rightarrow_p 0$ holds, regardless if there is exposure probability heterogeneity. However, when there is heterogeneity in both network effects and network exposure probabilities, $\hat \tau$ may be biased if $\hat{p}_{i1}^{\mathrm{emp}}(\gbar_t)$ is chosen as an estimator of $\pgone$. The bias is due to network confounding, where features of the network are correlated with both exposure probability and potential outcomes. To ameliorate this bias one may posit a model of the form $\pgone = f(U_n, i, \gbar_t)$, where $f$ is some summary of the network $U_n$ and position in the network $i$ for treatment history $\gbar_t$. Let $\hat{p}_{i1}^{\mathrm{net}}(\gbar_t)$ be some estimator of the model $\pgone = f(U_n, i, \gbar_t)$. One example of $f(U_n, i, \gbar_t)$ in the $T=1$ setting with $\gbar_1 = (0,1)$ is $p_{i1}(\gbar_1) = \mathrm{logit}^{-1}(\beta_0 + \beta n^{-1}\sum_{j=1}^{n} w_{ij1})$, i.e., a logistic regression on a summary of the interference matrix. The implications on inference from choosing $\hat{p}_{i1}^{\mathrm{emp}}(\gbar_t)$ or $\hat{p}_{i1}^{\mathrm{net}}(\gbar_t)$ are discussed in the next section. Note that assuming a parametric model for $\pgone$ implies a semiparametric model (assuming that no parametric assumptions are made on $\pi$ or $\mu$) so Proposition \ref{prop:eif} does not apply.

The outcome regression $\mmu$ and exposure propensity score $\pigx$ are other nuisance functions that must be estimated in $\hat \tau$. Network heterogeneity may exist for $\mmu$ and $\pigx$ (hence the indexing of the parameters by $i$); thus, models for these nuisance functions should include network features to account for this heterogeneity. The exposure propensity score $\pi_i(\boldX_i;\gbar_t) = \mathrm{P}(\Gbar_{it} = \gbar_t | \boldX_i)$ may be modeled directly, or alternatively, the treatment propensity score $\mathrm{P}(\Zbar_{t} = \zbar_{t} | \boldX_i)$ may be modeled first, followed by Monte Carlo integration. For example, in the unipartite setting, consider the scenario where $\Zbar_{jt} \indep \Zbar_{lt} | \boldX_i$ for $j \neq l$ and $j,l \in \mathcal{I}_{i0}$ and $\Zbar_{jt} \indep  \{ \boldX_l^{\mathrm{int}} \}_{l \neq j}$. Then, the exposure propensity score can be expressed as the following integral:
\begin{align*}
    \mathrm{P}(\Gbar_{it} = \gbar_{it} | \boldX_i) &= \int \mathbbm{1}(\gbar(\zbar_t; \wbar_{it}) = \gbar_{it}) \prod_{j \in \mathcal{I}_{it}} \mathrm{dF}(\Zbar_{jt} = \zbar_t | \boldX_j^{\mathrm{int}}),
\end{align*}
where $\mathrm{F}(\cdot)$ denotes the cumulative distribution function. Then, a Monte Carlo estimate of $\mathrm{P}(\Gbar_{it} = \gbar_{it} | \boldX_i)$ can be constructed by sampling from the estimated distribution of $(\Zbar_{jt}| \boldX_j^{\mathrm{int}})_{j \in \mathcal{I}_{it}}$ and taking the empirical average of the exposure indicator function computed from the samples. In the bipartite setting, replace $\boldX_j^{\mathrm{int}}$ in the above integral with $(\boldX_j^{\mathrm{int}}, \{\boldX_l^{\mathrm{out}}\}_{l \in \mathcal{I}^{*}_j})$ under the same conditional independence assumptions in addition to $\Zbar_{jt} \indep \{\boldX_l^{\mathrm{out}}\}_{l \notin \mathcal{I}^{*}_j}$.

\subsection{Inference} \label{sec:methods-inference}

In this section, sufficient conditions are provided to show that the proposed estimator of the AEE is root-$n$ consistent and asymptotically normal (CAN). A variance estimator is also proposed which is shown to be consistent under network effect homogeneity. 

Define the metric $d_n(i,k)$ to be the path distance between any two nodes $i,k \in \mathcal{N}_n$, where a path is defined as a sequence of edges connecting two nodes and path distance is defined as the shortest such sequence. Let $d_n(i,k) = \infty$ if there is no path connecting nodes $i$ and $k$ and $d_n(i,i)=0$. In the network model, covariance between data $\boldO_i$ and $\boldO_k$ is assumed to be a function of $d_n(i,k)$. Consider a sequence of network dependent processes $\{(\boldO_{1:n}, U_n)\}_{n \geq 1}$ as $n \rightarrow \infty$. In this section, asymptotic theory considers unweighted networks with path distance as the proximity metric governing dependency. However, the results hold for weighted networks with other proximity metrics such as the weighted path distance.

Assumption \ref{assump:lip_nuis} imposes a smoothness requirement on the nuisance functions. Since the composition of Lipschitz functions is also Lipschitz, Assumption \ref{assump:lip_nuis} implies that $\tau_i(\boldO_i)$ is a Lipschitz function of the data $\boldO_i$. 

\begin{assumption}[Smoothness of exposure propensity score and outcome regression]
    \label{assump:lip_nuis}
    The functions $\pi_i(\boldx_i;\gbar_t)$, $\pi_i(\boldx_i;\gbar_t')$, and $\mu_{i,\gbar_t,t^*}(\boldx_i)$ are Lipschitz functions of $\boldx_i$. 
\end{assumption}

Assumption \ref{assump:bounded} imposes a bound on the outcomes $Y_{it}$ and covariates $\boldX_i$, where $\|f\|_{L_2(\mathbb{P})}^2 = \int f(\boldo)^2 d\mathbb{P}(\boldo)$ denotes the squared $L_2(\mathbb{P})$ norm. 

\begin{assumption}[Boundedness]
    \label{assump:bounded}
    For $n \geq 1$, $\sup_{i \in \mathcal{N}_n, t \in \mathcal{T}} |Y_{it}(\zbar)| \leq Y^{\mathrm{max}} < \infty$ and \\
    $\sup_{i \in \mathcal{N}_n, t \in \mathcal{T}} ||\boldX_i||^2_{L_2(\mathbb{P})} < \infty$.
\end{assumption}

\noindent
Next, define the collection of two sets of nodes of sizes $a$ and $b$ with distance at least $s$ as $\mathcal{C}_n(a,b;s) = \{(A,B):A,B \subset \mathcal{N}_n, |A|=a, |B|=b, d_n(A,B) \geq s \},$ where $d_n(A,B) = \min_{i \in A, k \in B} d_n(i,k)$ is the shortest distance connecting a node in $A$ to a node in $B$. Then, following \citet{kojevnikov_limit_2021}, a notion of weak dependence called $\psi$-dependence is adopted, defined in Definition \ref{def:psi} where $\mathcal{L}_{\nu,a}$ is the set of real-valued Lipschitz functions $\{f: \mathbb{R}^{\nu \times a} \rightarrow \mathbb{R}: ||f||_{\infty} < \infty, \mathrm{Lip}(f) < \infty \}$ where $\|f\|_{\infty} = \sup_{\mathbf{q} \in \mathbb{R}^{\nu \times a}} |f(\mathbf{q})|$ and $\mathrm{Lip}(\cdot)$ is the Lipschitz constant.

\begin{definition}[Weak dependence \citep{kojevnikov_limit_2021}]
    \label{def:psi}
    Consider a triangular array $\{R_{n,i}\}_{i \in \mathcal{N}_n, n \geq 1}$, $R_{n,i} \in \mathbb{R}^{\nu}$. Let the collection of random variables $R_{n,i}$ in the set $A \subset \mathcal{N}_n$ be denoted $R_{n,A} = (R_{n,i})_{i \in A}$. Define the constants $\{\theta_{n,s}\}_{s \geq 0}$ where $\theta_{n,0} = 1$ and $\sup_n \theta_{n,s} \rightarrow 0$ as $s \rightarrow \infty$, and define the functionals $\{ \psi_{a,b} \}_{a,b \in \mathbb{N}}$ where $\psi_{a,b}: \mathcal{L}_{\nu,a} \times \mathcal{L}_{\nu,b} \rightarrow [0,\infty)$. Then, $\{R_{n,i}\}_{i \in \mathcal{N}_n, n \geq 1}$ is $\psi$-dependent if for all $n$, $(A, B) \in \mathcal{C}_n(a,b;s)$, $s > 0$, $f \in \mathcal{L}_{\nu,a}$, and $f' \in \mathcal{L}_{\nu,b}$,
    \begin{align*}
        |\Cov(f(R_{n,A}), f'(R_{n,B}))| \leq \psi_{a,b}(f,f') \theta_{n, s}.
    \end{align*}
\end{definition}

Definition \ref{def:psi} bounds the dependence of any two sets of data up to a functional term and constant that tends to zero as distance increases. In other words, nodes should have minimal dependence with nodes far away with respect to the distance metric. Assumption \ref{assump:psi-dep} assumes that the network dependent process $\{\boldO_i \}_{i \in \mathcal{N}_n}$ fulfills $\psi$-dependence and is the same as Assumption 2.1 in \citet{kojevnikov_limit_2021}. Further, $\tau_i(\boldO_i)$ is also $\psi$-dependent due to Assumption \ref{assump:lip_nuis}. 

\begin{assumption}[Weak dependence \citep{kojevnikov_limit_2021}]
    \label{assump:psi-dep}
    The triangular array $\{\boldO_i\}_{i \in \mathcal{N}_n, n \geq 1}$ is weakly dependent as defined in Definition \ref{def:psi} where $\psi_{a,b}(f,f') \leq C a b (\|f\|_{\infty} + \mathrm{Lip}(f))(\|f'\|_{\infty} + \mathrm{Lip}(f'))$ for some constant $C$ and $\sup_{n \geq 1} \max_{s \geq 1} \theta_{n,s} < \infty$ almost surely. 
\end{assumption}

\noindent
Weak dependence allows for different forms of network dependency, including dependency imposed by interference or latent variable dependence. For instance, suppose there is latent variable dependence in the outcomes $\{\deltay \}_{i \in \mathcal{N}_n}$ but no other network dependencies. Assumption \ref{assump:psi-dep} imposes that the covariance between outcomes $\deltay$ and $\Delta_{t^*} Y_k$, $i \neq k$, decays as the network distance between $i$ and $k$ increases. As discussed in \citet{kojevnikov_limit_2021}, many network dependent processes fulfill $\psi$-dependence. For example, define $\mathcal{N}_{n}(i, s) = \{k \in \mathcal{N}_n: d_n(i,k) < s\}$ as the set of units within $s$ distance of unit $i$. Then, the dependency structure termed $K$-locality states that data corresponding to a node $i$ depend only on data in other nodes within its $K$-neighborhood, $\mathcal{N}_{n}(i, K)$, for fixed $K$ that does not grow with $n$ \citep{leung_weak_2019}. In this scenario, $\psi$-dependence can be shown to be fulfilled with $\psi_{a,b}(f,f') = 2\|f\|_{\infty}\|f'\|_{\infty}$ and $\theta_{n,s} = \mathbbm{1}(s \leq 2\max\{K,1\})$ for all $n \in \mathbb{N}$ and $s > 0$.

Next, an assumption is made to restrict the density of the network as $n \rightarrow \infty$. Define the $s$-neighborhood shell of node $i$ to be the set of units exactly $s$ distance away from $i$, i.e., $\mathcal{N}_{n}^{\partial}(i, s) = \{k \in \mathcal{N}_n: d_n(i,k) = s\}$. Denote $M_{n}^{\partial}(s) = n^{-1}\sumin |\mathcal{N}_{n}^{\partial}(i, s)|$ as the average size of $s$-neighborhood shells. Network sparsity is imposed to limit the rate at which the average $s$-neighborhood shell sizes grow. In particular, Assumption \ref{assump:sparsity} stipulates that as $n$ increases, the dependence coefficient $\theta_{n,s}$ must decay to $0$ at a suitable rate compared to the network density, as represented by $M_{n}^{\partial}(s)$. 

\begin{assumption}[Asymptotic sparsity]
    \label{assump:sparsity}
    $\sum_{s=0}^{n} M_{n}^{\partial}(s) \theta_{n,s} = o(n)$.
\end{assumption}

Theorem \ref{theorem:consistency} shows that the estimator $\hat{\tau}$ is doubly robust in the sense that if either the propensity score or outcome regression nuisance models are consistently estimated, then the estimator converges in probability to the AEE. The nuisance function estimators are allowed to be data-adaptive and nonparametric, as long as the convergence conditions hold. 

\begin{theorem}
    \label{theorem:consistency}
    Let $\pgonehat = \hat{p}_{i1}^{\mathrm{emp}}(\gbar_t)$. If Assumptions \ref{assump:consistency} -- \ref{assump:sparsity} are satisfied and either \\
    (i) $\meanin \| \hat{\pi}_i(\boldX; \gbar_t) -\pi_i(\boldX; \gbar_t) \|^2_{L_2(\mathbb{P})} = o_{\mathbb{P}}(1)$ and $\meanin \| \hat{\pi}_i(\boldX; \gbar_t') -\pi_i(\boldX; \gbar_t') \|^2_{L_2(\mathbb{P})} = o_{\mathbb{P}}(1)$, or (ii) $\meanin \| \hat{\mu}_{\gbar_t',t^*}(\boldX) - \mu_{\gbar_t',t^*}(\boldX) \|^2_{L_2(\mathbb{P})} = o_{\mathbb{P}}(1)$ hold, then as $n \rightarrow \infty$, 
    \[ \hat{\tau} - \AEE_t(\gbar_t) - S_n^{(1)} \rightarrow_p 0, \]  
    where $S_n^{(1)} = \meanin [\pgone - \bar{p}(\gbar_t)] [\bar{p}(\gbar_t)]^{-1} \AEE_{it}(\gbar_t) = O_{\mathbb{P}}(1)$ and $\bar{p}(\gbar_t) = \meanin \pgone$. If additionally (a) there is either network effect homogeneity or exposure probability homogeneity, or (b) $\pgonehat$ is replaced with an estimator $\hat{p}_{i1}^{\mathrm{net}}(\gbar_t)$ such that $\meanin \| \hat{p}_{i1}^{\mathrm{net}}(\gbar_t) - \pgone \|^2_{L_2(\mathbb{P})} = o_{\mathbb{P}}(1)$, then 
    \[\hat{\tau} - \AEE_t(\gbar_t) \rightarrow_p 0.\]
\end{theorem}

The asymptotic bias term $S_n^{(1)}$ is equal to $\Cov_n(\AEE_{it}(\gbar_t), \pgone) / {\bar{p}(\gbar_t)}$, where $\Cov_n(\cdot, \cdot)$ is the population covariance function. Clearly, $S_n^{(1)} = 0$ exactly when there is either network effect homogeneity or exposure probability homogeneity. Consider that the causal estimand can be represented as 
\begin{align*}
    \AEE_{t}(\gbar_t) &= \AEE^{\dagger}_{t}(\gbar_t) - S_n^{(1)},
\end{align*} 
where 
\begin{align*}
    \AEE^{\dagger}_{t}(\gbar_t) &=  \meanin \frac{\PP(\Gbar_{it} = \gbar_t)}{n^{-1} \sum_{k=1}^{n} \PP(\Gbar_{kt} = \gbar_t)} \AEE_{it}(\gbar_t).
\end{align*}
The above representation show that when neither conditions (a) or (b) in Theorem \ref{theorem:consistency} are satisfied, $\hat \tau$ is actually a consistent estimator of $\AEE^{\dagger}_{t}(\gbar_t)$. In contrast to the causal estimand of interest which is a simple average of individual exposure effects, the $\AEE^{\dagger}_{t}(\gbar_t)$ is a weighted average of $\AEE_{it}(\gbar)$ where the weights sum to one and give more importance to units with higher exposure probabilities. This estimand may be of interest if likelihood of being exposed is an important consideration when evaluating exposure policies. Additionally, in many settings with possible network effect and exposure probability heterogeneity, it may not be feasible to propose an estimator $\hat{p}_{i1}^{\mathrm{net}}(\gbar_t)$ that is consistent. Instead, researchers may define their causal estimand of interest to be $\AEE^{\dagger}_{t}(\gbar_t)$. Also note that even if the exposure effects and probabilities are independent in the network, the bias term $S_n^{(1)}$ is still $O_{\mathbbm{P}}(1)$, though it would be expected to be small in practice for sufficiently large $n$. 

In the iid data setting, the $L_2(\mathbbm{P})$ convergence rate conditions stated in Theorem \ref{theorem:consistency} for estimators of $\pgone$, $\pigx$, and $\mmu$ are standard, and many machine learning or nonparametric estimators can achieve such rates \citep{kennedy_semiparametric_2023}. In the non-iid setting with weakly dependent data (as in Assumption \ref{assump:psi-dep}), determining convergence rates of nonparametric estimators is an open area of research but there has been some recent progress. For instance, \citet{saha_random_2023} showed that under $\beta$-mixing (a stronger weak dependence condition than Definition \ref{def:psi}), their proposed spatial random forest estimator is $L_2(\mathbbm{P})$ consistent. \citet{hansen_uniform_2008} proved that kernel-based regression estimators of data under various mixing conditions reach the same uniform convergence rate attained under iid data. Some promising estimators that specifically utilizes graph structure are graph neural networks \citep{leung_graph_2024} and Gaussian process regression with graph-based kernel functions \citep{borovitskiy_matern_2021}. 



Additional restrictions on nuisance function estimation are imposed to prove asymptotic normality. In many settings, data-adaptive nonparametric nuisance function estimators are used to help avoid model mis-specification. To avoid overfitting when using such nonparametric estimators, cross-fitting is often employed, where the target estimand and nuisance functions are estimated in different data splits \citep{chernozhukov_doubledebiased_2018}. However, this use of cross-fitting may not be justified in the presence of dependent data. Instead, Donsker conditions may be imposed as in Assumption \ref{assump:donsker} which restricts the complexity of the nuisance functions and their estimators but still allows for data-adaptive estimation. Deriving smoothness and complexity properties of machine learning estimators in non-iid settings is an active area of research; whereas, with iid data, the highly adaptive lasso (HAL) is one machine learning method that has been shown to fulfill both the Donsker and convergence rate conditions \citep{benkeser_highly_2016}. Nevertheless, in simulation experiments with network dependent data, presented in Section \ref{sec:simulation}, several machine learning nuisance function estimators including HAL were shown to perform well. 

\begin{assumption}[Donsker conditions on nuisance function estimators]
    \label{assump:donsker}
    Define a $\mathbb{P}$-Donsker class $\mathcal{F}$ of random functions $f_i: \mathbb{R}^{\nu} \mapsto \mathbb{R}$ for $i=1,\dots,n$ such that the sequence of processes $\{n^{1/2}\sumin (f_i - \E[f_i]) : f_i \in \mathcal{F} \}$ converges weakly to a tight, mean zero Gaussian process in the collection of all bounded functions of $\mathcal{F}$. Assume that the nuisance functions and their estimators are $\mathbb{P}$-Donsker. Specifically, $\pi, \hat{\pi} \in \mathcal{F}_{\pi}$ for $\gbar_t$ and $\gbar_t'$ and $\mu_{\gbar_t',t^*}, \hat{\mu}_{\gbar_t',t^*} \in \mathcal{F}_{\mu}$ where $\mathcal{F}_{\pi}$ and $\mathcal{F}_{\mu}$ are Donsker classes.
\end{assumption}

Theorem \ref{theorem:asymp_norm} provides the key asymptotic normality result of this paper. In particular, it can be shown that $\hat \tau$ admits an asymptotically linear representation of the form
\begin{align*}
    \sqrt{n} (\hat \tau - \tau) = n^{-1/2} \sumin (\varphi_i(\boldO_i) - \E[\varphi_i(\boldO_i)]) + o_{\mathbb{P}}(1),
\end{align*}
where $\varphi_i(\boldO_i)$ is the influence function of $\hat \tau$, provided that $\varphi_i(\boldO_i)$ has finite variance. Under a nonparametric model and network effect homogeneity or network exposure probability homogeneity, the above representation is shown to hold in Theorem \ref{theorem:asymp_norm} for $\varphi_i(\boldO_i) = \phi_i(\boldO_i)$. 

\begin{theorem}
    \label{theorem:asymp_norm}
    Let $\pgonehat = \hat{p}_{i1}^{\mathrm{emp}}(\gbar_t)$. Suppose that Assumptions \ref{assump:consistency} -- \ref{assump:donsker} and Assumption S1 in the Supplementary Material hold along with the following convergence rates for nuisance function estimators:
    \begin{enumerate}[label=(\roman*)]
        \item $\meanin \| \hat{\pi}_i(\boldX; \gbar_t) -\pi_i(\boldX; \gbar_t) \|^2_{L_2(\mathbb{P})} = o_{\mathbb{P}}(1)$
        \item $\meanin \| \hat{\pi}_i(\boldX; \gbar_t') -\pi_i(\boldX; \gbar_t') \|^2_{L_2(\mathbb{P})} = o_{\mathbb{P}}(1)$
        \item $\meanin \| \hat{\mu}_{i, \gbar_t',t^*}(\boldX) - \mu_{i, \gbar_t',t^*}(\boldX) \|^2_{L_2(\mathbb{P})} = o_{\mathbb{P}}(1)$
        \item $\left(\meanin \|\pighatX - \pigX\|^2_{L_2(\mathbb{P})} \right)^{1/2} \left( \meanin \|\hat{\mu}_{i, \gbar_t',t^*}(\boldX) - \mu_{i, \gbar_t',t^*}(\boldX) \|^2_{L_2(\mathbb{P})} \right)^{1/2} = o_{\mathbb{P}}(n^{-1/2})$
        \item $\left( \meanin \|\pigprimehatX - \pigprimeX\|^2_{L_2(\mathbb{P})} \right)^{1/2} \left( \meanin \|\hat{\mu}_{i, \gbar_t',t^*}(\boldX) - \mu_{i, \gbar_t',t^*}(\boldX)\|^2_{L_2(\mathbb{P})} \right)^{1/2} = o_{\mathbb{P}}(n^{-1/2})$.
    \end{enumerate}
    Then as $n \rightarrow \infty$,
    \begin{align*}
        \sigma_{n}^{-1} \sqrt{n} (\hat{\tau} - \tau) + S_n^{(2)} \rightarrow_d N(0,1),
    \end{align*}
     where $S_n^{(2)} =  \AEE_{t}(\gbar_t)\left(n^{-1/2}\sum_{i=1}^{n} \{\hat{p}_{i1}^{\mathrm{emp}}(\gbar_t) - \pgone \} [\hat{p}_{i1}^{\mathrm{emp}}(\gbar_t)]^{-1} \right) + n^{1/2} \hat{S}_n^{(1)} = O_{\mathbb{P}}(n^{-1/2})$, \\
     $\hat{S}_n^{(1)} = \meanin \{\pgone - \hat{p}_{i1}^{\mathrm{emp}}(\gbar_t) \} [\hat{p}_{i1}^{\mathrm{emp}}(\gbar_t)]^{-1} \AEE_{i,t}(\gbar_t)$, and $\sigma_n^2 / n = \Var(\phi^*(\boldO_i))$. If there is either network effect homogeneity or exposure probability homogeneity, then $S_n^{(2)} = o_{\mathbb{P}}(1)$ and $\hat \tau$ is nonparametric efficient with $\sigma_n^2/n$ the nonparametric efficiency bound. If there is both network effect heterogeneity and exposure probability heterogeneity, and a parametric model is assumed for $\pgone$ (Assumption S2 in the Supplementary Material), then for a RAL estimator $\hat{p}_{i1}^{\mathrm{net}}(\gbar_t)$ such that $\left( \meanin \{ \hat{p}_{i1}^{\mathrm{net}}(\gbar_t) - \pgone \}^2 \right)^{1/2} = O_{\mathbb{P}}(n^{-1/2})$,
     \begin{align*}
        \sigma_{n, \mathrm{adj}}^{-1} \sqrt{n} (\hat{\tau} - \tau) \rightarrow_d N(0,1),
    \end{align*}
    as $n \rightarrow \infty$, where $\sigma_{n, \mathrm{adj}}^2 / n = \Var(\meanin \phi_i(\boldO_i) - S_{\mathrm{adj}})$.
\end{theorem}

Similar to CAN results for doubly robust estimators in other settings, the conditions in Theorem \ref{theorem:asymp_norm} involve convergence of product terms of the form $O_{\mathbb{P}}(a_n) \times O_{\mathbb{P}}(b_n) = o_{\mathbb{P}}(n^{-1/2})$ which is fulfilled if, for instance, $a_n = b_n = o_{\mathbb{P}}(n^{-1/4})$. Theorem \ref{theorem:asymp_norm} also assumes (Assumption S1 in the Supplementary Material) a stronger network sparsity condition than Assumption \ref{assump:sparsity}. Assumption S2 is a regularity condition on the parametric model for $\pgone$. Similar to the consistency result of Theorem \ref{theorem:consistency}, the bias term $S_n^{(2)}$ appears if there is both network effect heterogeneity or exposure probability heterogeneity. The bias term may be eliminated by fitting a model for the exposure probabilities that satisfies $\left( \meanin \{ \hat{p}_{i1}^{\mathrm{net}}(\gbar_t) - \pgone \}^2 \right)^{1/2} = O_{\mathbb{P}}(n^{-1/2})$, which is typically only achieved by parametric models. An additional term, $S_{\mathrm{adj}}$, is then added to the to account for the uncertainty in estimating $\pgone$ must then be added to the influence function, i.e., $\meanin \varphi_i(\boldO_i) = \phi^*(\boldO) - S_{\mathrm{adj}}$. Under certain regularity conditions, the adjustment term $S_{\mathrm{adj}}$ is the projection of $\phi_i(\boldO_i)$ onto the score function of $\pgone$. See the Supplementary Material for more details on $S_{\mathrm{adj}}$ and the regularity conditions for $\hat{p}_{i1}^{\mathrm{net}}(\gbar_t)$ and $\pgone$. Table \ref{tab:inf_summ} summarizes the asymptotic results.

\begin{table}[!h]
\caption{Summary of asymptotic results for the proposed estimator $\hat \tau$}
\label{tab:inf_summ}
\begin{tabular}{c|c|c|c}
\hline
\textbf{\begin{tabular}[c]{@{}c@{}}Network effect \\ heterogeneity\end{tabular}} & \textbf{\begin{tabular}[c]{@{}c@{}}Exposure probability \\ heterogeneity\end{tabular}} & \textbf{Inference for $\boldsymbol{\hat \tau - \tau}$} & \textbf{\begin{tabular}[c]{@{}c@{}}Variance\\ estimation\end{tabular}} \\ \hline
\xmark & \xmark & $\sqrt{n}$-CAN, efficient & Consistent \\
\xmark & \cmark & $\sqrt{n}$-CAN, efficient & Consistent \\
\cmark & \xmark & $\sqrt{n}$-CAN & Conservative$^*$ \\
\cmark & \cmark & $\sqrt{n}$-CAN with RAL $\hat{p}_{i1}^{\mathrm{net}}(\gbar_t)$ & Conservative$^*$ \\ \hline
\end{tabular}

$^*$: conservative in large samples if unit-level treatment effects are not negatively correlated. CAN: consistent and asymptotically normal, RAL: regular and asymptotically linear.
\end{table}

To construct Wald-like confidence intervals using the result of Theorem \ref{theorem:asymp_norm}, consider the following network heteroskedasticity and autocorrelation consistent (HAC) variance estimator,
\begin{equation*} \label{eq:var_est}
    \hat{\sigma}^2_n = \frac{1}{n} \sum_{s \geq 0} \sum_{i \in \mathcal{N}_n} \sum_{j \in \mathcal{N}_n^{\partial}(i;s)} \hat{\phi}_i \hat{\phi}_j  \omega(s/b_n),
\end{equation*}
where $\hat{\phi}_i = \hat \tau_i - \hat{h}_{i1}(\Gbar_{it}) \hat{\tau}$, and the kernel weight $\omega$ is a function that maps $\{\mathbb{R}, -\infty, \infty \} \rightarrow [-1,1]$ with $\omega(0)=1$, $\omega(q)=0$ for $|q|>1$, and $\omega(q)=\omega(-q)$ for all $q \in \{\mathbb{R}, -\infty, \infty \}$. The term $b_n$ is a bandwidth parameter. When a parametric estimator $\hat{p}_{i1}^{\mathrm{net}}(\gbar_t)$ is used to estimate $\pgone$, consider the variance estimator $\hat{\sigma}^2_{n, \mathrm{adj}} = \hat{\sigma}^2_n - \hat S_{\mathrm{adj}}$, where $\hat S_{\mathrm{adj}}$ is a plug-in estimate of the adjustment term $S_{\mathrm{adj}}$. The consistency of the HAC variance estimator depends on $\omega(s/b_n)$; Assumption S3 in the Supplementary Material restricts choice of $\omega(s/b_n)$ based on network dependency and network asymptotics. 

Under network effect heterogeneity, the proposed variance estimators are not consistent for the true variance. This result is similar to the well-known result of conservative variance estimation in the design-based inference setting. Consider that the true variance $\sigma_n^2$ can be represented as
\begin{align*}
    \sigma_n^2 &= \frac{1}{n} \sum_{s \geq 0} \sum_{i \in \mathcal{N}_n} \sum_{k \in \mathcal{N}_n^{\partial}(i;s)} \E[\phi_i(\boldO_i) \phi_k(\boldO_k)] - V_n,
\end{align*}
where $V_n = \frac{1}{n} \sum_{s \geq 0} \sum_{i \in \mathcal{N}_n} \sum_{k \in \mathcal{N}_n^{\partial}(i;s)} (\AEE_{it}(\gbar_t) - \AEE_{t}(\gbar_t)) (\AEE_{kt}(\gbar_t) - \AEE_{t}(\gbar_t))$. The term $V_n$ is the sum of the population covariances of the individual exposure effects. If the exposure effects are independent, then $V_n$ is approximately the sample variance of the unit-level exposure effects. Since $V_n$ is not estimable in this setting, the plug-in estimator for $\E[\phi_i(\boldO_i) \phi_k(\boldO_k)]$, summed over all $i$ and $k$, will tend to be conservative, i.e., positively biased, where the bias is represented by $V_n$ and is non-zero when there is network effect heterogeneity. The result is formalized in Theorem \ref{theorem:var_consis}.

\begin{theorem}
    \label{theorem:var_consis}
    Suppose Assumptions \ref{assump:consistency} -- \ref{assump:donsker} and conditions (i) -- (iii) of Theorem \ref{theorem:asymp_norm} hold along with Assumption S3 in the Supplementary Material. In the absence of network effect heterogeneity, $\hat{\sigma}^2_n - \sigma^2_n \rightarrow_p 0$ as $n \rightarrow \infty$. Under network effect heterogeneity and network exposure probability homogeneity, $\hat{\sigma}^2_n - V_n - \sigma^2_n \rightarrow_p 0$. Similarly, under network effect heterogeneity and network exposure probability heterogeneity and a RAL estimator $\hat{p}_{i1}^{\mathrm{net}}(\gbar_t)$, $\hat{\sigma}^2_{n, \mathrm{adj}} - V_n - \sigma^2_n \rightarrow_p 0$. 
\end{theorem}

\noindent
In finite samples, the choice of $\omega$ may be influential in constructing an accurate variance estimator. Prior knowledge on the dependency structure of the data can be used to choose $\omega$. For instance, if $K$-neighborhood dependency is assumed, then a reasonable $\omega$ may be the uniform kernel that gives a weight of 1 to distances $s \leq K$ and weight 0 otherwise. In general, choosing $\omega$ is an open problem; in the Supplementary Material, the suggestion provided by \citet{leung_causal_2022} is reviewed. Additionally, in the Supplementary Material, the sources of dependence in $\phi_i(\boldO_i)$ are discussed under different assumptions on the data generating process, which may aid in choosing $\omega$.

\section{Simulation} \label{sec:simulation}

The proposed methods were evaluated by simulating datasets for both the unipartite and bipartite settings, with and without latent variable dependence. In the unipartite setting, simulated data were generated from a ring network where units are positioned in a circle; see Supplementary Figure S1 for an example of a ring network with 10 nodes. In the bipartite setting, the network was based on the data application described in Section \ref{sec:application}. The simulation datasets included $n = m = 5000$ units in the unipartite scenario and $n = 3105$ outcome units (representing counties in the contiguous United States) and $m = 484$ intervention units (representing coal power plants) in the bipartite scenario. For all scenarios, there were two time periods $t \in \{0,1\}$.

For the unipartite network, covariates $X_i$, treatments $Z_i$, exposures $G_i$, and outcome changes $\Delta_1 Y_i$ were generated as follows:
\begin{align*}
    &X_i \sim N(0,1), \\
    &Z_i | X_i \sim \mathrm{Bernoulli} \left\{ \mathrm{logit}^{-1}(0.5 \sin{(X_i-2)^2)}) \right\}, \\
    &G_i = \mathbbm{1}
    \left(\sum_{j=1}^{n} w_{ij} Z_j > 0.5\right), \\
    &\Delta_1 Y_i | \mathbf{X}, G_i \sim 5G_i + f^{\mathrm{ring}}(\mathbf{X}, i) + \epsilon_i,
\end{align*}
where $\mathbf{X} = (X_1,\dots,X_n)$, $f^{\mathrm{ring}}(\mathbf{X}, i) = X_{i-3} + 2X_{i-2}^2 + \mathbbm{1}(X_{i-1} > 0)\{\exp(X_{i-1}) - 1\} - X_{i}^3 + \mathrm{logit}^{-1}(X_{i+1}) - \sin(X_{i+2} X_{i+3})$, and $\epsilon_i$ is a mean-zero error term. The interference weights were $(w_{i,i-3},\dots, w_{i,i}, \dots, w_{i,i+3}) = (1/7, \dots, 1/7)$ with all other $w_{ij}$ equal to zero, where non-positive indices count down from $n$, e.g., $w_{i,-1} = w_{i,n-1}$. Thus, the exposure mapping for a unit $i$ equals one if at least four out of the seven units consisting of unit $i$ and its six closest neighbors had treatment $Z_i = 1$. Both independent and dependent outcome error terms were considered, where the latter implies latent variable dependence. In the independent error scenario $\epsilon_i \sim N(0,1)$ while in the dependent error scenario, $(\epsilon_{1},\dots, \epsilon_{n})^{\top} \sim \mathrm{MVN}(0, k^{\mathrm{ring}}(i,i'))$, where $\mathrm{MVN}(0, k(\cdot, \cdot))$ is a mean-zero multivariate normal distribution where the covariance between any units $i, i'$ is provided by the kernel function $k^{\mathrm{ring}}(i,i') = 0.6^{d(i,i')}$, and $d(i,i')$ is the path distance between units $i$ and $j$ according to the ring network. 

The data were generated in the bipartite setting as follows.
\begin{align*}
    &X_j \sim \mathrm{Unif}(-2, 2), \\
    &Z_j | X_j \sim \mathrm{Bernoulli} \left\{ \mathrm{logit}^{-1}(\sin{[(0.4X_j-2)^2)]}) \right\}, \\
    &G_i = \mathbbm{1}
    \left(\sum_{j=1}^{m} w_{ij} Z_j \geq 0.5\right), \\
    &\Delta_0 Y_i | \mathbf{X}, G_i \sim 5G_i + f^{\mathrm{bipart}}(\mathbf{X}, i) + \epsilon_i,
\end{align*}
where $f^{\mathrm{bipart}}(\mathbf{X}, i) = 4 - 2\mathbbm{1}(X_i^* < -1) + 2X_i^* \mathbbm{1}(-1 \leq X_i^* < -0.25) + (-0.1875 - 5(X_i^*)^2)\mathbbm{1}(-0.25 \leq X_i^* < 0.5) - 1.4375 \mathbbm{1}(X_i^* \geq 0.5)$, $\epsilon_i$ is a mean-zero error term, and $X_i^* = \sum_{j} \mathbbm{1}(\max_i w_{ij} = w_{ij}) X_j$. The interference matrix $\boldW$ was derived from the data application where $\boldW$ represents the possible influence of power plants on counties as computed from an atmospheric transport model (see Section \ref{sec:application} for more details). Rows in $\boldW$ were divided by their sum so that $\sum_{j} w_{ij} = 1$ for all $i$. In the scenario with independent outcome errors, $\epsilon_i \sim N(0,1)$. For the dependent error scenario, $(\epsilon_{1},\dots, \epsilon_{n})^{\top} \sim \mathrm{MVN}(0, k^{\mathrm{bipart}}(i,i'))$, where $k^{\mathrm{bipart}}(i,i')$ is a kernel function, with details provided in the Supplementary Material.

In both unipartite and bipartite settings, $\AEE(\gbar_1) = 5$ and was constant across units, where $\gbar_1 = (0,1)$ and $\gbar_t' = (0,0)$. Here, four simulation scenarios are considered that varied by network (ring or bipartite) and presence of latent variable dependence (independent or dependent outcome errors). In the Supplementary Material, four additional simulation scenarios based on heterogeneous network effects are presented. In each setting, 1000 datasets were simulated. Each dataset was evaluated using the proposed doubly robust estimator with different nuisance function estimators. In all cases, the exposure propensity score was estimated by first estimating the intervention propensity score then using Monte Carlo integration as described in Section \ref{sec:methods-estimation}. Parametric, nonparametric, and oracle estimators were compared. The parametric nuisance function estimators employed generalized linear models (GLM), specifically, logistic regression for the treatment propensity score and linear regression for the outcome model. Nonparametric nuisance function estimators for both the outcome and treatment propensity score functions included HAL, Bayesian additive regression trees (BART), and the Superlearner with GLMs, HAL, and BART included in the library of candidate prediction algorithms. As discussed earlier, HAL may satisfy the Donsker condition of Assumption \ref{assump:donsker}. BART is not in a Donsker class (even in the iid setting) but can avoid overfitting while allowing for flexible estimation. The Superlearner algorithm is an ensemble learner that generates predictions using weighted sums of predictions from the individual estimators in the specified library. The oracle nuisance function estimators employed the true data generating functions for the outcome and treatment propensity score models.

\begin{table}[!h]
\caption{Results from 1000 simulations.}
\label{tab:sim_results}
\begin{tabular}{ll|ll|rrrrc}
\hline
\multicolumn{2}{c|}{\textbf{Data generation}} & \multicolumn{2}{c|}{\textbf{Estimator parameters}} & \multicolumn{5}{c}{\textbf{Results}} \\ \hline
\textbf{Network} & \textbf{\begin{tabular}[c]{@{}l@{}}Outcome \\ errors\end{tabular}} & \textbf{\begin{tabular}[c]{@{}l@{}}Band-\\ width\end{tabular}} & \textbf{\begin{tabular}[c]{@{}l@{}}Nuisance\\ function\\ estimators\end{tabular}} & \textbf{Bias} & \textbf{MSE} & \textbf{ESE} & \textbf{ASE} & \textbf{\begin{tabular}[c]{@{}l@{}}Coverage (\%)\end{tabular}} \\ \hline
Ring & Ind. & 0 & GLM & 7.7 & 0.7 & 3.8 & 3.5 & 41.3 \\
 &  & 0 & BART & 0.0 & 0.1 & 2.9 & 2.8 & 94.0 \\
 &  & 0 & HAL & 0.0 & 0.1 & 2.9 & 2.9 & 94.9 \\
 &  & 0 & SuperLearner & 0.0 & 0.1 & 2.9 & 2.8 & 94.7 \\
 &  & 0 & Oracle & 0.1 & 0.1 & 2.8 & 2.9 & 96.1 \\ \hline
Ring & Dep. & 15 & GLM & 7.5 & 0.8 & 5.0 & 5.1 & 69.3 \\
 &  & 15 & BART & 0.0 & 0.2 & 4.6 & 4.4 & 94.3 \\
 &  & 15 & HAL & -0.1 & 0.2 & 4.6 & 4.5 & 95.3 \\
 &  & 15 & SuperLearner & -0.1 & 0.2 & 4.5 & 4.5 & 94.7 \\
 &  & 15 & Oracle & 0.0 & 0.2 & 4.5 & 4.6 & 96.4 \\
 &  & 0 & GLM & 7.5 & 0.8 & 5.0 & 3.5 & 45.9 \\
 &  & 0 & BART & 0.0 & 0.2 & 4.6 & 2.8 & 76.9 \\
 &  & 0 & HAL & -0.1 & 0.2 & 4.6 & 2.9 & 79.0 \\
 &  & 0 & SuperLearner & -0.1 & 0.2 & 4.5 & 2.8 & 77.5 \\
 &  & 0 & Oracle & 0.0 & 0.2 & 4.5 & 2.9 & 79.6 \\ \hline
Bipart & Ind. & 0 & GLM & -13.5 & 3.0 & 10.9 & 5.0 & 34.6 \\
 &  & 0 & BART & -0.5 & 0.3 & 5.7 & 4.7 & 91.0 \\
 &  & 0 & HAL & -0.3 & 0.3 & 5.9 & 4.8 & 92.8 \\
 &  & 0 & SuperLearner & -0.4 & 0.3 & 5.7 & 4.7 & 92.6 \\
 &  & 0 & Oracle & -0.1 & 0.2 & 4.7 & 4.6 & 95.2 \\ \hline
Bipart & Dep. & 1.1 & GLM & -1.1 & 2.8 & 16.7 & 7.4 & 63.7 \\
 &  & 1.1 & BART & -0.3 & 0.4 & 6.2 & 5.3 & 90.9 \\
 &  & 1.1 & HAL & -0.3 & 0.4 & 6.1 & 5.3 & 91.6 \\
 &  & 1.1 & SuperLearner & -0.3 & 0.4 & 6.1 & 5.3 & 91.7 \\
 &  & 1.1 & Oracle & -0.1 & 0.3 & 5.5 & 5.3 & 94.7 \\
 &  & 0 & GLM & -1.1 & 2.8 & 16.7 & 4.6 & 43.0 \\
 &  & 0 & BART & -0.3 & 0.4 & 6.2 & 4.2 & 81.9 \\
 &  & 0 & HAL & -0.3 & 0.4 & 6.1 & 4.2 & 82.3 \\
 &  & 0 & SuperLearner & -0.3 & 0.4 & 6.1 & 4.2 & 81.9 \\
 &  & 0 & Oracle & -0.1 & 0.3 & 5.5 & 4.2 & 85.4 \\ \hline
\end{tabular} \par
\smallskip
Ring: unipartite ring network, Bipart: bipartite network, Ind.: independent, Dep.: dependent, Bias: average bias ($\times 100$), MSE: mean squared error ($\times 100$), ASE: average standard error estimates ($\times 100$), ESE: empirical standard error ($\times 100$), Coverage (\%): 95\% confidence interval coverage.
\end{table}

The HAC variance estimator $\hat{\sigma}^2_n$ was used to compute standard errors and create Wald-like confidence intervals. The uniform kernel was used to give equal weight to all terms $\hat{\phi}_i \hat{\phi}_{i'}$ if $d(i, i') < b_n$ where $b_n$ is the bandwidth parameter. A bandwidth of $b_n=0$ ignores covariance terms between different units. In the dependent outcome error scenarios, the consequences of ignoring dependence were assessed by setting the bandwidth to zero. In the independent outcome error scenarios, a bandwidth of zero was appropriate due to treatment effect homogeneity (see Section S2 in the Supplementary Material for more details). 

\begin{figure}[!h]
    \centering
    \begin{subfigure}{0.4\textwidth}
        \centering
        \includegraphics[width=1\textwidth]{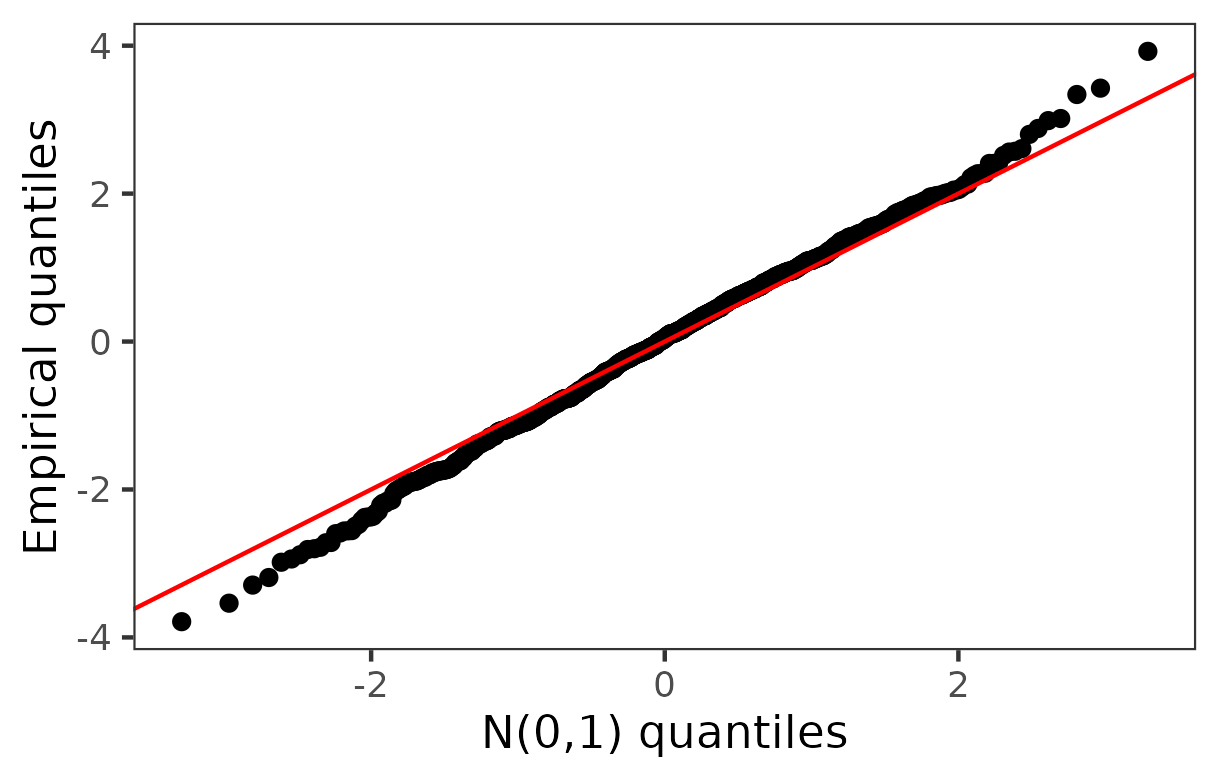}
        \caption{}
    \end{subfigure}
    ~
    \begin{subfigure}{0.4\textwidth}
        \centering
        \includegraphics[width=1\textwidth]{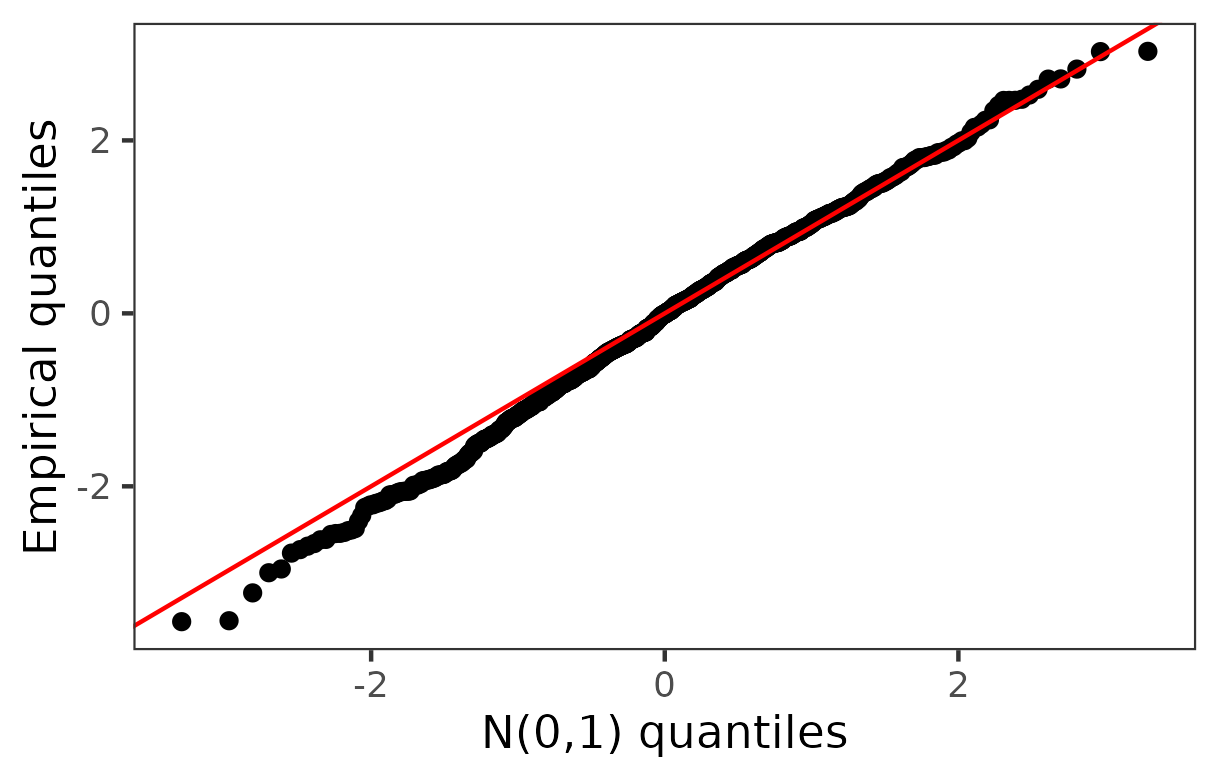}
        \caption{}
    \end{subfigure}
    \par \bigskip
    \begin{subfigure}{0.4\textwidth}
        \centering
        \includegraphics[width=1\textwidth]{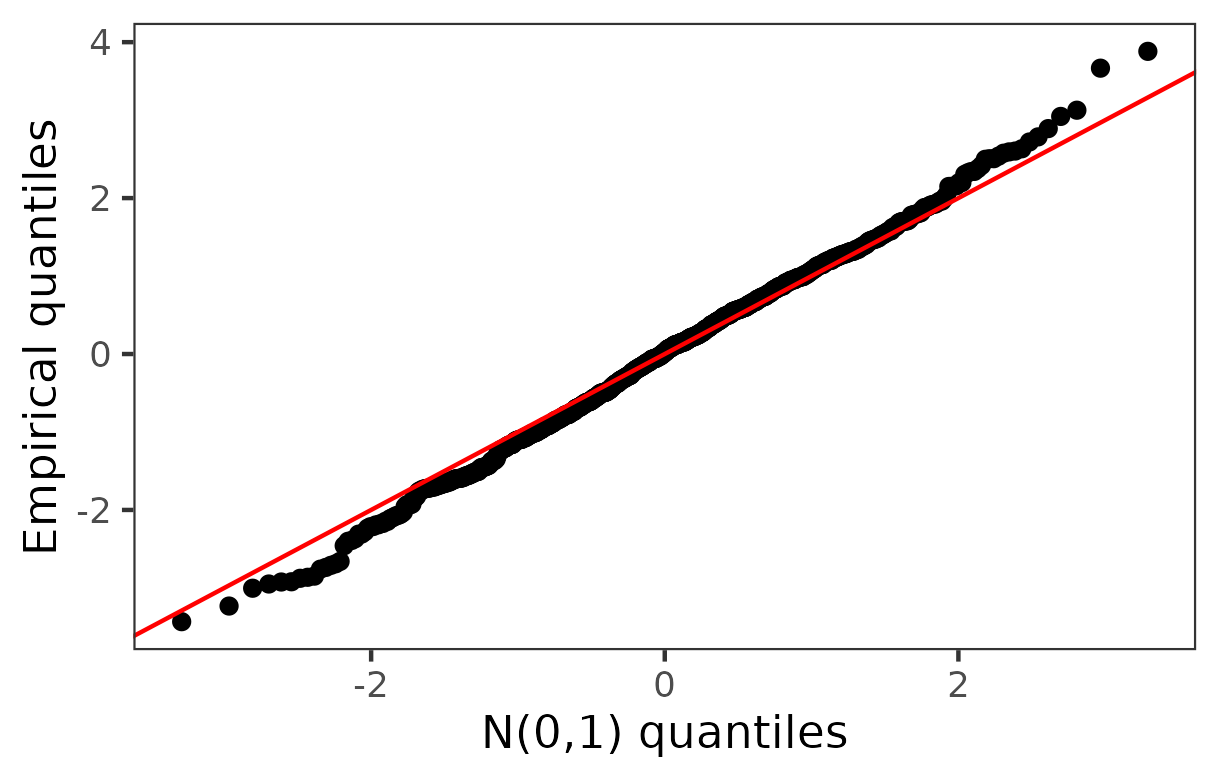}
        \caption{}
    \end{subfigure}
    ~
    \begin{subfigure}{0.4\textwidth}
        \centering
        \includegraphics[width=1\textwidth]{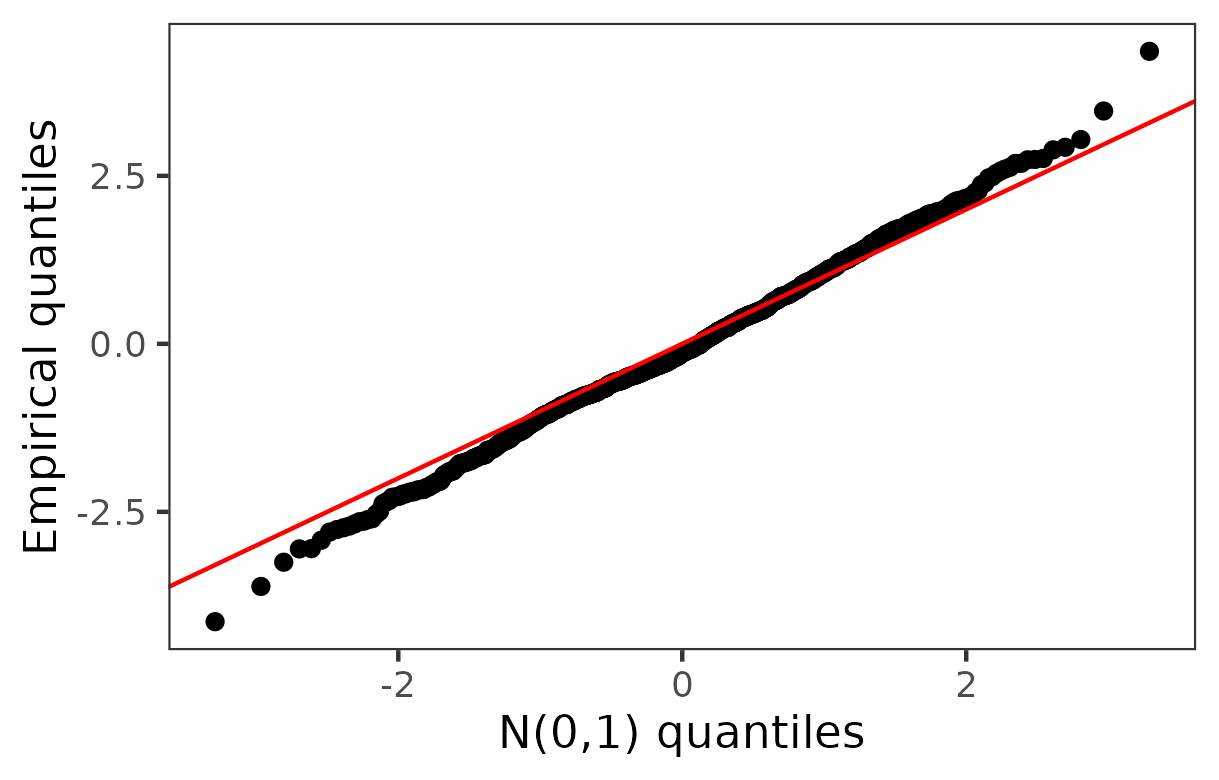}
        \caption{}
    \end{subfigure}
    
    \caption{Quantile-quantile plots comparing $\hat \sigma_n^{-1} \sqrt{n} (\hat \tau - \tau)$ with the $N(0,1)$ distribution where nuisance functions were estimated using the Superlearner under scenarios: (a) ring network, ind. errors, bandwidth 0; (b) ring network, dep. errors, bandwidth 15; (c) bipartite network, ind. errors, bandwidth 0; (d) bipartite network, dep. errors, bandwidth 1.1.}
    \label{fig:qqplots}
\end{figure}

Table \ref{tab:sim_results} provides results from 1000 simulations for each scenario. The estimators utilizing parametric nuisance function estimators (i.e., GLMs) performed poorly in all scenarios since the parametric models were mis-specified. In contrast, nonparametric estimation of nuisance functions (i.e., using HAL, BART, or the Superlearner) led to performance nearly on par with using oracle nuisance functions. Incorrectly setting the bandwidth to zero under the presence of latent variable dependence in the outcome yielded poor coverage rates. HAL, BART, and the Superlearner performed similarly. Overall, nonparametric estimation of nuisance functions led to estimators of the $\AEE$ that had low bias, low mean squared error, and confidence intervals with nominal or near nominal coverage rates when dependency was accounted for appropriately. In simulation settings using the bipartite network, coverage was slightly less than nominal, possibly due to a relatively small intervention unit sample size. Additional simulation results are presented in the Supplementary Material where the intervention unit sample size was increased and, subsequently, coverage was closer to nominal. Figure \ref{fig:qqplots} displays quantile-quantile plots comparing the scaled estimator $\hat \sigma_n^{-1} \sqrt{n} (\hat \tau - \tau)$ with the standard normal distribution where nuisance functions were estimated using the Superlearner, further confirming the theoretical results.

\section{Application} \label{sec:application}

In this section the proposed methods were used to assess the effect of implementing emission control technologies in coal power plants on county-level mortality in the United States. In particular, the binary treatment was the installation of flue-gas desulfurization scrubbers which reduce SO$_2$ emissions. Outcomes were annual county-level deaths per 100,000 due to any circulatory disease, as defined by ICD-10 codes I00-I99. 

Bipartite interference may have been present in this setting since county-level mortality due to cardiovascular diseases may have depended on scrubber installations in many coal power plants, possibly in different counties or states. HYSPLIT (Hybrid Single-Particle Lagrangian Integrated Trajectory) \citep{draxler_overview_1998, stein_noaas_2015}, an atmospheric transport model that estimates the movement of air parcels from point sources through three-dimensional space, was used to define the interference matrix $\boldW_t$ for every year $t$ from 2003 to 2013. Specifically, HyADS (HYSPLIT Average Dispersion) \citep{henneman_characterizing_2019} was used to create a transfer coefficient matrix (TCM) that associated the air parcel densities from power plants to counties, which was then used to define interference matrices. 

\begin{figure}[!h]
    \centering
    \includegraphics[width=1\textwidth]{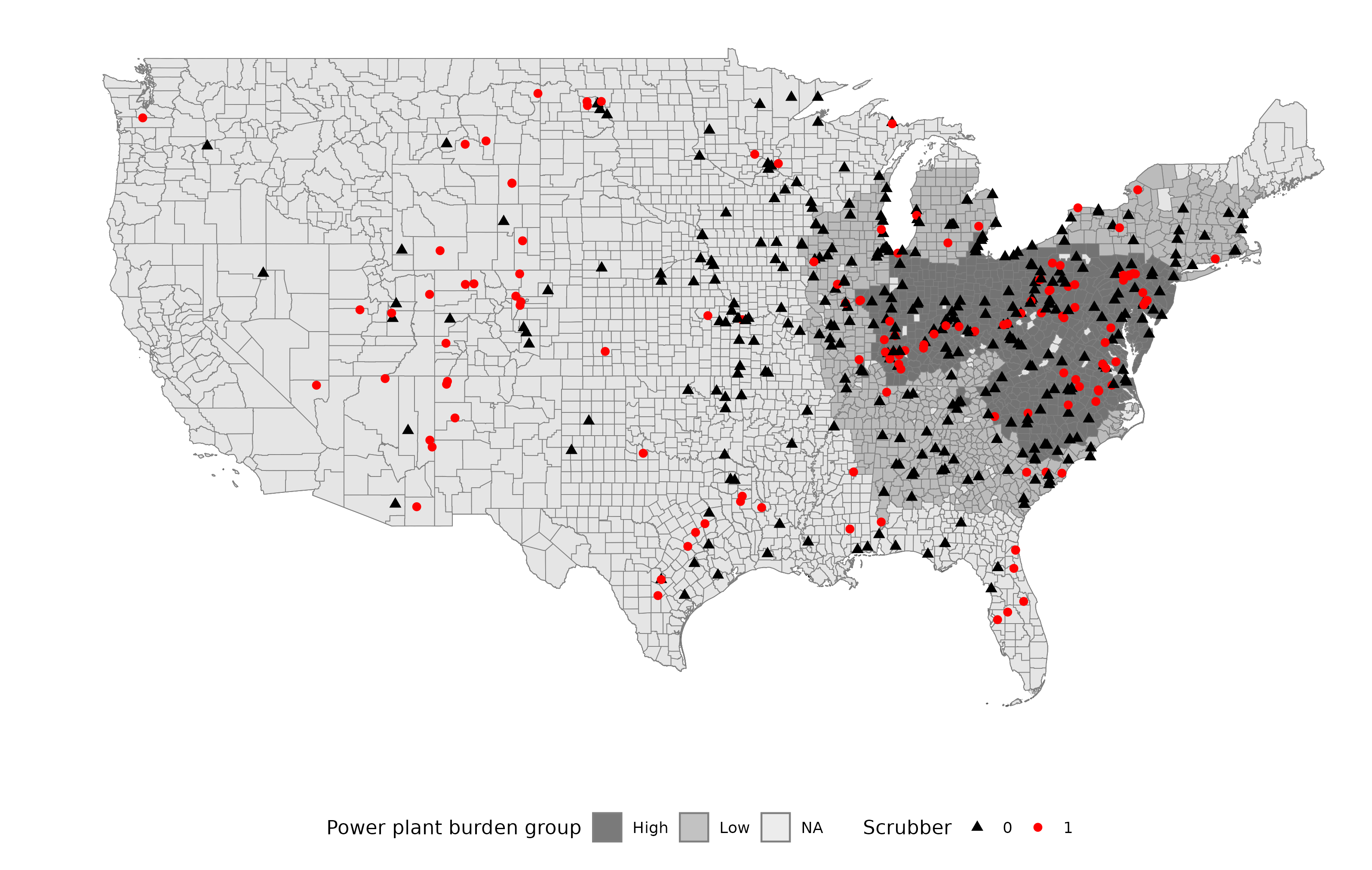}
    \caption{Counties by power plant burden group and power plants by scrubber status in 2007.}
    \label{fig:int_burden_map}
\end{figure}

The TCM provides a county's ``power plant burden," i.e., the cumulative HyADS contribution of all power plants to that county. Counties may have varied greatly in their power plant burden, and a county with small power plant burden was not necessarily comparable to a county with large power plant burden. To focus on the effect of scrubber installations, the data analysis was stratified by counties with similar power plant burdens. Formally, let $\boldW_t^*$ denote the TCM at time $t$ with elements $w_{ijt}^*$, and let power plant burden be denoted $w_{it}^* = \sum_j w_{ijt}^*$.  The power plant burdens $w_{i,2003}^*, \dots, w_{i,2013}^*$ were averaged across years, i.e., $w_{i,\mathrm{agg}}^* = (1/11) \sum_{t=2003}^{2013} w_{it}^*$. Then, counties were stratified based on quartiles of $\{w_{i,\mathrm{agg}}^* \}_{i=1}^{n}$. Counties in the lowest two quartiles were not analyzed since coal power plants had a relatively small effect on those counties, according to HyADS. Counties in the third and fourth quartile were analyzed separately and were labeled low and high power plant burden counties, respectively. These counties were located in the eastern United States, where most coal power plants operate. Figure \ref{fig:int_burden_map} shows a map displaying the power plant burden groups, power plant locations, and power plant scrubber statuses in 2007. 

Within each power plant burden group, the interference matrices were defined as the row-standardized TCM for each year, i.e., the interference weights were $w_{ijt} = (\sum_{j} w_{ijt}^*)^{-1} w_{ijt}^*$. The exposure mapping was defined to be $G_{it} = \mathbbm{1}\{ \sum_{j} w_{ijt} Z_{jt} > 0.25 \}$ where $Z_{jt}=1$ if power plant $j$ had a scrubber in year $t$ and $Z_{jt}=0$ otherwise. Thus, $G_{it} = 1$ indicates that many nearby power plants had scrubbers installed, where ``nearby" was determined by the atmospheric transport model. Scrubber installation followed a staggered adoption pattern, where once a power plant adopted a scrubber, the scrubber remained in place for the remainder of the study period. County-level exposures were not guaranteed to follow a staggered adoption pattern due to the time-varying interference weights and choice of exposure mapping; however, all but two counties in each power plant burden group had exposure histories that followed staggered adoption. These two counties were excluded from the analysis, leaving 643 and 644 counties in the low and high power plant burden groups, respectively. All coal power plants were included when analyzing either power plant burden group. In total, there were 517 power plants in the study period. 

Baseline covariates assumed to be sufficient in satisfying conditional parallel trends included county-level demographic information and power plant-level operating characteristics. Table S2 in the Supplementary Material provides summary statistics of these covariates in 2009. At the county-level, power plant covariates were summarized using a weighted average where the weights were from the interference matrix, e.g., $\sum_{j} w_{ijt} \boldX_{jt}^{\mathrm{int}}$. Further details on the data application including data processing are included in the Supplementary Material. 

Since exposure histories followed a staggered adoption pattern, counties' exposures were characterized by their exposure cohort, or the year a county changed from unexposed to exposed. For instance, the 2008 cohort has exposure history $\gbar_{2007:2010} = (0,1,1,1)$ and the corresponding reference history is $\gbar_{2007:2010}' = (0,\dots,0)$. For the low power plant burden counties, the effects of exposure adoption in years 2008 to 2010 were estimated. For the high power plant burden counties, the effects of exposure adoption in 2007 to 2009 were estimated. Other exposure cohorts were not studied due to low sample size (see Supplementary Material Table S1 for a summary of sample size by exposure cohort). For both power plant burden groups, lag effects of up to three years after the cohort year were estimated. Additionally, the two years preceding the cohort year was also studied as a negative control, where no effect was expected, provided identification assumptions held.

Nuisance functions were estimated using the Superlearner with GLMs, HAL, BART, and mean models (i.e., GLMs without covariates) in the library of candidate prediction algorithms. Additionally, an analysis assuming unconditional parallel trends was performed. Variance estimation employed the bandwidth $0$ since variance estimates using a uniform kernel with bandwidths at $1.1$ and $1.5$ tended to yield slightly smaller variance estimates than the variance estimates with bandwidth $0$. If latent variable dependence existed in this setting, negative correlations between outcomes in nearby counties would not be expected. Therefore, no latent variable dependence was assumed. Wald-like 95\% confidence intervals were estimated. 

\begin{figure}[!h]
    \centering
    \includegraphics[width=1\textwidth]{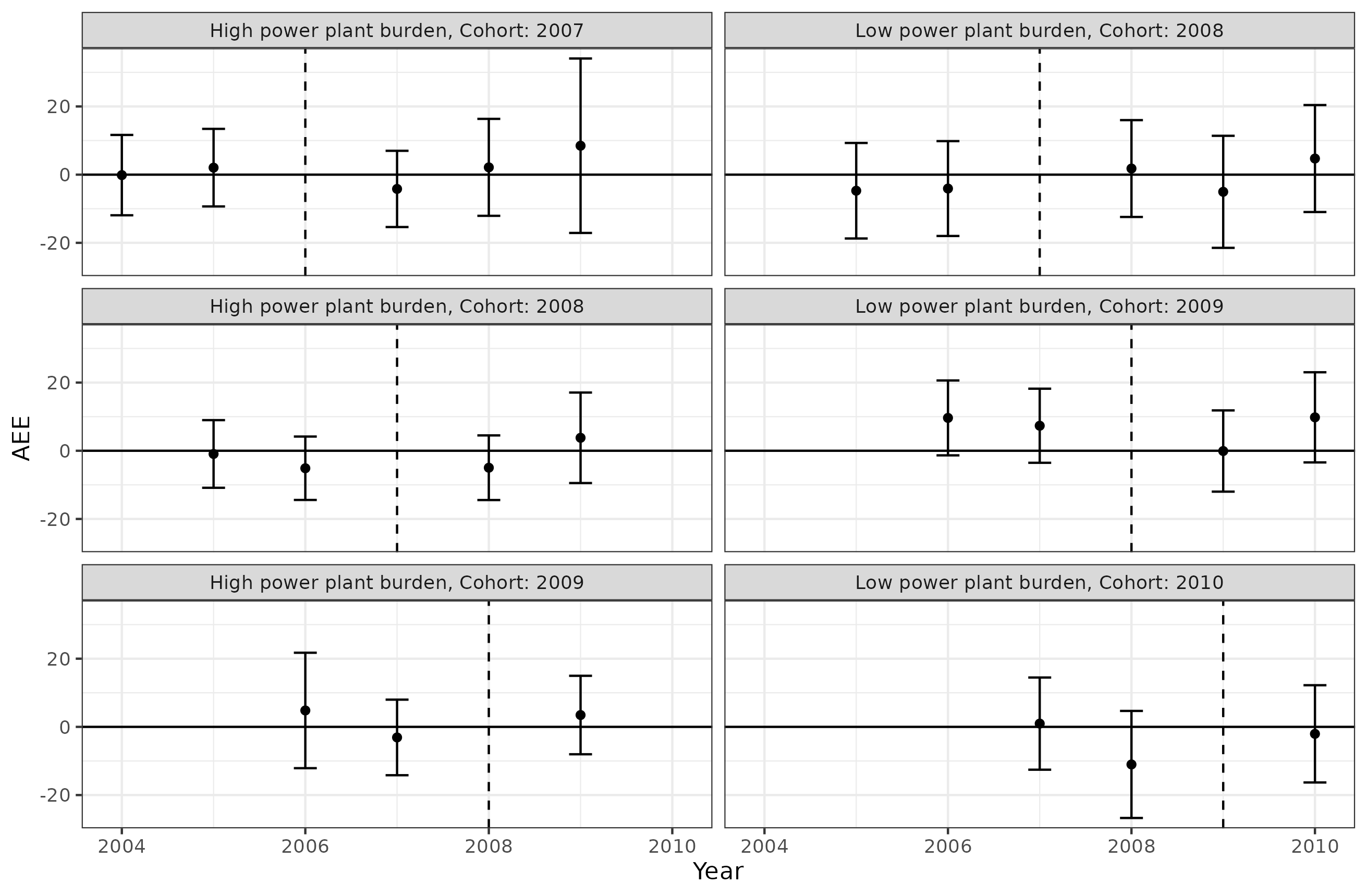}
    \caption{Estimated effects and corresponding 95\% confidence intervals of coal power plant scrubber exposure on county-level deaths due to cardiovascular diseases, per 100,000 individuals per year.}
    \label{fig:results_sl}
\end{figure}

The main results are shown in Figure \ref{fig:results_sl}. The dashed line denotes $t^*$, the reference year, equal to the exposure cohort year minus one. Estimates to the left of the dashed line correspond to the negative outcome control analysis while estimates to the right of the dashed line describe the average exposure effect from scrubber installations up to three years after initial exposure. Since the exact timing of scrubber installations within a year was not generally available in the data, a non-null effect was not necessarily anticipated. 

All negative outcome control estimates were near zero with confidence intervals containing zero, as expected from the identification assumptions. The analysis did not find an effect in a consistent direction for most years and cohorts. 
The analysis was repeated assuming unconditional parallel trends, with results shown in Supplementary Figure S3. Overall, the estimated effects were similar. 

Data on coal power plants and atmospheric transport were publicly available from the United States Environmental Protection Agency's Clean Air Markets Program \citep{united_states_environmental_protection_agency_epa_clean_2025}. County-level mortality data was publicly available from CDC WONDER \citep{centers_for_disease_control_and_prevention_compressed_2017}. County-level demographic data were obtained from the United States 2000 Census. Smoking prevalence estimates were derived from the Behavioral Risk Factor Surveillance System \citep{dwyer-lindgren_cigarette_2014}, and climate variables were computed as spatial averages of a validated gridded surface meteorological dataset \citep{abatzoglou_development_2013}.

\section{Discussion} \label{sec:discussion}

In this paper, a doubly robust DiD estimator was proposed for the finite population setting with network dependence, possible (bipartite) interference, and non-identically distributed data. Under assumptions on the network and interference structure, the DR estimator with data-adaptive nuisance function estimators was shown to be consistent for the AEE and asymptotically normal. The estimators were shown to perform well in finite samples through simulations in unipartite and bipartite settings. The proposed methods were also used to study the effect of scrubber installations in coal power plants on county-level deaths due to cardiovascular diseases. 

Future work may relax some assumptions made in this paper or extend the results to other settings. For example, this paper considered the finite population setting with the network considered fixed or known. In the case when the inferential target is not the sample but a population from which the sample is drawn, ignoring the network generation process may invalidate inference. Thus, future work may consider allowing for random networks and incorporate network models. It is also possible to make stronger assumptions such as multiple independent networks and compare nonparametric efficiency bounds and finite sample performance. The results of this paper can also be extended to the spatial setting by replacing the network topology with a spatial metric space. In general, many of the recent innovations in observational causal inference with non-iid data that rely on an ignorability assumption can be studied under an alternative set of conditional parallel trends assumptions. 

\section*{Acknowledgments and funding}
We thank Ye Wang, Chanhwa Lee, Brian D. Richardson, Bradley Saul, and Corwin Zigler for helpful comments. Michael Jetsupphasuk was supported by the National Institutes of Health (NIH) grant T32ES007018. Michael G. Hudgens was supported by NIH grant R01AI085073. The content in this article is solely the responsibility of the authors and does not necessarily represent the official views of the NIH. 

\section*{Disclosure statement}
The authors declare no potential conflict of interests.

\section*{References}

\printbibliography[heading=none]
\newpage

\title{Supplementary Material}
\author{}
\date{}
\affil{}

\emptythanks
\maketitle

\renewcommand\thesection{S\arabic{section}}
\renewcommand\thesubsection{\thesection.\arabic{subsection}}
\renewcommand\theequation{S\arabic{equation}}

\renewcommand\thetable{S\arabic{table}}
\renewcommand\thefigure{S\arabic{figure}}

\setcounter{section}{0}
\setcounter{figure}{0}
\setcounter{table}{0}
\setcounter{assumption}{0}

\section{Proofs}

\subsection{Additional assumptions}

\renewcommand\theassumption{S\arabic{assumption}}

Let $M_{n}^{\partial}(s;v) = n^{-1}\sumin |\mathcal{N}_{n}^{\partial}(i, s)|^v$ be a network density quantity that measures the average size of neighborhood shells $\mathcal{N}_{n}^{\partial}(i, s)$. Additionally, define the following notation:
\begin{align*}
    \zeta_n(s,m;v) = n^{-1} \sum_{i \in \mathcal{N}_n} \max_{k \in \mathcal{N}_n^{\partial}(i;s)} | \mathcal{N}_n(i;m) \setminus \mathcal{N}_n(k; s-1) |^v, \\
    c_n(s,m;v) = \inf_{\alpha > 1} [\zeta_n(s,m;v \alpha)]^{1/\alpha} \left[ M_n^{\partial}(s;\frac{\alpha}{1-\alpha})\right]^{1-(1/\alpha)},
\end{align*}
where $\zeta_n(s,m;v)$ is another network density quantity based on neighborhood sizes. The quantity $c_n(s,m;v)$ is an overall network density measure that combines information about average neighborhood sizes and average neighborhood shell sizes. Then, Assumption \ref{assump:sparsity2} bounds the large sample variance in relation to the neighborhood sizes and dependency coefficient.

\begin{assumption}
    \label{assump:sparsity2}
    (Limited asymptotic network dependency \citep[Assumption 3.4 in][]{kojevnikov_limit_2021}). 
    As $n \rightarrow \infty$, there exists a positive sequence $m_n \rightarrow \infty$ such that for $k=1,2$ and $v>4$,
    \begin{align*}
        \frac{n}{\sigma^{2+k}_n} \sum_{s \geq 0} c_n(s,m_n;k)\theta_{n,s}^{1-\frac{2+k}{v}} &\rightarrow_{a.s.} 0 \mbox{ and } \frac{n^2 \theta_{n,m_n}^{1-(1/v)}}{\sigma_n} \rightarrow_{a.s.} 0.
    \end{align*}
\end{assumption}

Assumption \ref{assump:reg-param} provides regularity conditions of the parametric model for $\PP(\Gbar_{it}=\gbar_t) \equiv \pgone$ in the case when there is network effect and exposure heterogeneity. These assumptions are standard assumptions for parametric models and estimators. 

\begin{assumption}[Regularity assumptions of parametric model for $\PP(\Gbar_{it}=\gbar_t)$]
    \label{assump:reg-param}
    Let $\pgone$ be indexed by a finite $d$-dimensional parameter $\eta \in E \subset \mathbb{R}^d$ where $E$ is compact. Assume that $p_{i1}(\gbar_t; \eta)$ is continuous and twice differentiable for all $\eta$. Further, assume that $\hat \eta$ has a $\sqrt{n}$ asymptotic linear representation:
    \begin{align*}
        \sqrt{n}(\hat \eta - \eta) = n^{-1/2} \sumin \varphi_{p_g}(\boldO_i; \eta) + o_{\mathbb{P}}(1),
    \end{align*}
    where $\varphi_{p_g}(\boldO_i; \eta)$ is mean zero with finite, positive definite variance and $n^{-1/2} \sumin \varphi_{p_g}(\boldO_i; \eta) \rightarrow_d N(0, \Var(\varphi_{p_g}(\boldO_i; \eta)))$. 
\end{assumption}

Assumption \ref{assump:var_consis}(a) restricts higher-level moments of the influence function, Assumption \ref{assump:var_consis}(b) ensures that the kernel weights $\omega$ converge to $1$ sufficiently fast, and Assumption \ref{assump:var_consis}(c) restricts the growth on the bandwidths $b_n$ as $n$ increases. 

\begin{assumption}
    \label{assump:var_consis}
    (Assumption 4.1 from \citet{kojevnikov_limit_2021}). There exists $v > 4$ such that
    \begin{enumerate}[label=\alph*)]
        \item $\sup_{n \geq 1} \max_{i \in \mathcal{N}_n} \|\phi_i(\boldO_i)\|_v < \infty$ a.s.,
        \item $\lim_{n \rightarrow \infty} \sum_{s \geq 1} | \omega(s/b_n)-1| M_n^{\partial}(s;1) \theta_{n,s}^{1-(2/v)} = 0$ a.s., and
        \item $\lim_{n \rightarrow \infty} n^{-1} \sum_{s \geq 1} c_n(s, b_n; 2) \theta_{n,s}^{1-(4/v)} = 0$ a.s.
    \end{enumerate}
\end{assumption}

\subsection{Proof of Proposition 1}

The proof for identification is similar to the standard setting with iid data, with the difference that an empirical average is considered here with the corresponding conditional parallel trends assumption:
\begin{align*}
    &\AEE_t(\gbar_t) \equiv \meanin \AEE_{it}(\gbar_t) \\
    &= \meanin \E[Y_{it}(\gbar_t) - Y_{it}(\gbar_t') | \Gbar_{it} = \gbar_t] \\
    &= \meanin \E[Y_{it}(\gbar_t) - Y_{it^*}(\gbar_t) | \Gbar_{it} = \gbar_t] + \E[ Y_{it^*}(\gbar_t)  - Y_{it}(\gbar_t') | \Gbar_{it} = \gbar_t] \\
    &= \meanin \E[\deltay | \Gbar_{it} = \gbar_t] - \meanin \E[Y_{it}(\gbar_t') -  Y_{it^*}(\gbar_t) | \Gbar_{it} = \gbar_t] \\
    &= \meanin \E[\deltay | \Gbar_{it} = \gbar_t] - \meanin \E[ \E[ Y_{it}(\gbar_t') -  Y_{it^*}(\gbar_t') | \boldX_i, \Gbar_{it} = \gbar_t]  \Gbar_{it} = \gbar_t] \\
    &= \meanin \E[\deltay | \Gbar_{it} = \gbar_t] - \E \left[ \meanin \frac{\Ig}{\PP(\Gbar_{it}=\gbar_t)}  \E[ Y_{it}(\gbar_t') -  Y_{it^*}(\gbar_t') | \boldX_i, \Gbar_{it} = \gbar_t] \right] \\
    &= \meanin \E[\deltay | \Gbar_{it} = \gbar_t] - \E \left[ \meanin \frac{\Ig}{\PP(\Gbar_{it}=\gbar_t)}  \E[ Y_{it}(\gbar_t') -  Y_{it^*}(\gbar_t') | \boldX_i, \Gbar_{it} = \gbar_t'] \right] \\
    &= \meanin \E[\deltay | \Gbar_{it} = \gbar_t] - \E \left[ \meanin \frac{\Ig}{\PP(\Gbar_{it}=\gbar_t)}  \E[\deltay | \boldX_i, \Gbar_{it} = \gbar_t'] \right] \\
    &= \meanin \left\{ \E[\deltay | \Gbar_{it} = \gbar_t] - \E[\E[\deltay | \boldX_i, \Gbar_{it} = \gbar_t'] | \Gbar_{it} = \gbar_t] \right\} \\
    &= \meanin \E[h_{i1}(\Gbar_{it}) (\Delta_{t^*} Y_i - \mu_{i,\gbar_t',t^*}(\boldX_i))],
\end{align*}
where the first equality follows from the definition of $\AEE_{it}(\gbar_t)$; the second equality adds and subtracts $\E[Y_{it^*}(\gbar_t) | \Gbar_{it} = \gbar_t]$; the third equality uses Assumptions 1 -- 2 (causal consistency through exposure mapping) and 3 (no anticipation); the fourth equality uses iterated expectations, the fifth equality re-arranges the expectations; the sixth equality uses Assumption 5 (conditional parallel trends); the seventh equality uses Assumption 1 -- 2 again; and the eighth and ninth equalities re-arrange the terms.

The above equalities provide the outcome regression based identification result of $\AEE_t(\gbar_t)$. Then, the same transformations used in the typical iid setting can be performed on the summand to arrive at the representation given in Proposition 1, i.e., following Theorem 1 in \citet{santanna_doubly_2020}. Specifically, it suffices to show that $\E[h_{i1}(\Gbar_{it}) (\Delta_{t^*} Y_i - \mu_{i,\gbar_t',t^*}(\boldX_i))] = \E[\tau_i(\boldO_i)]$, or equivalently, $\E[h_{i0}(\Delta_{t^*} Y_i - \mu_{i,\gbar_t',t^*}(\boldX_i))] = 0$:
\begin{align*}
    &\E[h_{i0}(\Delta_{t^*} Y_i - \mu_{i,\gbar_t',t^*}(\boldX_i))] \\
    &= \E \left[ \frac{\mathbbm{1}(\Gbar_{it}=\gbar_t')\pi_i(\boldX_i; \gbar_t)}{\pgtwo \pi_i(\boldX_i; \gbar_t')} (\Delta_{t^*} Y_i - \mu_{i,\gbar_t',t^*}(\boldX_i)) \right] \\
    &= \E \left[ \frac{\pi_i(\boldX_i; \gbar_t)}{\pgtwo \pi_i(\boldX_i; \gbar_t')} \E \left[ \mathbbm{1}(\Gbar_{it}=\gbar_t') (\Delta_{t^*} Y_i - \mu_{i,\gbar_t',t^*}(\boldX_i)) \bigg| \boldX_i \right] \right] \\
    &= \E \left[ \frac{\PP(\Gbar_{it}=\gbar_t')\pi_i(\boldX_i; \gbar_t)}{\pgtwo \pi_i(\boldX_i; \gbar_t')} \E \left[ \frac{\mathbbm{1}(\Gbar_{it}=\gbar_t')}{\PP(\Gbar_{it}=\gbar_t')} (\Delta_{t^*} Y_i - \mu_{i,\gbar_t',t^*}(\boldX_i)) \bigg| \boldX_i \right] \right] \\
    &= \E \left[ \frac{\PP(\Gbar_{it}=\gbar_t')\pi_i(\boldX_i; \gbar_t)}{\pgtwo \pi_i(\boldX_i; \gbar_t')} \{\mu_{i,\gbar_t',t^*}(\boldX_i)) - \mu_{i,\gbar_t',t^*}(\boldX_i)) \} \right] \\
    &= 0.
\end{align*}

\subsection{Notation}

As in the main text, let $\boldO_{1:n} \sim \mathbb{P}$. In this work, a nonparametric model $\mathcal{P}$ is assumed where $\mathbb{P} \in \mathcal{P}$. Define the function $\Psi: \mathcal{P} \mapsto \mathbb{R}$ so that $\Psi(\mathbb{P}) = \AEE_{t}(\gbar_t) = \tau$ and $\Psi(\hat{\mathbb{P}}) = \hat \tau$ where $\hat{\mathbb{P}}$ is the estimator distribution. Additionally, let $\Psi_i(\mathbb{P}) = \AEE_{it}(\gbar_t) = \E[\tau_i(\boldO)]$. Unless otherwise noted, all expectations $\E[\cdot]$ are over the distribution $\mathbb{P}$. Note that for the possibly random function $\hat f(O)$, $\E[\hat f(O)] = \int \hat f(o) d\mathbb{P}(o)$. To ease notation, let $\| \cdot \| = \| \cdot \|_{L_2(\mathbb{P})}$ denote the $L_2(\mathbb{P})$ norm, i.e., the ${L_2(\mathbb{P})}$ subscript is dropped for these proofs. Also, let $\mathbb{P}_n$ denote the empirical average, i.e., $\mathbb{P}_nf(O) = \meanin f(O_i)$. Finally, let $a \lesssim b$ denote $a \leq Cb$ where $C$ is a generic constant.

\subsection{Proof of Proposition 2}

\begin{proof}[Proof of Proposition 2]
    Let $\phi^*(\boldO; \mathbb{P})$ as stated in Proposition 2 be a conjectured efficient influence function. Note that in the iid data setting, $\phi^*(\boldO; \mathbb{P})$ is equal to the EIF proposed by \citet{santanna_doubly_2020}. Observe that $\phi^*(\boldO; \mathbb{P})$ is mean zero since 
    \begin{align*}
        \E[\phi^*(\boldO; \mathbb{P})] &= \E \left[\meanin \tau_i(\boldO_i) - \hOne \tau \right] \\
        &= \tau - \tau \\
        &= 0,
    \end{align*}
    and has finite variance since it is a Lipschitz function of bounded random variables. Let $\bar{\mathbb{P}} \in \mathcal{P}$ be a generic distribution and define
    \begin{align*}
        R_2(\bar{\mathbb{P}}, \mathbb{P}) = \Psi(\bar{\mathbb{P}}) - \Psi(\mathbb{P}) - \int \phi^*(\boldo; \bar{\mathbb{P}})d(\bar{\mathbb{P}} - \mathbb{P}),
    \end{align*}
    which is a re-arranged form of the von Mises expansion. By Lemma 2 of \citet{kennedy_semiparametric_2023-1}, if $R_2(\bar{\mathbb{P}}, \mathbb{P})$ is shown to be a second-order remainder term in the sense that it consists of products of differences between $\bar{\mathbb{P}}$ and $\mathbb{P}$, then $\phi^*(\boldO; \mathbb{P})$ is the EIF. The remainder of the proof showing that $R_2(\bar{\mathbb{P}}, \mathbb{P})$ is second-order when there is either network treatment effect homogeneity or network exposure probability homogeneity is given in the proof of Theorem 2. 
\end{proof}

\subsection{Proof of Theorem 1}

\begin{proof}[Proof of \textbf{Theorem 1}]

By Theorem 3.1 in \citet{kojevnikov_limit_2021} and Assumptions 7 -- 9, the following result holds as $n \rightarrow \infty$:
\begin{align*}
    n^{-1} \sumin \bigg\{& (\hhonehat - \hhzerohat)(\deltay - \mmuhat) - \\
    & \E[(\hhonehat - \hhzerohat)(\deltay - \mmuhat)] \bigg\} \rightarrow_p 0.
\end{align*}

Thus, to show $\meanin \hat \tau_i - \E[\tau_i] \rightarrow_p 0$, it suffices to show:
\begin{align}
    n^{-1} \sumin \bigg\{& \E[(\hhonehat - \hhzerohat)(\deltay - \mmuhat)] - \label{eq:consis} \\
    & \E[(\hhone - \hhzero)(\deltay - \mmu)] \bigg\} \rightarrow_p 0. \nonumber
\end{align}

The expression above can be decomposed into the following:
\begin{align} \label{eq:consis2}
    (*) \coloneqq \E \bigg[ n^{-1} \sumin \bigg\{& \underbrace{\big( \hhonehat \deltay - \hhone \deltay \big)}_{\circled{1}} \\
    &- \underbrace{\big( \hhonehat \mmuhat - \hhone \mmuhat \big)}_{\circled{2}}  \nonumber \\
    &- \underbrace{\big( \hhzerohat \deltay - \hhzero \deltay \big)}_{\circled{3}}  \nonumber \\
    &+ \underbrace{\big(\hhzerohat \mmuhat - \hhzero \mmu \big) \bigg\} \bigg]}_{\circled{4}}.  \nonumber
\end{align}
The first term, $\circled{1}$, in \eqref{eq:consis2} can be shown to be equal to:
\begin{align*}
    & \E \bigg[ n^{-1} \sumin \big( \hhonehat \deltay - \hhone \deltay \big) \bigg] \\
    &= \E \bigg[ n^{-1} \sumin \deltay \Ig \left( \frac{\pgone - \pgonehat}{\pgone \pgonehat} \right) \bigg].
\end{align*}
The second term, $\circled{2}$, in \eqref{eq:consis2} can be shown to be equal to:
\begin{align*}
    & \E \bigg[ n^{-1} \sumin \big( \hhonehat \mmuhat - \hhone \mmuhat \big) \bigg] \\
    &= \E \bigg[ \meanin \Ig \left\{ \frac{\mmuhat \pgone - \mmu \pgonehat}{\pgonehat \pgone} \right\} \bigg] \\
    &= \E \bigg[ \meanin \frac{\Ig}{\pgonehat \pgone} \left\{ \mmuhat [\pgone - \pgonehat] + \pgonehat[\mmuhat - \mmu] \right\} \bigg] \\
    &= \E \bigg[ \meanin \frac{\Ig}{\pgonehat \pgone}  \mmuhat [\pgone - \pgonehat] \\
    &\hspace{2em} + \meanin \frac{\Ig}{ \pgone} [\mmuhat - \mmu]  \bigg] .
\end{align*}
The third term, $\circled{3}$, in \eqref{eq:consis2} can be shown to be equal to:
\begin{align*}
    & \E \bigg[ \meanin \big( \hhzerohat \deltay - \hhzero \deltay \big) \bigg] \\
    &= \E \bigg[ \meanin \frac{\deltay \Igprime}{\pgtwohat \pigprimehat \pgtwo \pigprime} \\
    &\hspace{2em} \times \left\{  \pighat \pig \pigprime - \pig \pgtwohat \pigprimehat \right\} \bigg] \\
    &= \E \bigg[ \meanin \frac{\deltay \Igprime}{\pgtwohat \pigprimehat \pgtwo \pigprime} \\
    &\hspace{2em}\times \big\{ \pighat[\pgtwo \pigprime - \pgtwohat \pigprimehat] \\
    &\hspace{4em} + \pgtwohat \pigprimehat[\pighat - \pig] \big\} \bigg] \\
    &= \E \bigg[ \meanin \frac{\deltay \Igprime}{\pgtwohat \pigprimehat \pgtwo \pigprime} \\
    &\hspace{2em}\times \big\{ \pighat[ \pigprime[\pgtwo - \pgtwohat] + \pgtwohat[\pigprime - \pigprimehat] ] \\
    &\hspace{4em} + \pgtwohat \pigprimehat[\pighat - \pig] \big\} \bigg] \\
    &= \E \bigg[ \meanin \frac{\deltay \Igprime}{\pgtwohat \pgtwo}[\pgtwo - \pgtwohat] \\
    &\hspace{2em} + \meanin \frac{\deltay \Igprime}{ \pgtwo}[\pigprime - \pigprimehat] \\
    &\hspace{2em} + \meanin \frac{\deltay \Igprime}{ \pgtwo}[\pighat - \pig] \bigg].
\end{align*}

The fourth term, $\circled{4}$, in \eqref{eq:consis2} can be shown to be equal to:
\begin{align*}
    &\E \bigg[  \meanin \big(\hhzerohat \mmuhat - \hhzero \mmu \big) \bigg] \\
    &= \E \bigg[ \meanin \frac{\Igprime}{\pgtwohat \pigprimehat \pgtwo \pigprime} \\
    &\hspace{2em}\times \big\{ \pighat \mmuhat \pgtwo \pigprime - \pig \mmu \pgtwohat \pigprimehat \big\} \bigg] \\
    &= \E \bigg[ \meanin \frac{\Igprime}{\pgtwohat \pigprimehat \pgtwo \pigprime} \\
    &\hspace{2em}\times \big\{ \pighat \mmuhat[\pgtwo \pigprime - \pgtwohat \pigprimehat] \\
    &\hspace{4em} + \pgtwohat \pigprimehat[\pighat \mmuhat - \pig \mmu] \big\} \bigg] \\
    &= \E \bigg[ \meanin \frac{\Igprime}{\pgtwohat \pigprimehat \pgtwo \pigprime} \\
    &\hspace{2em}\times \big\{ \pighat \mmuhat[\pigprime[\pgtwo - \pgtwohat] \\
    &\hspace{4em} + \pgtwohat[\pigprime - \pigprimehat]] \\
    &\hspace{4em} + \pgtwohat \pigprimehat[\pighat[\mmuhat - \mmu] \\
    &\hspace{6em} + \mmu[\pighat - \pig]] \big\} \bigg] \\
    &= \E \bigg[ \meanin \frac{\Igprime}{\pgtwohat \pgtwo}\mmuhat [\pgtwo - \pgtwohat] \\
    &\hspace{2em} + \meanin \frac{\Igprime}{\pgtwo \pigprime}\mmuhat [\pigprime - \pigprimehat] \\
    &\hspace{2em} + \meanin \frac{\Igprime}{\pgtwo \pigprime} \pighat [\mmuhat - \mmu] \\
    &\hspace{2em} + \meanin \frac{\Igprime}{\pgtwo \pigprime}\mmu [\pighat - \pig] \bigg].
\end{align*}

Consider the expression $\E[\meanin f(\boldO_i, \hat{\mathbb{P}}) (\hat q_i(\boldO_i) - q_i(\boldO_i))] $ for a generic function $f$ and nuisance function estimators $\hat q_i$ and nuisance functions $q_i$, i.e., $q \in \{\pig, \pigprime, \mmu \}$. Then, note the following result,
\begin{align*}
    &\E[ \big| \meanin f(\boldO_i, \hat{\mathbb{P}}) (\hat q_i(\boldO_i) - q_i(\boldO_i)) \big| ] \\
    &\leq  \meanin \E[ \big|f(\boldO_i, \hat{\mathbb{P}})\big| \big|(\hat q_i(\boldO_i) - q_i(\boldO_i)) \big| ] \\
    &\leq \meanin \E[f(\boldO_i, \hat{\mathbb{P}})^2]^{1/2} \E[(\hat q_i(\boldO_i) - q_i(\boldO_i)^2]^{1/2} \\
    &\leq \left( \meanin \E[f(\boldO_i, \hat{\mathbb{P}})^2] \right)^{1/2} \left( \meanin \E[(\hat q_i(\boldO_i) - q_i(\boldO_i)^2] \right)^{1/2} \\
    &= \left( \meanin \E[f(\boldO_i, \hat{\mathbb{P}})^2] \right)^{1/2} \left( \meanin \| \hat q_i(\boldO_i) - q_i(\boldO_i) \|^2 \right)^{1/2},
\end{align*}
where the first inequality uses triangle inequality, second inequality uses the Cauchy-Schwarz inequality, and the third inequality uses H\"older's inequality. Then, under boundedness of $f(\boldO_i, \hat{\mathbb{P}})$ and $\meanin \| \hat q_i(\boldO_i) - q_i(\boldO_i) \|^2 = o_{\mathbb{P}}(1)$, then $\E[\meanin f(\boldO_i, \hat{\mathbb{P}}) (\hat q_i(\boldO_i) - q_i(\boldO_i))] \rightarrow_p 0$ as $n \rightarrow \infty$. 

Applying the above result with the conditions $\meanin \| \hat q_i(\boldO_i) - q_i(\boldO_i) \|^2$ for \\
$q \in \{\pig, \pigprime, \mmu \}$ to the decomposed expression \eqref{eq:consis2} leaves the following:
\begin{align*}
    (*) &= \E \bigg[ n^{-1} \sumin \deltay \Ig \left( \frac{\pgone - \pgonehat}{\pgone \pgonehat} \right) \bigg] \\
    &\hspace{2em} - \E \bigg[ \meanin \frac{\Ig}{\pgonehat \pgone}  \mmuhat [\pgone - \pgonehat] \bigg] \\
    &\hspace{2em} - \E \bigg[ \meanin \frac{\deltay \Igprime}{\pgtwohat \pgtwo}[\pgtwo - \pgtwohat] \bigg] \\
    &\hspace{2em} + \E \bigg[ \meanin \frac{\Igprime}{\pgtwohat \pgtwo}\mmuhat [\pgtwo - \pgtwohat] \bigg] \\
    &= \meanin \frac{\pgone - \pgonehat}{\pgonehat} \E \left[\frac{\Ig}{\pgone} (\deltay - \mmu) \right] \\
    &\hspace{2em} + \meanin \frac{\pgtwo - \pgtwohat}{\pgtwohat \pgtwo}\E \left[ \mmuhat \Igprime - \deltay \Igprime \right] \\
    &= \meanin \frac{\pgone - \pgonehat}{\pgonehat} \E \left[\frac{\Ig}{\pgone} (\deltay - \mmu + \mmu - \mmuhat) \right] \\
    &\hspace{2em} + \meanin \frac{\pgtwo - \pgtwohat}{\pgtwohat \pgtwo}\E \left[ \mmuhat \Igprime - \mmu \Igprime \right] \\
    &= \meanin \frac{\pgone - \pgonehat}{\pgonehat} \Psi_i(\mathbb{P}) \\
    &\hspace{2em} + \meanin \frac{\Ig}{\pgonehat \pgone} (\pgone - \pgonehat) \E \left[(\mmu - \mmuhat) \right] + o_{\mathbb{P}}(1) \\
    &= \meanin \frac{\pgone - \pgonehat}{\pgonehat} \Psi_i(\mathbb{P}) + o_{\mathbb{P}}(1) \\
    &\lesssim \meanin (\pgone - \pgonehat) \Psi_i(\mathbb{P}) + o_{\mathbb{P}}(1) \\
    &\leq \left( \meanin (\pgone - \pgonehat)^2 \right)^{1/2} \left( \meanin \Psi_i(\mathbb{P})^2 \right)^{1/2} + o_{\mathbb{P}}(1) \\
    &\lesssim \left( \meanin (\pgone - \pgonehat)^2 \right)^{1/2} + o_{\mathbb{P}}(1) .
\end{align*}

The remaining term above can be shown to be $o_{\mathbb{P}}(1)$ provided that $\meanin \| \pgone - \pgonehat \|^2 = o_{\mathbb{P}}(1)$. If $\pgonehat = \meanin \Ig$, then $(*)$ can be shown to be consistent for the bias term $S_n^{(1)}$ provided in the main text. Recall that $\bar p(\gbar_t) = \meanin p_1(\gbar_t)$. Then,
\begin{align*}
    &\meanin \Psi_i(\mathbb{P}) \frac{\pgone - \pgonehat}{\pgonehat} - S_n^{(1)} \\
    &= \meanin \Psi_i(\mathbb{P}) \left(\frac{\pgone - \pgonehat}{\pgonehat} - \frac{\pgone - \bar p(\gbar_t)}{\bar p(\gbar_t)} \right) \\
    &\lesssim \meanin \Psi_i(\mathbb{P}) \left\{ \bar p(\gbar_t)[\pgone - \pgonehat] - \pgonehat[\pgone - \bar p(\gbar_t)] \right\} \\
    &= \meanin \Psi_i(\mathbb{P})  [\bar p(\gbar_t) - \pgonehat] \pgone \\
    &= [\bar p(\gbar_t) - \pgonehat]  \meanin \Psi_i(\mathbb{P}) \pgone \\
    &= o_{\mathbb{P}}(1),
\end{align*}
since $\pgonehat - \bar p(\gbar_t) = o_{\mathbb{P}}(1)$ by the dependent data law of large numbers of \citet{kojevnikov_limit_2021}. Thus, \eqref{eq:consis} is proved.

Next, the double robustness property of $\hat{\tau}$ is shown. Consider the estimators:
\begin{align*}
    \meanin \{ \hat{\pi}_i^{0}(\boldX_i;\gbar_t) - \pi_i^{0}(\boldX_i;\gbar_t) \} &\rightarrow_p 0, \\
    \meanin \{ \hat{\pi}_i^{0}(\boldX_i;\gbar_t') - \pi_i^{0}(\boldX_i;\gbar_t') \} &\rightarrow_p 0,
\end{align*}
where $\pi_i^{0}(\boldX_i;\gbar_t') \neq \pig$, i.e., the exposure propensity score model is incorrectly specified. Define the following:
\begin{align*}
    \tau(\pi^{0}) &= \meanin \E[ (\hhone - h_0(\Gbar_t, \boldX_i; \pi_i^{0}))(\deltay - \mmu) ], \\
    \hat{\tau}(\pi^{0}) &= \meanin (\hhonehat - \hat{h}_0(\Gbar_t, \boldX_i; \hat{\pi}_i^{0}))(\deltay - \mmuhat)
\end{align*}
Theorem 1 in \citet{santanna_doubly_2020} shows that $\tau(\pi^{0}) = \AEE$ under Assumptions 1 -- 5. Using the same arguments above, $\hat{\tau}(\pi^{0})-\tau(\pi^{0}) \rightarrow_p 0$. A similar argument can be made by assuming a mis-specified outcome regression for $\mmu$. Thus, $\hat{\tau}$ is a doubly robust estimator in the sense that only consistent estimation of one of: (i) $\pig$ and $\pigprime$, or (ii) $\mmu$ is needed for $\hat{\tau} - \AEE \rightarrow_p 0$. 

\end{proof}

\subsection{Proof of Theorem 2}

Let $\varphi(\boldO_{1:n}; \mathbb{P}) = \meanin \varphi(\boldO_{i}; \mathbb{P})$ denote the influence function of $\Psi(\hat{\mathbb{P}})$. As stated in Proposition 2, we will show that if there is either network effect homogeneity or exposure probability homogeneity, then $\varphi(\boldO_{1:n}; \mathbb{P}) = \phi^*(\boldO_{1:n}; \mathbb{P})$, the EIF of $\Psi(\mathbb{P})$. In the absence of both network effect heterogeneity and exposure probability heterogeneity, $\varphi(\boldO_{1:n}; \mathbb{P}) = \phi_i(\boldO_{i}; \mathbb{P})$ and is equal to the EIF discussed in \citet{santanna_doubly_2020}. 

In the following proof, we generally consider $\varphi(\boldO_{1:n}; \mathbb{P}) = \phi^*(\boldO_{1:n}; \mathbb{P})$ since the extensions for a non-zero adjustment term $S^{\mathrm{adj}}$ for the parametric model of $\pgone$ are trivial for most parts of the proof. Instead, by developing the asymptotic results for $\varphi(\boldO_{1:n}; \mathbb{P}) = \phi^*(\boldO_{1:n}; \mathbb{P})$, the bias stemming from the existence of both network effect and exposure probability heterogeneity is made explicit, which also shows the derivation of $S^{\mathrm{adj}}$. 


\begin{proof}[Proof of \textbf{Theorem 2}]

We show that the proposed estimator follows the form of a von Mises expansion of $\Psi(\hat{\mathbb{P}})$ about $\Psi(\mathbb{P})$,
\begin{align*}
    \Psi(\hat{\mathbb{P}}) - \Psi(\mathbb{P}) &= - \mathbb{P} \{ \phi^*(\hat{\mathbb{P}}) \} + R_2(\hat{\mathbb{P}}, \mathbb{P}).
\end{align*}
\noindent
In the absence of network effect heterogeneity, the one-step influence function-based estimator is then $\Psi(\hat{\mathbb{P}})$ plus an estimate of the so-called drift term $\mathbb{P} \{ \phi^*(\hat{\mathbb{P}}) \}$ \citep{kennedy_semiparametric_2023}. Below, $\phi^*(\hat{\mathbb{P}})$ is shown to equal $0$ so that the one-step estimator is equal to $\hat{\tau}$.
\begin{align*}
    \phi^*(\hat{\mathbb{P}}) &= \meanin \hat{\tau}_i(\boldO_i) - \meanin \frac{\mathbbm{1}(\Gbar_{it}=\gbar_t)}{\mathbb{P}_n(\mathbbm{1}(\Gbar_{it}=\gbar_t))} \hat \tau \\
    &= \meanin \hat{\tau}_i(\boldO_i) - \hat \tau \times \meanin \frac{\mathbbm{1}(\Gbar_{it}=\gbar_t)}{\mathbb{P}_n(\mathbbm{1}(\Gbar_{it}=\gbar_t))} \\
    &= \meanin \hat{\tau}_i(\boldO_i) - \hat \tau \\
    &= 0,
\end{align*}
\noindent
where the third equality follows from $\mathbb{P}_n \{ \frac{\mathbbm{1}(\Gbar_t=\gbar_t)}{\mathbb{P}_n \{\mathbbm{1}(\Gbar_t=\gbar_t)\}} \} = 1$. Then, $\Psi(\hat{\mathbb{P}}) - \Psi(\mathbb{P})$ can be further decomposed (where $\mathbb{P}_n \phi^* = \phi^*$ since $\phi^*$ is already a sample average),
\begin{align*}
    \Psi(\hat{\mathbb{P}}) - \Psi(\mathbb{P}) &= - \mathbb{P} \{ \phi^*(\hat{\mathbb{P}}) \} + R_2(\hat{\mathbb{P}}, \mathbb{P}), \\
    &= (\mathbb{P}_n - \mathbb{P})\phi^*(\hat{\mathbb{P}}) - (\mathbb{P}_n - \mathbb{P})\phi^*(\mathbb{P}) + (\mathbb{P}_n - \mathbb{P})\phi^*(\mathbb{P}) + R_2(\hat{\mathbb{P}}, \mathbb{P}) \\
    &= (\mathbb{P}_n - \mathbb{P})\phi^*(\mathbb{P}) + (\mathbb{P}_n - \mathbb{P})(\phi^*(\hat{\mathbb{P}}) - \phi^*(\mathbb{P})) + R_2(\hat{\mathbb{P}}, \mathbb{P}),
\end{align*}
\noindent
where the second equality uses $\mathbb{P}_n \phi^*(\hat{\mathbb{P}}) = 0$ and adds and subtracts $(\mathbb{P}_n - \mathbb{P})\phi^*(\mathbb{P})$; and the third equality re-arranges terms. The root-$n$ scaled first term, $\sqrt{n}(\mathbb{P}_n - \mathbb{P})\phi^*(\mathbb{P})$ can be shown to converge to $N(0, 1)$ by the central limit theorem (Theorem 3.2) from \citet{kojevnikov_limit_2021} and Assumptions 6 -- 9: 
\begin{align*}
    \frac{\sqrt{n}(\mathbb{P}_n - \mathbb{P})\phi^*(\mathbb{P})}{\sqrt{\Var(n^{-1/2}\sumin \phi_i(\mathbb{P}))}} &= \frac{n^{-1/2} \sumin \phi_i(\mathbb{P})}{\sqrt{\Var(n^{-1/2}\sumin \phi_i(\mathbb{P}))}} \\
    &= \frac{\sumin \phi_i(\mathbb{P})}{\sqrt{\Var(\sumin \phi_i(\mathbb{P}))}} \\
    &\rightarrow_d N(0,1).
\end{align*}
\noindent
Below it is shown that the second term, $(\mathbb{P}_n - \mathbb{P})(\phi^*(\hat{\mathbb{P}}) - \phi^*(\mathbb{P}))$, (empirical process term) and the third term, $R_2(\hat{\mathbb{P}}$, (remainder term) go to zero at the root-$n$ rate under suitable conditions.

The root-$n$ empirical process term $\sqrt{n}(\mathbb{P}_n - \mathbb{P})(\phi^*(\hat{\mathbb{P}}) - \phi^*(\mathbb{P}))$ is shown to be equal to $o_{\mathbb{P}}(1)$. By Assumption 11, nuisance functions and their estimators are in Donsker classes, implying that $\phi^*(\mathbb{P})$ and $\phi^*(\hat{\mathbb{P}})$ are also in the Donsker class since Lipschitz transformations of functions in the Donsker class and indicator functions are in the Donsker class \citep{kennedy_semiparametric_2016}. The following proof follows the proof of Lemma 19.24 of \citet{van_der_vaart_asymptotic_1998}.

First, let $\mathcal{F}$ be a semimetric space with $L_2(\bar{\mathbb{P}})$ metric where $\| \cdot \|_{L_2(\bar{\mathbb{P}})} = \meanin \| \cdot \|_{L_2(\mathbb{P})}$. Denote the empirical process $\mathbb{G}_n f = n^{1/2} \sumin (f(O_i) - \E[f(O_i)])$. By Assumption 11, $\mathbb{G}_n$ converges weakly to a tight, mean zero Gaussian process. Then, using the arguments in Lemma 19.24 of \citet{van_der_vaart_asymptotic_1998}, $\mathbb{G}_n (\hat{f} - f) \rightarrow_p 0$ provided that $\hat{f} - f$ converges to zero in $L_2(\bar{\mathbb{P}})$-norm. 

Thus, to complete the proof, we show that $\meanin \| \phi_i(\hat{\mathbb{P}}) - \phi_i(\mathbb{P}) \|^2 \rightarrow_p 0$. To simplify notation, for the remainder of the proof let $\pi_g = \pi_i(\boldX_i;\gbar_t)$, $\hat{\pi}_g = \hat{\pi}_i(\boldX_i;\gbar_t)$, $\mu = \mu_{i,g',t^*}(\boldX_i)$, $\hat{\mu} = \hat{\mu}_{i,g',t^*}(\boldX_i)$, $p_g = \mathbb{P}(\Ig)$, $\hat{p}_g = \mathbb{P}_n(\Ig)$, $I_g = \Ig$, $h_1 = h_{i1}(\Gbar_{it})$, $\hat{h}_1 = \hat{h}_{i1}(\Gbar_t)$, $h_0 = h_{i0}(\Gbar_{it}, \boldX_i; \pi_i)$, and $\hat{h}_0 = \hat{h}_{i0}(\Gbar_{it}, \boldX_i; \hat{\pi_i})$. 
\begin{align*}
    &\E \left[ \meanin ( \phi_i(\hat{\mathbb{P}}) - \phi_i(\mathbb{P}) )^2 \right] \\
    &= \E \left[ \meanin ( (\hat \tau_i - \tau_i) - (\hat h_1 \Psi(\hat{\mathbb{P}}) - h_1 \Psi(\mathbb{P})) )^2 \right] \\
    &= \meanin \| \hat \tau_i - \tau_i \|^2 - 2 \E \left[ \meanin (\hat \tau_i - \tau_i)(\hat h_1 \Psi(\hat{\mathbb{P}}) - h_1 \Psi(\mathbb{P})) \right] + \E \left[ \meanin (\hat h_1 \Psi(\hat{\mathbb{P}}) - h_1 \Psi(\mathbb{P}))^2 \right].
\end{align*}
Using $\meanin \| \hat \tau_i - \tau_i \|^2 = o(1)$, proved in Section \ref{sec:theorem3}, the first term above goes to zero as does the second term after an application of H\"older's inequality. The last term can also be shown to converge to zero,
\begin{align*}
    & \E \left[ \meanin (\hat h_1 \Psi(\hat{\mathbb{P}}) - h_1 \Psi(\mathbb{P}))^2 \right] \\
    &= \E \left[ \meanin \hat h_1^2 \Psi(\hat{\mathbb{P}})^2 - 2 \hat h_1 h_1 \Psi(\hat{\mathbb{P}}) \Psi(\mathbb{P}) + h_1^2 \Psi(\mathbb{P})^2 \right] \\
    &= \frac{\Psi(\hat{\mathbb{P}})^2}{\hat p_g} - \frac{2}{\hat p_g} \Psi(\hat{\mathbb{P}}) \Psi(\mathbb{P}) + \frac{\Psi(\mathbb{P})^2}{p_g} \\
    &= \frac{\Psi(\hat{\mathbb{P}})}{\hat p_g}(\Psi(\hat{\mathbb{P}}) - \Psi(\mathbb{P})) + \Psi(\mathbb{P})^2(\frac{1}{p_g} - \frac{1}{\hat p_g}) - \frac{\Psi(\mathbb{P})}{\hat p_g}(\Psi(\hat{\mathbb{P}}) - \Psi(\mathbb{P})),
\end{align*}
where the first and last terms converge to zero by boundedness and consistency (Theorem 1), and the second term goes to zero by boundedness and consistency of $\hat p_g$. Thus, $\sqrt{n}(\mathbb{P}_n - \mathbb{P})(\phi^*(\hat{\mathbb{P}}) - \phi^*(\mathbb{P})) = o_{\mathbb{P}}(1)$. 

Next, the remainder term $R_2(\hat{\mathbb{P}}, \mathbb{P})$ is shown to converge to zero. 
\begin{align*}
    R_2(\hat{\mathbb{P}}, \mathbb{P}) &= \Psi(\hat{\mathbb{P}}) + \int \phi^*(\hat{\mathbb{P}}) d\mathbb{P} - \Psi(\mathbb{P}) \\
    &= \Psi(\hat{\mathbb{P}}) + \int \meanin \left[ (\hat{h}_1 - \hat{h}_0)(\Delta_{t^*} Y - \hat{\mu}) - \hat{h}_1 \Psi(\hat{\mathbb{P}}) \right] d\mathbb{P} - \Psi(\mathbb{P}) \\
    &= \Psi(\hat{\mathbb{P}}) - \frac{\Psi(\hat{\mathbb{P}})}{\hat{p}_g} \int \meanin \Ig   d\mathbb{P} + \int \meanin \left[ (\hat{h}_1 - \hat{h}_0)(\Delta_{t^*} Y - \hat{\mu}) \right] d\mathbb{P} - \Psi(\mathbb{P}) \\
    &= \Psi(\hat{\mathbb{P}})\left(\meanin \frac{\hat{p}_g - p_g}{\hat{p}_g} \right) + \int \meanin \left[ (\hat{h}_1 - \hat{h}_0)(\Delta_{t^*} Y - \hat{\mu}) \right] d\mathbb{P} - \Psi(\mathbb{P}) \\
    &= \Psi(\hat{\mathbb{P}})\left(\meanin \frac{\hat{p}_g - p_g}{\hat{p}_g} \right) + \int \meanin \left[ (\hat{h}_1 - \hat{h}_0)(\Delta_{t^*} Y - \hat{\mu}) - (h_1 - h_0)(\Delta_{t^*} Y - \mu) \right] d\mathbb{P} \\
    &= \Psi(\hat{\mathbb{P}})\left(\meanin \frac{\hat{p}_g - p_g}{\hat{p}_g} \right) \\
    &\hspace{2em} + \int \meanin \left[ (\hat{h}_1-h_1)\Delta_{t^*} Y - (\hat{h}_0 - h_0) \Delta_{t^*} Y - (\hat{h}_1 - \hat{h}_0) \hat{\mu} + (h_1 - h_0) \mu \right] d\mathbb{P} \\
    &= \Psi(\hat{\mathbb{P}})\left(\meanin \frac{\hat{p}_g - p_g}{\hat{p}_g} \right) \\
    &\hspace{2em} + \int \meanin \left[ (\hat{h}_1-h_1)\Delta_{t^*} Y - (\hat{h}_0 - h_0) \Delta_{t^*} Y + (h_1 \mu - \hat{h}_1 \hat{\mu}) - (h_0 \mu - \hat{h}_0 \hat{\mu}) \right] d\mathbb{P}.
\end{align*}

\noindent
Recall that: 
\begin{align*}
    &h_1 = \hOne \\
    &h_0 = \hZero,
\end{align*}
and observe that $p_g \equiv \E[\mathbbm{1}(\Gbar_{it}=\gbar_t)] = \mathrm{E}[(\mathbbm{1}(\Gbar_{it}=\gbar_t'){\pi}_i(\boldX_i; \gbar_t)/{\pi}_i(\boldX_i; \gbar_t')]$. The estimator $\mathbb{P}_n[(\mathbbm{1}(\Gbar_{it}=\gbar_t')\hat{\pi}_i(\boldX_i; \gbar_t)/\hat{\pi}_i(\boldX_i; \gbar_t')]$ is asymptotically equivalent at the $\sqrt{n}$-rate to $\mathbb{P}_n[\mathbbm{1}(\Gbar_{it}=\gbar_t]$ so to simplify the proof we let
\begin{align*}
    &h_0 = \frac{I_{g'} \pi_g}{\pi_{g'} p_g},~~\hat h_0 = \frac{I_{g'} \hat \pi_g}{\hat \pi_{g'} \hat p_g} 
\end{align*}
for the remainder of this section. Now, each term in the integral above is analyzed separately. The first term is equal to:
\begin{align*}
    \meanin (\hat{h}_1-h_1)\Delta_{t^*} Y &= \meanin \left[ \frac{I_g}{\hat{p}_g} - \frac{I_g}{p_g} \right] \Delta_{t^*} Y \\
    &= \meanin \left[ \frac{I_g (p_g - \hat{p}_g)}{\hat{p}_g p_g} \right] \Delta_{t^*} Y.
\end{align*}
where the first equality follows from definition and the second equality re-arranges terms. Next,
\begin{align*}
    (\hat{h}_0 - h_0) \Delta_{t^*} Y &= \left[ \frac{I_{g'} \hat \pi_g}{\hat \pi_{g'} \hat p_g} - \frac{I_{g'} \pi_g}{\pi_{g'} p_g} \right] \Delta_{t^*} Y \\
    &= \bigg[ \frac{I_{g'} \hat \pi_g}{\hat \pi_{g'} \hat p_g} - \frac{I_{g'} \hat \pi_g (p_g - \hat p_g)}{\hat \pi_{g'} \hat p_g p_g} + \frac{I_{g'} \hat \pi_g (p_g - \hat p_g)}{\hat \pi_{g'} \hat p_g p_g} - \frac{I_{g'} \pi_g}{\pi_{g'} p_g} \bigg] \Delta_{t^*} Y \\
    &= \bigg[ \frac{I_{g'} \hat \pi_g }{\hat \pi_{g'} p_g } - \frac{I_{g'} \pi_g}{\pi_{g'} p_g} + \frac{I_{g'} \hat \pi_g (p_g - \hat p_g)}{\hat \pi_{g'} \hat p_g p_g} \bigg] \Delta_{t^*} Y \\
    &= \bigg[ \frac{I_{g'} \hat \pi_g }{\hat \pi_{g'} p_g } - \frac{I_{g'} \hat \pi_g (\pi_{g'} - \hat \pi_{g'})}{\hat \pi_{g'} p_g \pi_{g'}} + \frac{I_{g'} \hat \pi_g (\pi_{g'} - \hat \pi_{g'})}{\hat \pi_{g'} p_g \pi_{g'}} - \frac{I_{g'} \pi_g}{\pi_{g'} p_g} + \frac{I_{g'} \hat \pi_g (p_g - \hat p_g)}{\hat \pi_{g'} \hat p_g p_g} \bigg] \Delta_{t^*} Y \\
    &= \bigg[ \frac{I_{g'} \hat \pi_g}{ p_g \pi_{g'}} + \frac{I_{g'} \hat \pi_g (\pi_{g'} - \hat \pi_{g'})}{\hat \pi_{g'} p_g \pi_{g'}} - \frac{I_{g'} \pi_g}{\pi_{g'} p_g} + \frac{I_{g'} \hat \pi_g (p_g - \hat p_g)}{\hat \pi_{g'} \hat p_g p_g} \bigg] \Delta_{t^*} Y \\
    &= \bigg[ \frac{I_{g'} (\hat \pi_g - \pi_g)}{ \pi_{g'} p_g } + \frac{I_{g'} \hat \pi_g (\pi_{g'} - \hat \pi_{g'})}{\hat \pi_{g'} \pi_{g'} p_g} + \frac{I_{g'} \hat \pi_g (p_g - \hat p_g)}{\hat \pi_{g'} \hat p_g p_g} \bigg] \Delta_{t^*} Y.
\end{align*}
Next, the third term equals
\begin{align*}
    h_1 \mu - \hat{h}_1 \hat{\mu} &= \frac{I_g \mu}{p_g} - \frac{I_g \hat \mu}{\hat p_g} \\
    &= \frac{I_g \mu}{p_g} - \frac{I_g \hat \mu}{\hat p_g} + \frac{I_g \hat \mu (p_g \hat p_g)}{\hat p_g p_g} - \frac{I_g \hat \mu (p_g \hat p_g)}{\hat p_g p_g} \\
    &= \frac{I_g (\mu - \hat \mu)}{p_g} - \frac{I_g \hat \mu (p_g - \hat p_g)}{\hat p_g p_g}.
\end{align*}
The fourth term equals
\begin{align*}
    &h_0 \mu - \hat{h}_0 \hat{\mu} \\
    &= \frac{I_{g'} \pi_g \mu}{\pi_{g'} p_g} - \frac{I_{g'} \hat \pi_g \hat \mu}{\hat \pi_{g'} \hat p_g} \\
    &= \frac{I_{g'} \pi_g \mu}{\pi_{g'} p_g} - \frac{I_{g'} \hat \pi_g \hat \mu}{\hat \pi_{g'} \hat p_g} + \frac{I_{g'} \hat \pi_g (p_g - \hat p_g) \hat \mu}{\hat \pi_{g'} p_g \hat p_g} - \frac{I_{g'} \hat \pi_g (p_g - \hat p_g) \hat \mu}{\hat \pi_{g'} p_g \hat p_g} \\
    &= \frac{I_{g'} \pi_g \mu}{\pi_{g'} p_g} - \frac{I_{g'} \hat \pi_g \hat \mu}{\hat \pi_{g'} p_g} - \frac{I_{g'} \hat \pi_g (p_g - \hat p_g) \hat \mu}{\hat \pi_{g'} p_g \hat p_g} \\
    &= \frac{I_{g'} \pi_g \mu}{\pi_{g'} p_g} - \frac{I_{g'} \hat \pi_g \hat \mu}{\hat \pi_{g'} p_g} + \frac{I_{g'} \hat \pi_g \hat \mu (\pi_{g'} - \hat \pi_{g'})}{\hat \pi_{g'} \pi_{g'} p_g} - \frac{I_{g'} \hat \pi_g \hat \mu (\pi_{g'} - \hat \pi_{g'})}{\hat \pi_{g'} \pi_{g'} p_g} - \frac{I_{g'} \hat \pi_g (p_g - \hat p_g) \hat \mu}{\hat \pi_{g'} p_g \hat p_g} \\
    &= \frac{I_{g'} (\pi_g \mu - \hat \pi_g \hat \mu)}{\pi_{g'} p_g} - \frac{I_{g'} \hat \pi_g \hat \mu (\pi_{g'} - \hat \pi_{g'})}{\hat \pi_{g'} \pi_{g'} p_g} - \frac{I_{g'} \hat \pi_g (p_g - \hat p_g) \hat \mu}{\hat \pi_{g'} p_g \hat p_g} \\
    &= \frac{I_{g'} (\pi_g(\mu - \hat \mu) + (\hat \mu - \mu)(\pi_g - \hat \pi_g) + \mu(\pi_g - \hat \pi_g))}{\pi_{g'} p_g} - \frac{I_{g'} \hat \pi_g \hat \mu (\pi_{g'} - \hat \pi_{g'})}{\hat \pi_{g'} \pi_{g'} p_g} - \frac{I_{g'} \hat \pi_g (p_g - \hat p_g) \hat \mu}{\hat \pi_{g'} p_g \hat p_g}.
\end{align*}
Now, the remainder term can be written as:
\begin{align*}
    &R_2(\hat{\mathbb{P}}, \mathbb{P}) = \Psi(\hat{\mathbb{P}})\left(\meanin \frac{\hat{p}_g - p_g}{\hat{p}_g} \right) \\
    &\hspace{2em} + \meanin \bigg\{ \E \left[ \left( \frac{I_g (p_g - \hat{p}_g)}{\hat{p}_g p_g} \right) \Delta_{t^*} Y \right]  \\
    &\hspace{2em} - \E \left[ \left( \frac{I_{g'} (\hat \pi_g - \pi_g)}{ \pi_{g'} p_g } + \frac{I_{g'} \hat \pi_g (\pi_{g'} - \hat \pi_{g'})}{\hat \pi_{g'} \pi_{g'} p_g} + \frac{I_{g'} \hat \pi_g (p_g - \hat p_g)}{\hat \pi_{g'} \hat p_g p_g} \right) \Delta_{t^*} Y \right]  \\
    &\hspace{2em} + \E \left[ \frac{I_g (\mu - \hat \mu)}{p_g} - \frac{I_g \hat \mu (p_g - \hat p_g)}{\hat p_g p_g} \right] \\
    &\hspace{2em} - \E \bigg[ \frac{I_{g'} (\pi_g(\mu - \hat \mu) + (\hat \mu - \mu)(\pi_g - \hat \pi_g) + \mu(\pi_g - \hat \pi_g))}{\pi_{g'} p_g}  \\
    &\hspace{4em} - \frac{I_{g'} \hat \pi_g \hat \mu (\pi_{g'} - \hat \pi_{g'})}{\hat \pi_{g'} \pi_{g'} p_g} - \frac{I_{g'} \hat \pi_g (p_g - \hat p_g) \hat \mu}{\hat \pi_{g'} p_g \hat p_g} \bigg] \bigg \} \\
    &= \Psi(\hat{\mathbb{P}})\left(\meanin \frac{\hat{p}_g - p_g}{\hat{p}_g} \right) \\
    &\hspace{2em} \meanin \bigg\{ -\E \left[ \frac{I_{g'}(\hat \pi_g - \pi_g) \Delta Y}{p_g \pi_{g'}} - \frac{I_{g'} \mu (\pi_g - \hat \pi_g)}{\pi_{g'} p_g} \right] \\
    &\hspace{2em} - \E \left[ \frac{I_{g'}}{\pi_{g'} p_g}(\hat \mu - \mu)(\pi_g - \hat \pi_g) \right] \\
    &\hspace{2em} - \E \left[\frac{I_{g'} \hat \pi_g (\pi_{g'} - \hat \pi_{g'}) \Delta Y}{\hat \pi_{g'} p_g \pi_{g'}} - \frac{I_{g'} \hat \pi_g \hat \mu (\pi_{g'} - \hat \pi_{g'})}{\hat \pi_{g'} \pi_{g'} p_g} \right] \\
    &\hspace{2em} + \E \left[\frac{I_g (\mu - \hat \mu)}{p_g} - \frac{I_{g'} \pi_g (\mu - \hat \mu)}{\pi_{g'} p_g} \right] \\
    &\hspace{2em} + \E \left[\frac{I_g (p_g - \hat p_g) \Delta Y}{p_g \hat p_g} - \frac{I_g \hat \mu (p_g - \hat p_g)}{\hat p_g p_g} \right] \\
    &\hspace{2em} - \E \left[\frac{I_{g'} \hat \pi_g (p_g - \hat p_g) \Delta Y}{\hat \pi_{g'} \hat p_g p_g} - \frac{I_{g'} \hat \pi_g (p_g - \hat p_g) \hat \mu}{\hat \pi_{g'} p_g \hat p_g} \right] \bigg\} \\
    &= \Psi(\hat{\mathbb{P}})\left(\meanin \frac{\hat{p}_g - p_g}{\hat{p}_g} \right) \\
    &\hspace{2em} \meanin \bigg\{ \E \left[ \frac{1}{p_g}(\hat \mu - \mu)(\hat \pi_g - \pi_g) +  \frac{I_g(p_g - \hat p_g)}{p_g \hat p_g}(\Delta Y - \hat \mu) + \frac{I_{g'} \hat \pi_g (\hat p_g - p_g)}{\hat \pi_{g'} \hat p_g p_g} (\Delta Y - \hat \mu) \right] \bigg\},
\end{align*}
where the first equality substitutes in the previous results, the second equality re-arranges terms, and the third equality uses the following results:
\begin{align*}
    \E \left[ \frac{I_{g'}(\hat \pi_g - \pi_g) \Delta Y}{p_g \pi_{g'}} \right] &= \E \left[ \frac{(\hat \pi_g - \pi_g)}{p_g \pi_{g'}} \E[I_{g'} \Delta Y | \boldX] \right] \\
    &= \E \left[ \frac{(\hat \pi_g - \pi_g)}{p_g \pi_{g'}} \mu \pi_{g'} \right] \\
    &= \E \left[ \frac{(\hat \pi_g - \pi_g)}{p_g} \mu  \right] \\
    \E \left[\frac{I_{g'} \mu (\pi_g - \hat \pi_g)}{\pi_{g'} p_g} \right] &= \E \left[\frac{\mu (\pi_g - \hat \pi_g)}{\pi_{g'} p_g} \E[I_{g'} | \boldX] \right] \\
    &= \E \left[\frac{ \mu (\pi_g - \hat \pi_g)}{p_g} \right],
\end{align*}
which shows that $\E \left[ \frac{I_{g'}(\hat \pi_g - \pi_g) \Delta Y}{p_g \pi_{g'}} \right] - \E \left[\frac{I_{g'} \mu (\pi_g - \hat \pi_g)}{\pi_{g'} p_g} \right] = 0$. A similar strategy can be used to show that $\E \left[\frac{I_{g'} \hat \pi_g (\pi_{g'} - \hat \pi_{g'}) \Delta Y}{\hat \pi_{g'} p_g \pi_{g'}} - \frac{I_{g'} \hat \pi_g \hat \mu (\pi_{g'} - \hat \pi_{g'})}{\hat \pi_{g'} \pi_{g'} p_g} \right] = 0$ and $\E \left[\frac{I_g (\mu - \hat \mu)}{p_g} - \frac{I_{g'} \pi_g (\mu - \hat \mu)}{\pi_{g'} p_g} \right] = 0$. The same result is used to simplify the other quantities in the last equality. 

Next, the remaining terms are simplified as follows:
\begin{align*}
    \E \left[ \frac{I_g(p_g - \hat p_g)}{p_g \hat p_g}(\Delta Y - \hat \mu) \right] &=  \frac{p_g - \hat p_g}{\hat p_g} \E \left[ \frac{I_g}{p_g}(\Delta Y - \mu + \mu - \hat \mu) \right] \\
    &= \frac{p_g - \hat p_g}{\hat p_g} \left\{ \E \left[ \frac{I_g}{p_g}(\Delta Y - \mu) \right] - \E \left[ \frac{I_g}{p_g} (\hat \mu - \mu) \right] \right\} \\
    &= \frac{p_g - \hat p_g}{\hat p_g} \left\{ \Psi_i(\mathbb{P}) - \E \left[ \frac{I_g}{p_g} (\hat \mu - \mu) \right] \right\}, \\
    \E \left[ \frac{I_{g'} \hat \pi_g (\hat p_g - p_g)}{\hat \pi_{g'} \hat p_g p_g} (\Delta Y - \hat \mu) \right] &= \frac{\hat p_g - p_g}{ \hat p_g} \E \left[ \frac{I_{g'} \hat \pi_g }{\hat \pi_{g'} p_g} (\Delta Y - \hat \mu) \right] \\
    &= \frac{\hat p_g - p_g}{ \hat p_g} \E \left[ \frac{\hat \pi_g }{\hat \pi_{g'} p_g}\E[\Delta Y I_{g'} | \boldX] - \frac{\pi_{g'} \hat \pi_g }{\hat \pi_{g'} p_g} \hat \mu \right] \\
    &= \frac{\hat p_g - p_g}{ \hat p_g} \E \left[ \frac{\hat \pi_g }{\hat \pi_{g'} p_g} \mu \pi_{g'} - \frac{\pi_{g'} \hat \pi_g }{\hat \pi_{g'} p_g} \hat \mu \right] \\
    &= \frac{\hat p_g - p_g}{ \hat p_g} \E \left[ \frac{\pi_{g'} \hat \pi_g }{\hat \pi_{g'} p_g} (\mu - \hat \mu) \right] \\
    &= \frac{\hat p_g - p_g}{ \hat p_g} \E \bigg[ \frac{\pi_{g'} \hat \pi_g }{\hat \pi_{g'} p_g} (\mu - \hat \mu) - \frac{\pi_{g'} \hat \pi_g (\mu - \hat \mu)(\pi_{g'} - \hat \pi_{g'})}{\hat \pi_{g'} \pi_{g'} p_g} \\
    &\hspace{4em} + \frac{\pi_{g'} \hat \pi_g (\mu - \hat \mu)(\pi_{g'} - \hat \pi_{g'})}{\hat \pi_{g'} \pi_{g'} p_g} \bigg] \\
    &= \frac{\hat p_g - p_g}{ \hat p_g} \E \left[ \frac{\hat \pi_g (\mu - \hat \mu)}{p_g} + \frac{\hat \pi_g}{\hat \pi_{g'} p_g}(\mu - \hat \mu)(\pi_{g'} - \hat \pi_{g'})  \right].
\end{align*}

Using the above results and substituting into the remainder:
\begin{align*}
    &R_2(\hat{\mathbb{P}}, \mathbb{P}) = \Psi(\hat{\mathbb{P}})\left(\meanin \frac{\hat{p}_g - p_g}{\hat{p}_g} \right) - \meanin \frac{\hat p_g - p_g}{\hat p_g} \Psi_i(\mathbb{P}) \\
    &\hspace{2em} + \meanin \bigg\{ \E \left[ \frac{1}{p_g}(\hat \mu - \mu)(\hat \pi_g - \pi_g) \right] - \frac{p_g - \hat p_g}{\hat p_g}  \E \left[ \frac{I_g}{p_g} (\hat \mu - \mu) \right] \\
    &\hspace{4em} - \frac{p_g - \hat p_g}{\hat p_g} \E \left[ \frac{\hat \pi_g (\mu - \hat \mu)}{p_g} + \frac{\hat \pi_g}{\hat \pi_{g'} p_g}(\mu - \hat \mu)(\pi_{g'} - \hat \pi_{g'}) \right]  \bigg\} \\
    &= \Psi(\hat{\mathbb{P}})\left(\meanin \frac{\hat{p}_g - p_g}{\hat{p}_g} \right) - \meanin \frac{\hat p_g - p_g}{\hat p_g} \Psi_i(\mathbb{P}) \\
    &\hspace{2em} + \meanin \bigg\{ \E \left[ \frac{1}{p_g}(\hat \mu - \mu)(\hat \pi_g - \pi_g) \right] + \frac{p_g - \hat p_g}{\hat p_g}  \E \left[\frac{\hat \pi_g (\hat \mu -  \mu)}{p_g} - \frac{I_g}{p_g} (\hat \mu - \mu) \right] \\
    &\hspace{4em} - \frac{p_g - \hat p_g}{\hat p_g} \E \left[ \frac{\hat \pi_g}{\hat \pi_{g'} p_g}(\mu - \hat \mu)(\pi_{g'} - \hat \pi_{g'}) \right]  \bigg\} \\
    &= \Psi(\hat{\mathbb{P}})\left(\meanin \frac{\hat{p}_g - p_g}{\hat{p}_g} \right) - \meanin \frac{\hat p_g - p_g}{\hat p_g} \Psi_i(\mathbb{P}) \\
    &\hspace{2em} + \meanin \bigg\{ \E \left[ \frac{1}{p_g}(\hat \mu - \mu)(\hat \pi_g - \pi_g) \right] + \frac{p_g - \hat p_g}{\hat p_g}  \E \left[\frac{1}{p_g}(\hat \pi_g - \pi_g)(\hat \mu -  \mu) \right] \\
    &\hspace{4em} - \frac{p_g - \hat p_g}{\hat p_g} \E \left[ \frac{\hat \pi_g}{\hat \pi_{g'} p_g}(\mu - \hat \mu)(\pi_{g'} - \hat \pi_{g'}) \right]  \bigg\} \\
    &= \Psi(\hat{\mathbb{P}})\left(\meanin \frac{\hat{p}_g - p_g}{\hat{p}_g} \right) - \meanin \frac{\hat p_g - p_g}{\hat p_g} \Psi_i(\mathbb{P}) \\
    &\hspace{2em} + \meanin \bigg\{ \frac{1}{\hat p_g}  \E \left[ (\hat \pi_g - \pi_g)(\hat \mu -  \mu) \right]  - \frac{p_g - \hat p_g}{\hat p_g} \E \left[ \frac{\hat \pi_g}{\hat \pi_{g'} p_g}(\mu - \hat \mu)(\pi_{g'} - \hat \pi_{g'}) \right]  \bigg\} \\
    &= \Psi(\hat{\mathbb{P}})\left(\meanin \frac{\hat{p}_g - p_g}{\hat{p}_g} \right) - \meanin \frac{\hat p_g - p_g}{\hat p_g} \Psi_i(\mathbb{P}) \\
    &\hspace{2em} + \meanin \bigg\{ \frac{1}{\hat p_g}  \E \left[ (\hat \pi_g - \pi_g)(\hat \mu -  \mu) \right]  - \frac{p_g - \hat p_g}{\hat p_g} \E \left[ \frac{\hat \pi_g}{\hat \pi_{g'} p_g}(\mu - \hat \mu)(\pi_{g'} - \hat \pi_{g'}) \right]  \bigg\}.
\end{align*}

From the expression above, it can be seen that if there is network treatment effect homogeneity or network exposure probability homogeneity then $R_2(\hat{\mathbb{P}}, \mathbb{P})$ is second-order in the sense that is can be represented by a sum of products of differences between distributions $\hat{\mathbb{P}}$ and $ \mathbb{P}$, thus completing the proof of Proposition 2. The remainder of this proof shows that $\sqrt{n} R_2(\hat{\mathbb{P}}, \mathbb{P}) = o_{\mathbb{P}}(1)$ under suitable conditions. 

Since $p_g$ and $\hat p_g$ are bounded, it suffices to analyze the root-$n$ convergence properties of the following product terms: 
\begin{align} 
    &\Psi(\hat{\mathbb{P}})\left(\meanin (\frac{\hat{p}_g - p_g}{\hat p_g}) \right) - \meanin (\frac{\hat{p}_g - p_g}{\hat p_g}) \Psi_i(\mathbb{P}), \label{eq:prod1} \\
    &\meanin \E[(\hat \mu - \mu)(\hat \pi_g - \pi_g)], \label{eq:prod2} \\
    &\meanin \E[(\hat \mu - \mu)(\hat \pi_{g'} - \pi_{g'})]. \label{eq:prod3}
\end{align}
The expression \eqref{eq:prod1} can be decomposed as:
\begin{align*}
    &\Psi(\hat{\mathbb{P}})\left(\meanin (\frac{\hat{p}_g - p_g}{\hat p_g}) \right) - \meanin (\frac{\hat{p}_g - p_g}{\hat p_g}) \Psi_i(\mathbb{P}) \\
    &= \left(\meanin \Psi_i(\hat{\mathbb{P}}) \right) \left(\meanin (\frac{\hat{p}_g - p_g}{\hat p_g}) \right) - \left(\meanin \Psi_i(\mathbb{P}) \right) \left(\meanin (\frac{\hat{p}_g - p_g}{\hat p_g}) \right) \\
    &\hspace{2em} + \left(\meanin \Psi_i(\mathbb{P}) \right) \left(\meanin (\frac{\hat{p}_g - p_g}{\hat p_g}) \right)  - \meanin (\frac{\hat{p}_g - p_g}{\hat p_g}) \Psi_i(\mathbb{P}) \\
    &= \left(\meanin \Psi_i(\hat{\mathbb{P}}) - \Psi_i(\mathbb{P}) \right) \left(\meanin (\frac{\hat{p}_g - p_g}{\hat p_g}) \right) \\
    &\hspace{2em} + \left(\meanin \Psi_i(\mathbb{P}) \right) \left(\meanin (\frac{\hat{p}_g - p_g}{\hat p_g}) \right)  - \meanin (\frac{\hat{p}_g - p_g}{\hat p_g}) \Psi_i(\mathbb{P}) \\
    &= \left(\meanin \Psi_i(\mathbb{P}) \right) \left(\meanin (\frac{\hat{p}_g - p_g}{\hat p_g}) \right)  - \meanin (\frac{\hat{p}_g - p_g}{\hat p_g}) \Psi_i(\mathbb{P}) + o_{\mathbb{P}}(n^{-1/2}) \\
    &= \meanin (\frac{\hat{p}_g - p_g}{\hat p_g}) (\Psi(\mathbb{P}) - \Psi_i(\mathbb{P})) + o_{\mathbb{P}}(n^{-1/2}).
\end{align*}
If there is homogeneity in either the unconditional exposure probability or exposure effect, i.e., $\PP(\Gbar_{it}=\gbar_t) = \PP(\Gbar_{kt}=\gbar_t)$ or $\E[\tau_i] = \E[\tau_k]$ for any $i,k$, then the above term is equal to $0 + o_{\mathbb{P}}(1) = o_{\mathbb{P}}(1)$. However, if there is network heterogeneity in both exposure probability and exposure effect then the remaining term is $O_{\mathbb{P}}(1)$ and is equal to $\sqrt{n}\Cov_n(\Psi_i(\mathbb{P}), (\hat p_g - p_g)/\hat p_g)$ where $\Cov_n$ denotes the sample covariance function. With an application of H\"older's inequality, the remaining term can be shown to be bounded by $\left(\meanin (\hat p_g - p_g)^2 \right)^{1/2}$. To adjust for this remaining term, assume a parametric model for $p_g$ such as a logistic regression, indexed by finite dimensional parameters $\eta$. By Assumption \ref{assump:reg-param}, $\left(\meanin (\hat p_g - p_g)^2 \right)^{1/2} = \meanin \varphi_{p_g}(\boldO_i) + o_{\mathbb{P}}(1)$ and that a CLT holds for $\meanin \varphi_{p_g}(\boldO_i)$ where $\varphi_{p_g}(\boldO_i)$ is an influence function of $\hat p_g$. Then, a simple adjustment to the influence function is $\varphi(\boldO_i) = \phi_i(\boldO_i) + \varphi_{p_g}(\boldO_i)$. Consider if $\varphi_{p_g}(\boldO_i) = I(\eta)^{-1} S_{\eta}(\boldO)$ where $S_{\eta}(\boldO)$ is the score function of the parametric model and $I(\eta)$ is the Fisher information. Then, the influence function can be adjusted with the projection of $\phi_i$ on the score function $S_{\eta}(\boldO)$, i.e., $\varphi_i = \phi_i - \E[\boldsymbol{\phi} S_{\eta}^{\top}] I(\eta)^{-1} S_{\eta}(\boldO)$. 

Expressions \eqref{eq:prod2} and \eqref{eq:prod3} can be analyzed using the triangle inequality and the Cauchy-Schwarz inequality:
\begin{align*}
    \big| \E \left[ n^{-1}\sum_{i=1}^{n} (\hat \mu - \mu)(\hat \pi_g - \pi_g) \right] \big| &\leq n^{-1}\sum_{i=1}^{n}  \big| \E \left[ (\hat \mu - \mu)(\hat \pi_g - \pi_g) \right] \big| \\
    &\leq n^{-1}\sum_{i=1}^{n} \| \hat \mu - \mu \| \times \| \hat \pi_g - \pi_g \| \\
    &\leq \left( n^{-1}\sum_{i=1}^{n} \| \hat \mu - \mu \|^2  \right)^{1/2} \left( n^{-1}\sum_{i=1}^{n} \| \hat \pi_g - \pi_g \|^2  \right)^{1/2}.
\end{align*}
Thus, a sufficient condition for the term to converge to 0 is \\
$\left( n^{-1}\sum_{i=1}^{n} \| \hat \mu - \mu \|^2  \right)^{1/2} \left( n^{-1}\sum_{i=1}^{n} \| \hat \pi_g - \pi_g \|^2  \right)^{1/2} = o_{\mathbb{P}}(n^{-1/2})$, which can be satisfied if \\
$\left( n^{-1}\sum_{i=1}^{n} \| \hat \mu - \mu \|^2  \right)^{1/2} =  o_{\mathbb{P}}(n^{-1/4})$ and $\left( n^{-1}\sum_{i=1}^{n} \| \hat \pi_g - \pi_g \|^2  \right)^{1/2} =  o_{\mathbb{P}}(n^{-1/4})$.

\end{proof}

\subsection{Proof of Theorem 3} \label{sec:theorem3}

We first prove a stronger consistency result $\meanin \| \hat \tau_i - \tau_i \|^2 \rightarrow_p 0$. Using the inequality $(a+b+c)^2 \leq 3(a^2 + b^2 + c^2)$ and the results of Theorem 1, assuming homogeneity in either exposure effects or exposure probabilities, observe that
\begin{align*}
    \meanin \| \hat \tau_i - \tau_i \|^2 &\leq \max_i \{ f_1(\boldO_i, \mathbb{P}) \}^2 \meanin \| \hat \mu - \mu \|^2 \\
    &\hspace{2em} + \max_i \{ f_2(\boldO_i, \mathbb{P}) \}^2 \meanin \| \hat \pi_g - \pi_g \|^2 \\
    &\hspace{2em} + \max_i \{ f_3(\boldO_i, \mathbb{P}) \}^2 \meanin \| \hat \pi_{g'} - \pi_{g'} \|^2 + o_{\mathbb{P}}(1).
\end{align*}
Then, by $L_2(\mathbb{P})$ consistency of the nuisance functions, $\meanin \| \hat \tau_i - \tau_i \|^2 \rightarrow_p 0$. 

\begin{proof}[Proof of \textbf{Theorem 3}]
    
Let $\phi^*(\mathbb{P}) = \meanin \phi_i(\mathbb{P})$. We decompose the variance $\sigma_n^2/n$ as follows,
\begin{align*}
    \sigma_n^2 &= n \Var(\phi^*(\mathbb{P})) \\
    &= n \Var(\meanin \phi_i(\mathbb{P})) \\
    &= n^{-1} \sum_{ik} \E[(\phi_i(\mathbb{P}) - \E[\phi_i(\mathbb{P})]) (\phi_k(\mathbb{P}) - \E[\phi_k(\mathbb{P})]) ] \\
    &= n^{-1} \sum_{ik} \E[(\phi_i(\mathbb{P}) - \E[\phi^*(\mathbb{P})])(\phi_k(\mathbb{P}) - \E[\phi^*(\mathbb{P})])] \\
    &\hspace{2em} - n^{-1} \sum_{ik} (\E[\phi_i(\mathbb{P})] - \E[\phi^*(\mathbb{P})])(\E[\phi_k(\mathbb{P})] - \E[\phi^*(\mathbb{P})]) \\
    &= n^{-1} \sum_{ik} \E[\phi_i(\mathbb{P})\phi_k(\mathbb{P})] - n^{-1} \sum_{ik} \E[\phi_k(\mathbb{P})]\E[\phi_i(\mathbb{P})] \\
    &= \frac{1}{n} \sum_{s \geq 0} \sum_{i \in \mathcal{N}_n} \sum_{k \in \mathcal{N}_n^{\partial}(i;s)} \E[\phi_i(\mathbb{P}) \phi_k(\mathbb{P})] - \frac{1}{n} \sum_{s \geq 0} \sum_{i \in \mathcal{N}_n} \sum_{k \in \mathcal{N}_n^{\partial}(i;s)} \E[\phi_k(\mathbb{P})]\E[\phi_i(\mathbb{P})] \\
    &= \frac{1}{n} \sum_{s \geq 0} \sum_{i \in \mathcal{N}_n} \sum_{k \in \mathcal{N}_n^{\partial}(i;s)} \E[\phi_i(\mathbb{P}) \phi_k(\mathbb{P})] - \frac{1}{n} \sum_{s \geq 0} \sum_{i \in \mathcal{N}_n} \sum_{k \in \mathcal{N}_n^{\partial}(i;s)} (\Psi_i(\mathbb{P}) - \Psi(\mathbb{P})) (\Psi_k(\mathbb{P}) - \Psi(\mathbb{P})), 
\end{align*}
where we have used that $\E[\phi_i] = \Psi_i(\mathbb{P}) - \Psi(\mathbb{P})$ and thus $\E[\phi^*] = 0$. The remainder of the proof shows that the variance estimator is consistent for the first term \\
$\sigma_n^{2,*} = \frac{1}{n} \sum_{s \geq 0} \sum_{i \in \mathcal{N}_n} \sum_{k \in \mathcal{N}_n^{\partial}(i;s)} \E[\phi_i(\mathbb{P}) \phi_k(\mathbb{P})]$ and is therefore conservative for $\sigma_n^2$ with the bias equal to $V_n = \frac{1}{n} \sum_{s \geq 0} \sum_{i \in \mathcal{N}_n} \sum_{k \in \mathcal{N}_n^{\partial}(i;s)} (\Psi_i(\mathbb{P}) - \Psi(\mathbb{P})) (\Psi_k(\mathbb{P}) - \Psi(\mathbb{P}))$, or the sample covariance of the network exposure effects. 

Consider the following,
\begin{align*}
    \Tilde{\sigma}_n^{* 2} &= \frac{1}{n} \sum_{s \geq 0} \sum_{i \in \mathcal{N}_n} \sum_{k \in \mathcal{N}_n^{\partial}(i;s)} \E[\phi_i(\hat{\mathbb{P}}) \phi_k(\hat{\mathbb{P}}) ] \omega(s/b_n), \\
    \hat{\sigma}_n^2 &= \frac{1}{n} \sum_{s \geq 0} \sum_{i \in \mathcal{N}_n} \sum_{k \in \mathcal{N}_n^{\partial}(i;s)} \phi_i(\hat{\mathbb{P}}) \phi_k(\hat{\mathbb{P}}) \omega(s/b_n) .
\end{align*}
Then, by Proposition 4.1 of \citet{kojevnikov_limit_2021}, $\hat \sigma_n^2 - \Tilde{\sigma}_n^{* 2} \rightarrow_p 0$ under Assumptions 8 and 10. Thus, to show $\hat{\sigma}_n^2 - \sigma_n^{*2} \rightarrow_p 0$, it suffices to prove $\Tilde{\sigma}_n^{*2} - \sigma_n^{*2} \rightarrow_p 0$. For simplicity, we consider uniform weights $\omega(s/b_n)$ that gave weight $1$ if the corresponding expectation is non-zero and $0$ otherwise. First, consider the squared terms.
\begin{align} \label{eq:var1}
    &\E \bigg[ \meanin [\phi_i(\hat{\mathbb{P}})^2 - \phi_i(\mathbb{P})^2] \bigg] \\
    &= \E \bigg[ \meanin \bigg\{ \underbrace{[(\hat h_1 - \hat h_0)(\Delta Y - \hat \mu)]^2 - [(h_1 - h_0)(\Delta Y - \mu)]^2}_{\circled{1}} \nonumber \\
    &\hspace{2em} \underbrace{- 2[(\hat h_1 - \hat h_0)(\Delta Y - \hat \mu)\hat h_1 \Psi(\hat{\mathbb{P}})] + 2[( h_1 -  h_0)(\Delta Y -  \mu) h_1 \Psi(\mathbb{P})]}_{\circled{2}} \nonumber \\
    &\hspace{2em} + \underbrace{\hat h_1^2 \Psi(\hat{\mathbb{P}})^2 - h_1^2 \Psi(\mathbb{P})^2 \bigg\} \bigg]}_{\circled{3}} \nonumber.
\end{align}
Term $\circled{1}$ in \eqref{eq:var1} can be expressed
\begin{align*}
    &\E \bigg[  \meanin [(\hat h_1 - \hat h_0)(\Delta Y - \hat \mu)]^2 - [(h_1 - h_0)(\Delta Y - \mu)]^2 \bigg] \\
    &= \E \bigg[ \meanin \hat \tau_i^2 - \tau_i^2 \bigg] \\
    &= \E \bigg[ \meanin (\hat \tau_i - \tau_i)^2 + 2(\hat \tau_i - \tau_i)\tau_i \bigg] \\
    &= \meanin \E[(\hat \tau_i - \tau_i)^2] + 2\E[(\hat \tau_i - \tau_i)\tau_i] \\
    &\leq \meanin \|\hat \tau_i - \tau_i \|^2 + \meanin \E[(\hat \tau_i - \tau_i)^2]^{1/2} \E[4\tau_i^2]^{1/2} \\
    &\leq \meanin \|\hat \tau_i - \tau_i \|^2 + \left\{ \meanin \|\hat \tau_i - \tau_i \|^2 \right\}^{1/2} \left\{ \meanin \E[4\tau_i^2] \right\}^{1/2} \\
    &=  o_{\mathbb{P}}(1).
\end{align*}
Term $\circled{2}$ in \eqref{eq:var1} can be expressed (ignoring the $-2$ scaling)
\begin{align*}
    &\E \bigg[  \meanin [(\hat h_1 - \hat h_0)(\Delta Y - \hat \mu)\hat h_1 \Psi(\hat{\mathbb{P}})] - [( h_1 -  h_0)(\Delta Y -  \mu) h_1 \Psi(\mathbb{P})] \bigg] \\
    &= \E \bigg[ \meanin \hat \tau_i \hat h_1 \Psi(\hat{\mathbb{P}}) - \tau_i  h_1  \Psi(\mathbb{P}) \bigg] \\
    &= \E \bigg[ \meanin \hat \tau_i \hat h_1 (\Psi(\hat{\mathbb{P}}) - \Psi(\mathbb{P})) + \Psi(\mathbb{P})(\hat \tau_i \hat h_1 - \tau_i  h_1) \bigg] \\
    &= \E \bigg[ \meanin \Psi(\mathbb{P})(\hat \tau_i \hat h_1 - \tau_i  h_1) \bigg] + o_{\mathbb{P}}(1) \\
    &= \E \bigg[ \meanin \Psi(\mathbb{P})(\hat h_1 (\hat \tau_i - \tau_i) - \tau_i(\hat h_1 - h_1)) \bigg] + o_{\mathbb{P}}(1) \\
    &= \E \bigg[ \frac{\Psi(\mathbb{P})}{\hat p_g} \meanin I_g (\hat \tau_i - \tau_i) -  \Psi(\mathbb{P}) \meanin \tau_i(\hat h_1 - h_1) \bigg] + o_{\mathbb{P}}(1) \\
    &= - \Psi(\mathbb{P}) \E \bigg[\meanin \tau_i(\frac{I_g}{\hat p_g} - \frac{I_g}{p_g}) \bigg] + o_{\mathbb{P}}(1) \\
    &= - \Psi(\mathbb{P}) \meanin \Psi_i(\mathbb{P})(\frac{1}{\hat p_g} - \frac{1}{p_g})  + o_{\mathbb{P}}(1) \\
    &= - \Psi(\mathbb{P}) \left( \frac{1}{\hat p_g} \Psi(\mathbb{P}) - \meanin \frac{1}{p_g} \Psi_i(\mathbb{P}) \right) + o_{\mathbb{P}}(1),
\end{align*}
where we have used $\E[\tau_i I_g] = \E[h_1(\deltay - \mu)] = \E[\deltay|\Gbar=\gbar] - \E[\mu|\Gbar=\gbar]$, which is the outcome regression based representation of the $i$-th target estimand $\Psi_i(\mathbb{P})$. The term in the last equality is $o_{\mathbb{P}}(1)$, which follows from homogeneity of either the exposure effects or exposure probabilities and the dependent data law of large numbers so that $\meanin \hat p_g - p_g \rightarrow_p 0$. 
Finally, term $\circled{3}$ in \eqref{eq:var1} can be expressed
\begin{align*}
    &\E \bigg[  \meanin \hat h_1^2 \Psi(\hat{\mathbb{P}})^2 - h_1^2 \Psi(\mathbb{P})^2 \bigg]\\
    &= \E \bigg[ \meanin \frac{I_g}{\hat p_g^2} \Psi(\hat{\mathbb{P}})^2 - \frac{I_g}{p_g^2} \Psi(\mathbb{P})^2 \bigg] \\
    &= \frac{1}{\hat p_g} \Psi(\hat{\mathbb{P}})^2 - \meanin \frac{1}{p_g} \Psi(\mathbb{P})^2 \\
    &= \frac{1}{\hat p_g} \Psi(\hat{\mathbb{P}})^2 - \meanin \left\{ \frac{1}{p_g}\Psi(\hat{\mathbb{P}})^2 - \frac{1}{p_g}\Psi(\hat{\mathbb{P}})^2 \right\} - \meanin \frac{1}{p_g} \Psi(\mathbb{P})^2 \\
    &= \Psi(\hat{\mathbb{P}})^2 \left( \meanin \frac{1}{\hat p_g} - \frac{1}{p_g}\right) + \meanin \frac{1}{p_g} (\Psi(\hat{\mathbb{P}})^2 - \Psi(\mathbb{P})^2) \\
    &= \Psi(\hat{\mathbb{P}})^2 \left( \meanin \frac{p_g - \hat p_g}{\hat p_g p_g}\right) + (\Psi(\hat{\mathbb{P}})^2 - \Psi(\mathbb{P})^2) \meanin \frac{1}{p_g} \\
    &= o_{\mathbb{P}}(1),
\end{align*}
where the first term in the second to last equality follows from boundedness and strict positivity of $p_g$ along with the dependent data law of large numbers, and the second term follows from the continuous mapping theorem. 

The cross-terms are handled similarly. Using the same decomposition,
\begin{align*}
    &\E \bigg[ n^{-1}\sum_{ik} \hat \tau_i \hat \tau_k - \tau_i \tau_k \bigg] \\
    &= \E \bigg[ n^{-1}\sum_{ik} \hat \tau_i (\hat \tau_k - \tau_k) + \tau_k (\hat \tau_i - \tau_i) \bigg] \\
    &= n^{-1}\sum_{ik} \E[\hat \tau_i (\hat \tau_k - \tau_k)] + n^{-1}\sum_{ik}\E[\tau_k (\hat \tau_i - \tau_i)] \\
    &\lesssim \left( n^{-1}\sum_{ik} \| \hat \tau_k - \tau_k \|^2 \right)^{1/2} + \left( n^{-1}\sum_{ik} \| \hat \tau_i - \tau_i \|^2 \right)^{1/2} \\
    &= o_{\mathbb{P}}(1),
\end{align*}
and 
\begin{align*}
    &\E \bigg[ n^{-1}\sum_{ik} \hat \tau_i \hat h_{1k} \Psi(\hat{\mathbb{P}}) - \tau_i h_{1k} \Psi(\mathbb{P})  \bigg] \\
    &= (\Psi(\hat{\mathbb{P}}) - \Psi(\mathbb{P})) \E \bigg[n^{-1}\sum_{ik} \hat \tau_i \hat h_{1k} \bigg] + \Psi(\mathbb{P})  \E \bigg[ n^{-1}\sum_{ik} \hat \tau_i \hat h_{1k} - \tau_i h_{1k} \bigg] \\
    &= \Psi(\mathbb{P})  \E \bigg[ n^{-1}\sum_{ik} \hat h_{1k} (\hat \tau_i - \tau_i) - \tau_i (\hat h_{1k} - h_{1k} )\bigg]  + o_{\mathbb{P}}(1) \\
    &= o_{\mathbb{P}}(1),
\end{align*}
and
\begin{align*}
    &\E \bigg[ n^{-1}\sum_{ik} \hat h_{1i} \hat h_{1k} \Psi(\hat{\mathbb{P}})^2 - h_{1i} h_{1k} \Psi(\mathbb{P})^2 \bigg] \\
    &= [\Psi(\hat{\mathbb{P}})^2 - \Psi(\mathbb{P})^2] \E \bigg[ n^{-1}\sum_{ik} \hat h_{1i} \hat h_{1k} \bigg] + \Psi(\mathbb{P})^2  \E \bigg[ n^{-1}\sum_{ik} \hat h_{1i} \hat h_{1k} - h_{1i} h_{1k}  \bigg] \\
    &=  \Psi(\mathbb{P})^2  \E \bigg[ n^{-1}\sum_{ik} \hat h_{1i} \hat h_{1k} - h_{1i} h_{1k}  \bigg]  + o_{\mathbb{P}}(1) \\
    &\leq  \Psi(\mathbb{P})^2 \left\{ n^{-1}\sum_{ik} p_i(p_k - \hat p) + \hat p(p_i - \hat p) \right\} + o_{\mathbb{P}}(1) \\
    &= o_{\mathbb{P}}(1).
\end{align*}
Finally, in the case where there is network effect heterogeneity and exposure probability heterogeneity, one may estimate the exposure probabilities with a parametric model. Then, the variance must account for this estimation, e.g., $\Var(\meanin \varphi_i(\boldO_i; \mathbb{P})) = \Var(\meanin \phi_i(\mathbb{P})) - \meanin \E[\phi_i S_{\eta}^{\top}] I(\eta)^{-1} \E[S_{\eta} \phi_i]$. By standard theory, a plug-in estimator of the latter term is consistent. This concludes the proof.
\end{proof}

\section{Further discussion on variance estimation}

\subsection{Sources of network dependency}

The network dependency conditions considered in this work are primarily imposed on the influence function. For instance, the main consistency (Theorem 1) and asymptotic normality results (Theorem 2) relied on limit theorems for the influence functions. Similarly, the network HAC variance estimator (Theorem 3) is a function of the influence function. Recall that the proposed network HAC variance estimate utilizes a kernel weight function $\omega$, where choice of $\omega$ may affect finite sample performance (even if $\omega$ satisfies conditions for consistency of the variance estimator). For example, if it is known that $\E[\phi_i \phi_k] = 0$ for $i \neq k$, then the terms $n^{-1} \sum_{i,k} \hat{\phi}_i \hat{\phi}_k$ in the network HAC estimator would contribute to finite sample bias. Here, we examine dependency in the influence function $\phi_i(\boldO_i)$ where dependency across $\boldO_i$ arises only from interference.

Consider a data generating process (DGP) where $\deltay - \mmu = \mathbbm{1}(\Gbar_{it}=\gbar_t) f(\boldX_i, i) + \epsilon_i$ for a generic function $f(\cdot)$. This DGP is often assumed; the commonly used two-way fixed effects model, for example, implies such a result. Assume here that $\epsilon_i$ is a mean zero and independent error term, that is also independent of the data $\boldO_i$. 

First, observe the general result under the additive error DGP,
\begin{align*}
    \phi_i &= \tau_i(\boldO_i) - h_{i1}(\Gbar_{it}) \tau \\
    &= (h_1 - h_0)(\deltay - \mu_{\gbar_t'}(\boldX_i)) - h_1 \tau \\
    &= (h_1 - h_0)(\mathbbm{1}(\Gbar_{it}=\gbar_t) f(\boldX_i, i) + \epsilon_i) - h_1 \tau \\
    &= \frac{\mathbbm{1}(G_i=g)}{\PP(G_i=g)}(f(\boldX_i, i) - \tau) + (h_1 - h_0) \epsilon_i.
\end{align*}
Consider the scenario where there is no treatment effect heterogeneity, i.e., $f(\boldX_i, i) = \tau$. Then, the first term is equal to zero since $\E[\tau_i] = \tau$ under no treatment effect heterogeneity. Thus, $\Cov(\phi_i, \phi_k) = 0$ for $i \neq k$ since $\epsilon$ is independent and mean zero, implying that the variance estimator $\meanin \hat \phi_i^2$ is consistent (provided that the other conditions in Theorem 3 hold). 

Next, suppose $f(X_i, i) = \tau + \theta_i$, where $\theta_i$ is a non-random term that controls treatment effect heterogeneity. Then,
\begin{align*}
    \Cov(\phi_i, \phi_k) &= \Cov\left( \frac{\mathbbm{1}(G_i=g)}{\PP(G_i=g)}\theta_i, \frac{\mathbbm{1}(G_k=g)}{\PP(G_k=g)}\theta_k \right) \\
    &= \frac{\theta_i \theta_k}{\PP(G_i=g) \PP(G_k=g)} \Cov(\mathbbm{1}(G_i=g), \mathbbm{1}(G_k=g)),
\end{align*}
implying that the covariance of the exposure mappings drives the covariance of the influence functions. 

Next, we examine the setting where the exposure mappings are conditionally independent, i.e., $G_i \indep G_k | \boldX_i$ and $G_i \indep G_k | \boldX_k$, a reasonable assumption in many exposure mappings. To simplify notation, let $\mathbbm{1}(G_i=g)$ be denoted $G_i$ (i.e., letting the exposures be binary) and $\PP(G_i=g) = p_i$. Then,
\begin{align*}
    \Cov(\phi_i, \phi_k) &= \Cov \left( \frac{G_i}{p_i}(f(\boldX_i) - \tau), \frac{G_k}{p_k}(f(\boldX_k) - \tau) \right) \\
    &= \Cov \left( \frac{G_i}{p_i}f(\boldX_i), \frac{G_k}{p_k}f(\boldX_k) \right) - \Cov \left( \frac{G_i}{p_i}f(\boldX_i), \frac{G_k}{p_k}\tau \right) \\
    &\hspace{2em} - \Cov \left( \frac{G_i}{p_i}\tau, \frac{G_k}{p_k}f(\boldX_k) \right) + \Cov \left( \frac{G_i}{p_i}\tau, \frac{G_k}{p_k}\tau \right).
\end{align*}
The fourth term is equal to $\tau^2 / (p_i p_k) \Cov(G_i, G_k)$. The first term can be decomposed as,
\begin{align*}
     \Cov \left( \frac{G_i}{p_i}f(\boldX_i), \frac{G_k}{p_k}f(\boldX_k) \right) &= \E \left[ f(\boldX_i) f(\boldX_k) \Cov(\frac{G_i}{p_i}, \frac{G_k}{p_k} | \boldX_i, \boldX_k) \right] \\
     &\hspace{2em}+ (p_i p_k)^{-1} \Cov \left( f(\boldX_i) \E[G_i | \boldX_i, \boldX_k], f(\boldX_k) \E[G_k | \boldX_i, \boldX_k] \right) \\
     &= (p_i p_k)^{-1} \Cov \left( f(\boldX_i) \E[G_i | \boldX_i, \boldX_k], f(\boldX_k) \E[G_k | \boldX_i, \boldX_k] \right),
\end{align*}
where the first equality follows from the law of total covariance and the second equality follows from assuming $G_i \indep G_k | \boldX_i$. If one were to assume that $f(\cdot)$ and $\E[G_i | \boldX_i, \boldX_k]$ are Lipschitz functions, then the second equality shows that $\Cov \left( \frac{G_i}{p_i}f(\boldX_i), \frac{G_k}{p_k}f(\boldX_k) \right)$ is proportional to $\Cov\left(r(\boldX_i), r(\boldX_k) \right)$ for a Lipschitz function $r$. Thus, this covariance term can be bounded using weak dependence assumptions on $\boldX_i$ and $\boldX_k$. A similar result can be shown for the remaining two terms. Though $\Cov(\phi_i, \phi_k) \neq 0$ in general when there is exposure effect heterogeneity in covariates, the above equalities provide a way one could conjecture choosing the regularization parameter $\omega$ in the variance estimator. 

\subsection{Kernel and bandwidth recommendations}

\citet{kojevnikov_limit_2021} recommended the following bandwidth for the Parzen kernel:
\begin{align*}
    b_n = \frac{2 \log n}{\log( \max \{ M_{n}^{\partial}(1), 1.05 \})}.
\end{align*}

Suppose $K$-neighborhood dependence holds. \citet{leung_causal_2022} recommended the following bandwidth criterion for a uniform kernel in a network-HAC variance estimator,
\begin{align*}
    \hat{\sigma}^2_n &= \frac{1}{n} \sum_{s \geq 0} \sum_{i \in \mathcal{N}_n} \sum_{j \in \mathcal{N}_n^{\partial}(i;s)} \hat{\phi}_i \hat{\phi}_j  \mathbbm{1}(d_n(i,j) \leq b_n),
\end{align*}
where $b_n = \lfloor \max \{\Tilde{b}_n, 2K \} \rceil$, $\Tilde{b}_n = 0.5 L \mathbbm{1}(L < 2 \frac{\log n}{\log M_{n}^{\partial}(1)}) + L^{1/3}\mathbbm{1}(L \geq 2 \frac{\log n}{\log M_{n}^{\partial}(1)})$, and $L$ is the average path length. 

\subsection{Positive semi-definite variance estimation}

\citet{gao_causal_2023} proposed an improved HAC variance estimator that is guaranteed to be positive semi-definite in finite samples. Let $\Tilde{\phi}(\boldO) = (\phi(\boldO_1), \dots, \phi(\boldO_n))^{\top}$. Then, the proposed HAC variance estimator can be expressed as:
\begin{align*}
    \hat \sigma^2 &= \Tilde{\phi}(\boldO) K_n \Tilde{\phi}(\boldO)^{\top},
\end{align*}
where $K_n$ is a symmetric kernel matrix. 
Let $Q_n \Lambda_n Q_n^{\top}$ be the eigendecomposition of $K_n$ and $K_n^{+} = Q_n \max \{\Lambda_n, 0\} Q_n^{\top}$ where the max is taken element-wise. Then,
\begin{align*}
    \hat \sigma^2_{\mathrm{GD}} = \Tilde{\phi}(\boldO) K_n^{+} \Tilde{\phi}(\boldO)^{\top}
\end{align*}
is the improved HAC variance estimator.

\section{Additional simulations}

\begin{figure}[!h]
    \centering
    \includegraphics[width=0.75\textwidth]{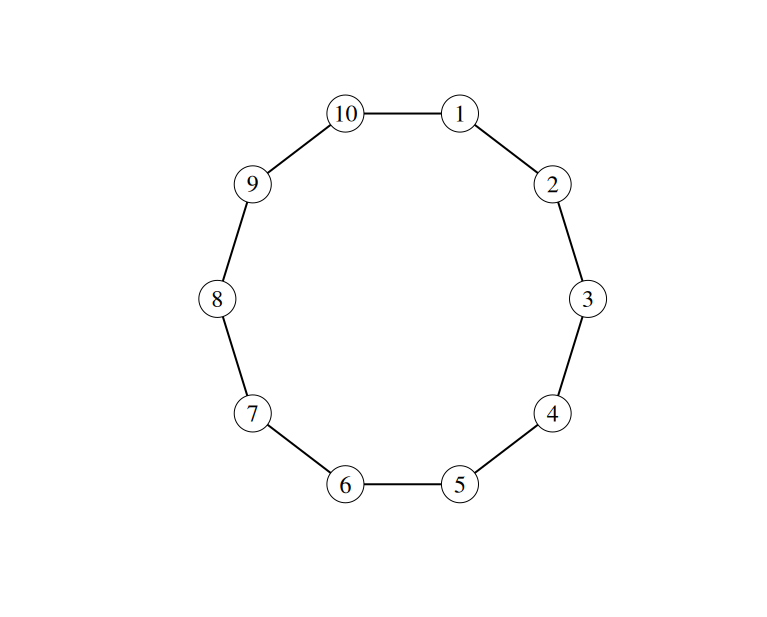}
    \caption{Example ring network with 10 nodes.}
    \label{fig:ring_graph}
\end{figure}

Figure \ref{fig:ring_graph} shows an example ring network for $n=10$ nodes. 

\subsection{Network effect and exposure heterogeneity}

The consequences of network effect and exposure heterogeneity are illustrated here with simulations. Consider the unipartite, ring network considered in the main text. In these simulations, covariates $X_i$, exposures $G_i$, and outcome changes $\Delta Y_i$ were generated as follows.
\begin{align*}
    &X_i \sim \mathrm{Uniform}(-0.1, 0.1), \\
    &G_i | \boldX \sim \mathrm{Bernoulli}(0.5 + \alpha_i + f_G(\boldX, i)) \\
    &\Delta_{0} Y_i | \mathbf{X}, G_i \sim G_i(5 + \theta_i) + f_{Y}(\mathbf{X}, i) + \epsilon_i,
\end{align*}
where $\mathbf{X} = (X_1,\dots,X_n)$, $f_G(\boldX, i) = 0.1X_{i-3} + 0.25X_{i-2} + 0.5X_{i-1} + X_i + 0.5X_{i+1} + 0.25X_{i+2} + 0.1X_{i+3}$, $f_{Y}(\mathbf{X}, i) = \sum_{k=i-3}^{i+3} X_k$ and $\epsilon_i \sim N(0,1)$ and independent. The network heterogeneity terms are $\alpha_i$ and $\theta_i$ and were fixed across simulations. 

\begin{table}[!h]
\caption{Summary of additional simulation scenarios}
\label{tab:sim_summ}
\begin{tabular}{lll|ll}
\hline
\textbf{\begin{tabular}[c]{@{}l@{}}Network effect\\heterogeneity\end{tabular}}  & \textbf{\begin{tabular}[c]{@{}l@{}}Exposure probability\\heterogeneity\end{tabular}} & \textbf{\begin{tabular}[c]{@{}l@{}}Network\\correlation\end{tabular}} & $\boldsymbol{\alpha_i}$ & $\boldsymbol{\theta_i}$ \\ \hline
Yes & No & NA & $0$ & $N(2,2^2)$ \\
Yes & Yes & No & $\mathrm{Unif}(-0.05, 0.15)$ & $N(2,2^2)$ \\
Yes & Yes & Yes & $D_i \times \mathrm{Unif}(-0.05, 0.15)$ & $D_i \times N(2,2^2)$ \\ \hline
\end{tabular}
\end{table}

Three different simulation scenarios were considered, summarized in Table \ref{tab:sim_summ}, that varied by the network heterogeneity terms $\alpha_i$ and $\theta_i$. In the first scenario, there is network effect heterogeneity but not exposure effect heterogeneity. Exposure effect heterogeneity is added in the second scenario. In the third scenario, $D_i \sim \mathrm{Bernoulli}(1/3)$ was generated. With a slight abuse in notation, here $D_i \times \mathrm{Unif}(-0.05, 0.15)$ was used to denote that if $D_i = 1$, then $\alpha_i$ was generated as $\mathrm{Unif}(-0.05, 0.15)$, and $\alpha_i = 0$ (i.e., constant), otherwise. Similarly for $\theta_i$. Thus, in the third scenario, there is a strong, positive correlation between $\alpha_i$ and $\theta_i$. The true exposure probabilities were $\PP(G_i =1) = 0.5 + \alpha_i$, the true individual effects were $\AEE_i = 5 + \theta_i$, and the true total effect was $\AEE = \meanin (5 + \theta_i)$.

\begin{table}[!h]
\caption{Results from 1000 simulations.}
\label{tab:sim_results_supp}
\begin{tabular}{ll|ll|llll}
\hline
\multicolumn{2}{c|}{\textbf{Data generation}} & \multicolumn{2}{c|}{\textbf{Estimator parameters}} & \multicolumn{4}{c}{\textbf{Results}} \\ \hline
\textbf{\begin{tabular}[c]{@{}l@{}}Exposure\\prob.\\het.\end{tabular}} & \textbf{\begin{tabular}[c]{@{}l@{}}Network\\corr.\end{tabular}} & \textbf{\begin{tabular}[c]{@{}l@{}}Exposure\\ prob.\\ estimator\end{tabular}} & \textbf{\begin{tabular}[c]{@{}l@{}}Bias +\\ variance\\ correction\end{tabular}} & \textbf{Bias} & \textbf{ESE} & \textbf{ASE} & \textbf{\begin{tabular}[c]{@{}l@{}}Coverage\\ (\%)\end{tabular}} \\ \hline
No & NA & Sample average & No & 0.0 & 4.0 & 4.8 & 98.6  \\
No & NA & Sample average & Var only & 0.0 & 4.0 & 3.9 & 94.2 \\ \hline
Yes & No & Sample average & No & -0.2 & 4.0 & 4.9 & 98.6 \\
Yes & No & Sample average & Bias + var & -0.7 & 4.0 & 3.9 & 94.2 \\ \hline
Yes & Yes & Sample average & No & 8.4  & 3.7 & 4.1 & 46.6 \\
Yes & Yes & Sample average & Bias + var & -0.1 & 3.7 & 3.5 & 93.4 \\
Yes & Yes & Parametric model & Var only & 0.5 & 3.2 & 3.1 & 93.1 \\
\hline
\end{tabular} \par
\smallskip
Het.: heterogeneity, Prob.: probability, Corr.: correlation, Bias: average bias ($\times 100$) ASE: average standard error estimates ($\times 100$), ESE: empirical standard error ($\times 100$), Coverage (\%): 95\% confidence interval coverage.
\end{table}

To focus attention on the consequences of network heterogeneity, only results from estimators using BART for the outcome regression and propensity score nuisance functions are shown. Results using other nuisance function estimators were similar. In the scenario where there was correlation between the network effects and exposure probabilities, a parametric model was implemented to model the exposure probabilities $\PP(G_i = 1)$. In particular, logistic regression was performed with outcome $G_i$ and covariate $D_i$. In all other cases, the sample average was used, i.e., $\meanin G_i$. Additionally, results are shown with ``corrections" for the unfeasible bias and variance terms described in the main text. These corrections are not feasible in practice since they are unobserved in real data settings, but they are shown here to support the theory presented in the main text. For each simulation scenario, 1000 simulation datasets were generated with sample sizes of $n=5000$ each. Code to replicate the simulations is provided on Github. 

\begin{figure}[!h]
    \centering
    \begin{subfigure}{0.4\textwidth}
        \centering
        \includegraphics[width=1\textwidth]{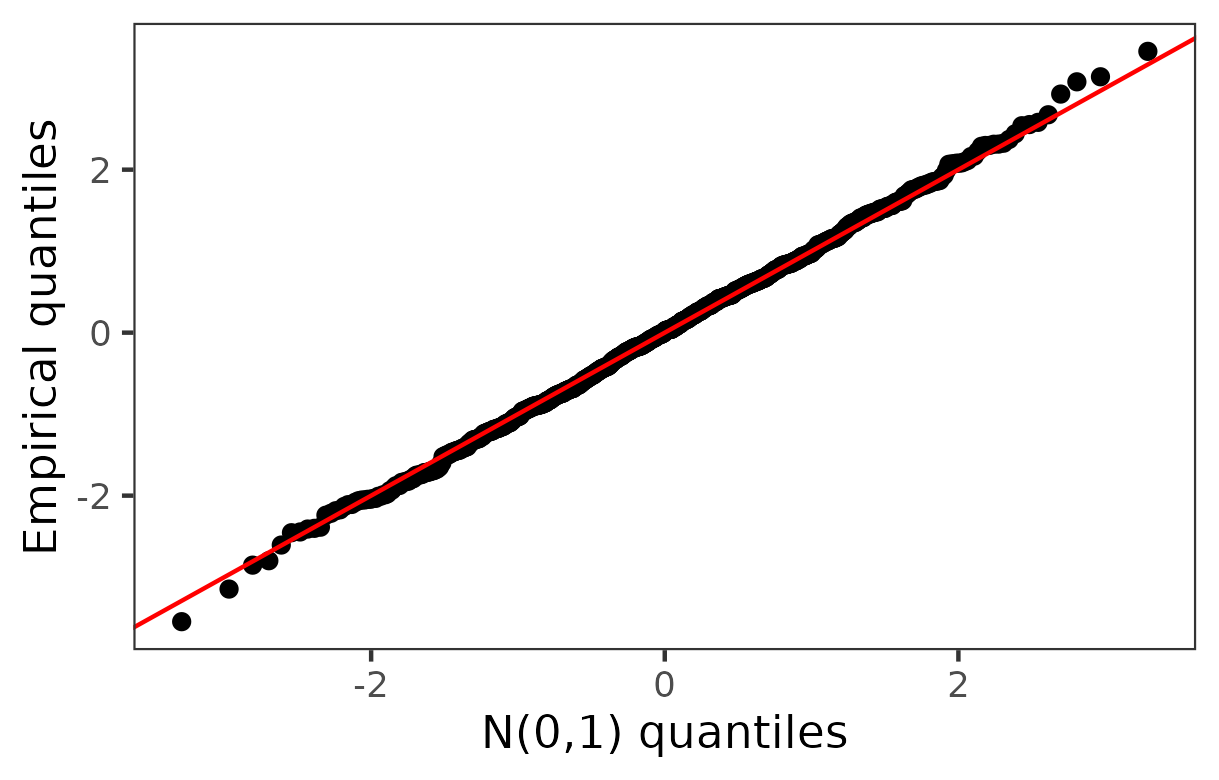}
        \caption{}
    \end{subfigure}
    ~
    \begin{subfigure}{0.4\textwidth}
        \centering
        \includegraphics[width=1\textwidth]{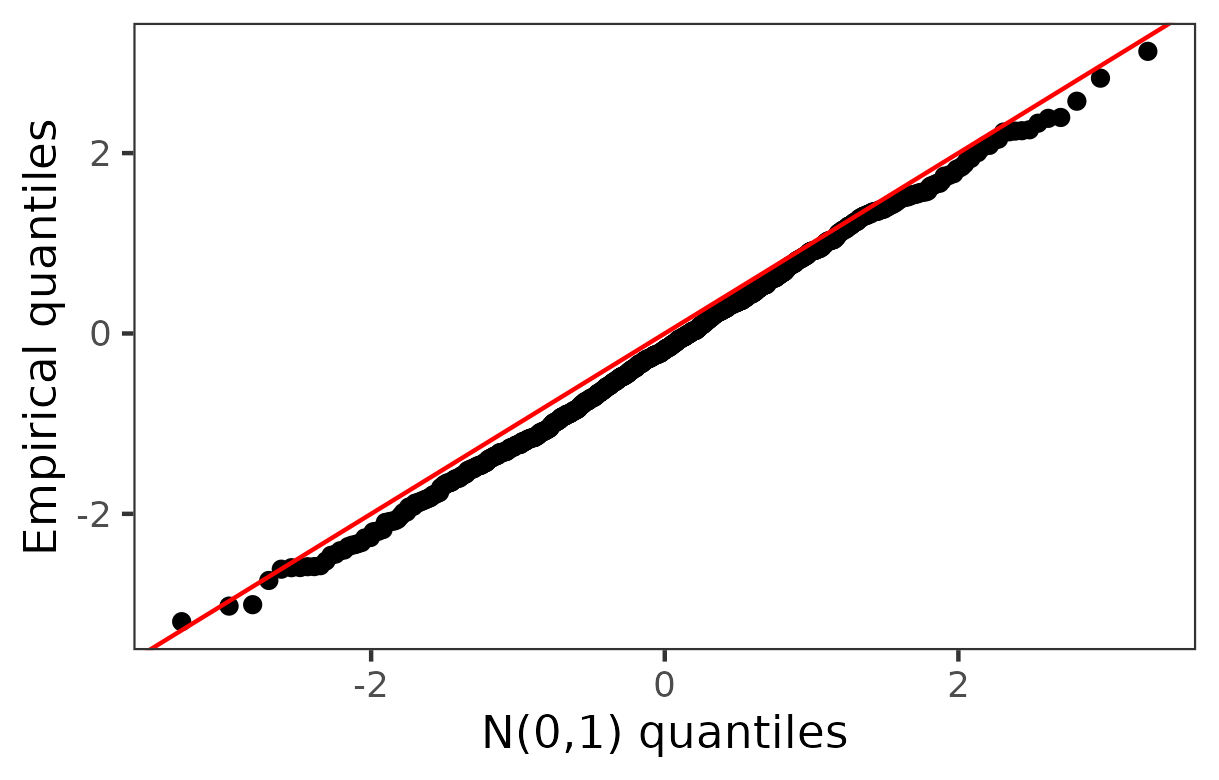}
        \caption{}
    \end{subfigure}
    
    \begin{subfigure}{0.4\textwidth}
        \centering
        \includegraphics[width=1\textwidth]{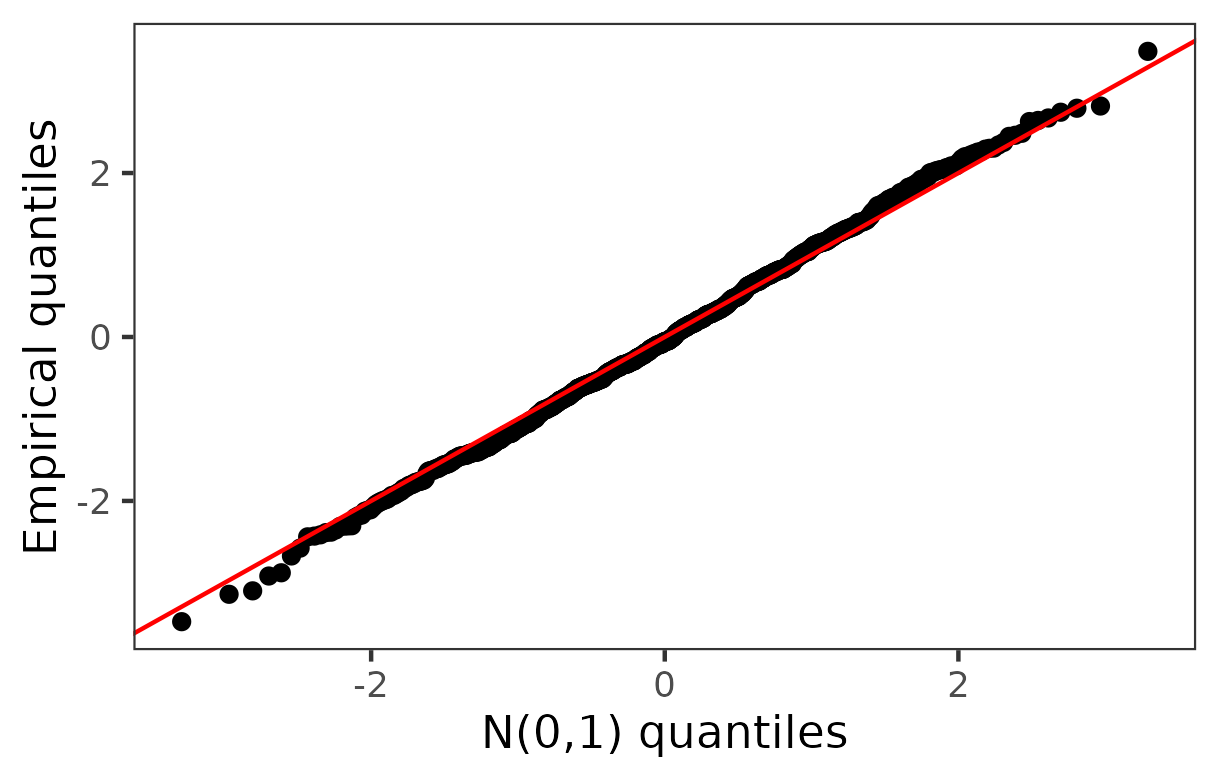}
        \caption{}
    \end{subfigure}
    ~
    \begin{subfigure}{0.4\textwidth}
        \centering
        \includegraphics[width=1\textwidth]{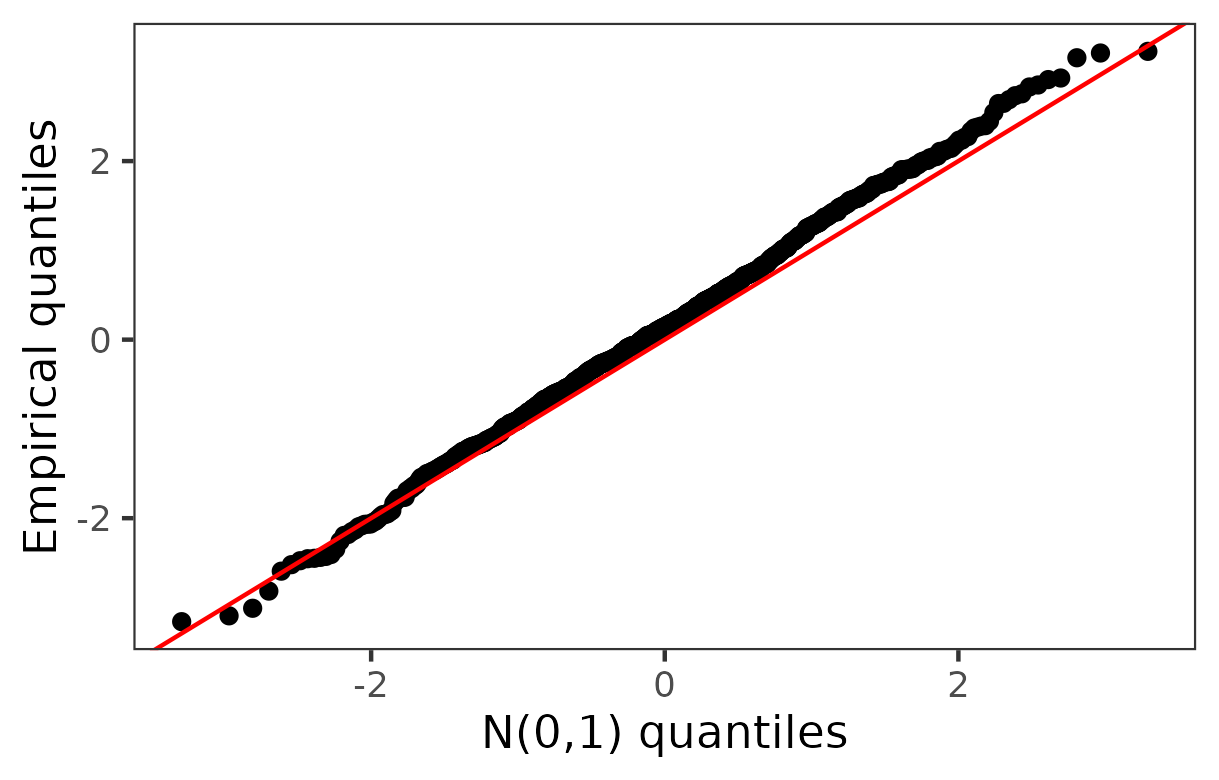}
        \caption{}
    \end{subfigure}
    
    \caption{Quantile-quantile plots comparing $\hat \sigma_n^{-1} \sqrt{n} (\hat \tau - \tau)$ with the $N(0,1)$ distribution where nuisance functions were estimated using BART and bias and variance corrections were applied (if applicable) under scenarios: (a) no exposure probability heterogeneity; (b) exposure probability heterogeneity without network correlation; (c) exposure probability heterogeneity with network correlation; (d) exposure probability heterogeneity with network correlation and parametric modeling of exposure probability.}
    \label{fig:qqplots-supp}
\end{figure}

Table \ref{tab:sim_results_supp} summarizes the results. When there is network effect heterogeneity but no exposure probability heterogeneity, the point estimate is unbiased but the variance estimator captures the heterogeneity in the exposure effects and is thus conservative. However, when the (unfeasible) variance correction is added in, coverage is approximately nominal. In the second simulation scenario, there is exposure probability heterogeneity but the heterogeneity is independent from the network effects so the bias is small and negligible; otherwise the results are similar to the first scenario. Finally, in the third scenario when there is substantial correlation between exposure effects and exposure probabilities, the bias of the point estimator is non-negligible. However, parametrically modeling the exposure probabilities eliminates the bias and consequently, the point estimator is shown to perform as well as adding the unfeasible bias correction to the estimator utilizing the sample average to estimate exposure probability. The variance estimator performs similarly as in the other scenarios. Figure \ref{fig:qqplots-supp} further supports the theoretical plots by showing quantile-quantile plots comparing the scaled estimator $\hat \sigma_n^{-1} \sqrt{n} (\hat \tau - \tau)$ with bias and variance corrections to the standard normal distribution.

\subsection{Additional simulations with a bipartite network}

The bipartite simulation scenarios presented in the main text relied on the interference matrix $\boldW$ with $n = 3105$ rows (outcome units) and $m = 484$ columns (intervention units). In this section, an additional simulation scenario is described where the intervention unit sample size was increased. To preserve a similar bipartite structure while increasing $m$, the $n \times 2m$ interference matrix $\boldW^{\dagger} = (\boldW \hspace{0.5em} \boldW^{*}) $ was used, where $\boldW^*$ has elements $w_{ij}^*$ which are re-samples, with replacement, from the elements of $\boldW$, i.e., $\{w_{ij}: i = 1,\dots,n; j=1,\dots,m \}$. All other simulation settings are the same as described in the main text. Table \ref{tab:sim_results_bipartm} summarizes the results. With the increased intervention unit sample size, the average standard error estimates are closer to the empirical standard errors and coverage is nearly nominal. 

\begin{table}[!h]
\caption{Results from 1000 simulations with increased intervention unit sample size.}
\label{tab:sim_results_bipartm}
\begin{tabular}{ll|ll|rrrrc}
\hline
\multicolumn{2}{c|}{\textbf{Data generation}} & \multicolumn{2}{c|}{\textbf{Estimator parameters}} & \multicolumn{5}{c}{\textbf{Results}} \\ \hline
\textbf{Network} & \textbf{\begin{tabular}[c]{@{}l@{}}Outcome \\ errors\end{tabular}} & \textbf{\begin{tabular}[c]{@{}l@{}}Band-\\ width\end{tabular}} & \textbf{\begin{tabular}[c]{@{}l@{}}Nuisance\\ function\\ estimators\end{tabular}} & \textbf{Bias} & \textbf{MSE} & \textbf{ESE} & \textbf{ASE} & \textbf{\begin{tabular}[c]{@{}l@{}}Coverage (\%)\end{tabular}} \\ \hline
Bipart & Ind. & 0 & GLM & -11.5 & 1.7 & 6.0 & 4.8 & 35.2 \\
 &  & 0 & BART & 0.1 & 0.3 & 5.0 & 4.6 & 93.7 \\
 &  & 0 & HAL & 0.1 & 0.2 & 5.0 & 4.6 & 94.2 \\
 &  & 0 & SuperLearner & 0.1 & 0.3 & 5.0 & 4.6 & 93.9 \\
 &  & 0 & Oracle & 0.1 & 0.2 & 4.8 & 4.6 & 94.2 \\ \hline
\end{tabular} \par
\smallskip
Bipart: bipartite network, Ind.: independent, Dep.: dependent, Bias: average bias ($\times 100$), MSE: mean squared error ($\times 100$), ASE: average standard error estimates ($\times 100$), ESE: empirical standard error ($\times 100$), Coverage (\%): 95\% confidence interval coverage.
\end{table}

\section{Additional details on the power plant emission control analysis}

\subsection{Data processing}

The \texttt{disperseR} R \href{https://github.com/lhenneman/disperseR}{package} \citep{henneman_characterizing_2019} was used to calculate the interference matrices. We used the recommended parameters per the package vignettes, running the HYSPLIT model daily starting at hours 0, 6, 12, 18, with duration 240 hours for all study years. The species for the model was specified as SO$_2$. The HYSPLIT output was then linked with counties. We considered counties in the contiguous United States, excluding islands and territories. 

Climate data were downloaded from gridMET as a raster. The data was converted to the county-level by taking a spatial average, that is, the weighted average of the raster cell data where the weights were the fraction of each raster cell overlapping with the county polygon. 

County-level mortality data, downloaded from CDC WONDER, was excluded if the data was noted as unreliable, suppressed, or missing. Additionally, counties were excluded unless mortality data was available for all study years. 

Data processing for the coal power plant data followed other recent studies \citep{wikle_causal_2024}. Power plants are composed of at least one electricity generating unit (EGU) at a particular facility. Coal EGUs were defined as any EGU whose primary fuel type was coal, i.e., EGUs whose secondary fuel type was coal (but not primary) were excluded. Any SO$_2$ control was included as a scrubber, including flue-gas desulfurization, dual alkali, magnesium oxide, sodium based, wet limestone, fluidized bed limestone injection, and dry sorbent injection emission controls. Max heat input rate was missing for some EGUs within power plants. These were imputed using the maximum for that EGU across all years. If data were not available for other years, then max heat input rate was imputed using the maximum of all other EGUs at that power plant facility across all years. Otherwise, max heat input rate was imputed as the average across all EGUs and facilities across all years. Max operating time for any EGU was defined to be the maximum annual operating time over the study years. Capacity was then defined as max operating time multiplied by the max hourly heat input rate (mmbtu). Heat input was similarly imputed. EGUs with missing or zero operating time were considered non-operational and excluded. 

\begin{figure}[!h]
    \centering
    \includegraphics[width=1\textwidth]{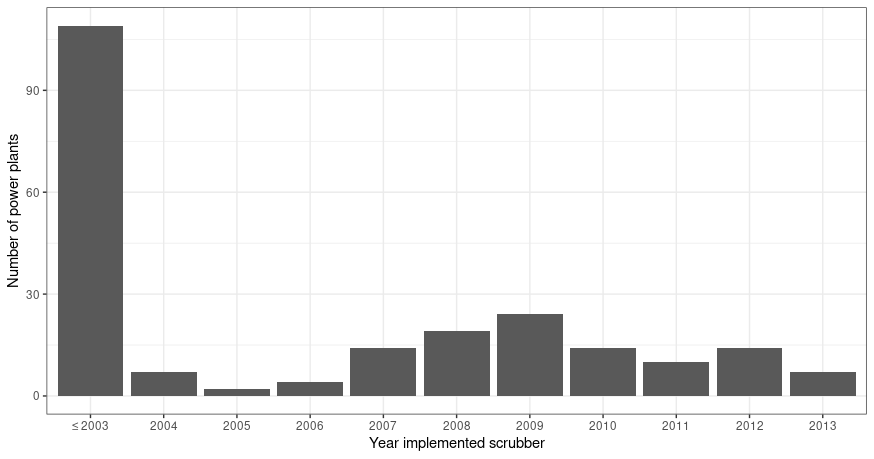}
    \caption{Frequency of scrubber installation timing among coal power plants.}
    \label{fig:scrubber_begin}
\end{figure}

EGU-level data was then aggregated to the facility (power plant) level. Power plants were considered to have scrubbers if all EGUs had scrubbers. Heat input and operating time were summed. Percent capacity was defined as the summed heat input over the summed capacity. A power plant was considered to have a selective non-catalytic reduction (SNCR) emission control (typically controls NOx emissions) if at least half of EGUs in the power plant had SNCRs.

\subsection{Data descriptions}

\begin{table}[!h]
\caption{Number (percent) of counties within each exposure cohort.}
\label{tab:sample_size}
\centering
\begin{tabular}{lll}
\hline
\textbf{Exposure cohort} & \textbf{Low power plant burden} & \textbf{High power plant burden} \\ \hline
\textless 2007 & 0 (0.0\%) & 27 (4.2\%) \\
2007 & 5 (0.8\%) & 98 (15.2\%) \\
2008 & 70 (10.9\%) & 198 (30.7\%) \\
2009 & 367 (57.1\%) & 288 (44.7\%) \\
2010 & 147 (22.9\%) & 27 (4.2\%) \\
\textgreater 2010 & 54 (8.4\%) & 6 (0.9\%) \\ \hline
\end{tabular}
\end{table}

Coal power plants implemented scrubbers at various times over the study period. Figure \ref{fig:scrubber_begin} shows the frequency of scrubber adoption for each year. Most power plants installed scrubbers in 2003 or before. Table \ref{tab:sample_size} shows the sample size (number of counties) for the different exposure cohorts, split by low and high power plant burden groups. Finally, Table \ref{tab:cov_summ} shows descriptive statistics of county and power plant covariates in 2009. These summary statistics were similar for other study years. 

\begin{table}[!h]
\caption{Summaries of county-level and power plant-level covariates by power plant burden group, 2009.}
\label{tab:cov_summ}
\begin{tabular}{lll}
\hline
\multirow{2}{*}{\textbf{Covariate}} & \multicolumn{2}{c}{\textbf{Mean (SD)}} \\
& \textbf{Low burden} & \textbf{High burden} \\ \hline
\textbf{County} \\ \hline 
Proportion White & 0.849 (0.163) & 0.857 (0.162) \\
Proportion Black & 0.118 (0.161) & 0.108 (0.146) \\
Proportion Hispanic & 0.026 (0.029) & 0.025 (0.04) \\
Proportion female & 0.508 (0.015) & 0.51 (0.016) \\
Median age & 36.8 (2.9) & 37.2 (3.0) \\
Average household size & 2.5 (0.1) & 2.5 (0.1) \\
Proportion urban & 0.419 (0.287) & 0.468 (0.304) \\
Proportion in poverty & 0.132 (0.064) & 0.121 (0.054) \\
Proportion high school graduate & 0.497 (0.065) & 0.51 (0.054) \\
log(Population) & 10.8 (1.1) & 10.8 (1.2) \\
log(Population / mi$^2$) & 4.5 (1.1) & 5 (1.3) \\
Smoking prevalence & 25.2 (3.9) & 25.7 (4.1) \\
Average daily precipitation, mm & 3.8 (0.9) & 3.3 (0.5) \\
Average daily relative humidity, \% & 90.4 (2.5) & 88.9 (2.8) \\
Average daily maximum temperature, $^\circ$C & 18.0 (4.1) & 17.6 (2.3) \\
Power plant burden & 408,689 (69405.5) & 592,738.8 (68,944.6) \\ \hline
\textbf{Power plant} \\ \hline 
Scrubber & \multicolumn{2}{c}{0.362 (0.481)} \\
log(Heat input), mmbtu & \multicolumn{2}{c}{16.5 (1.8)} \\
log(Operating time), hours & \multicolumn{2}{c}{9.3 (1.0)} \\
Percent capacity & \multicolumn{2}{c}{50.3 (24.1)} \\
Proportion with selective non-catalytic reduction & \multicolumn{2}{c}{0.324 (0.469)} \\
Participation in ARP Phase II & \multicolumn{2}{c}{0.709 (0.455)} \\ \hline
\end{tabular} \par
\smallskip
ARP: Acid Rain Program.
\end{table}

\subsection{Additional analysis}

\begin{figure}[!h]
    \centering
    \includegraphics[width=1\textwidth]{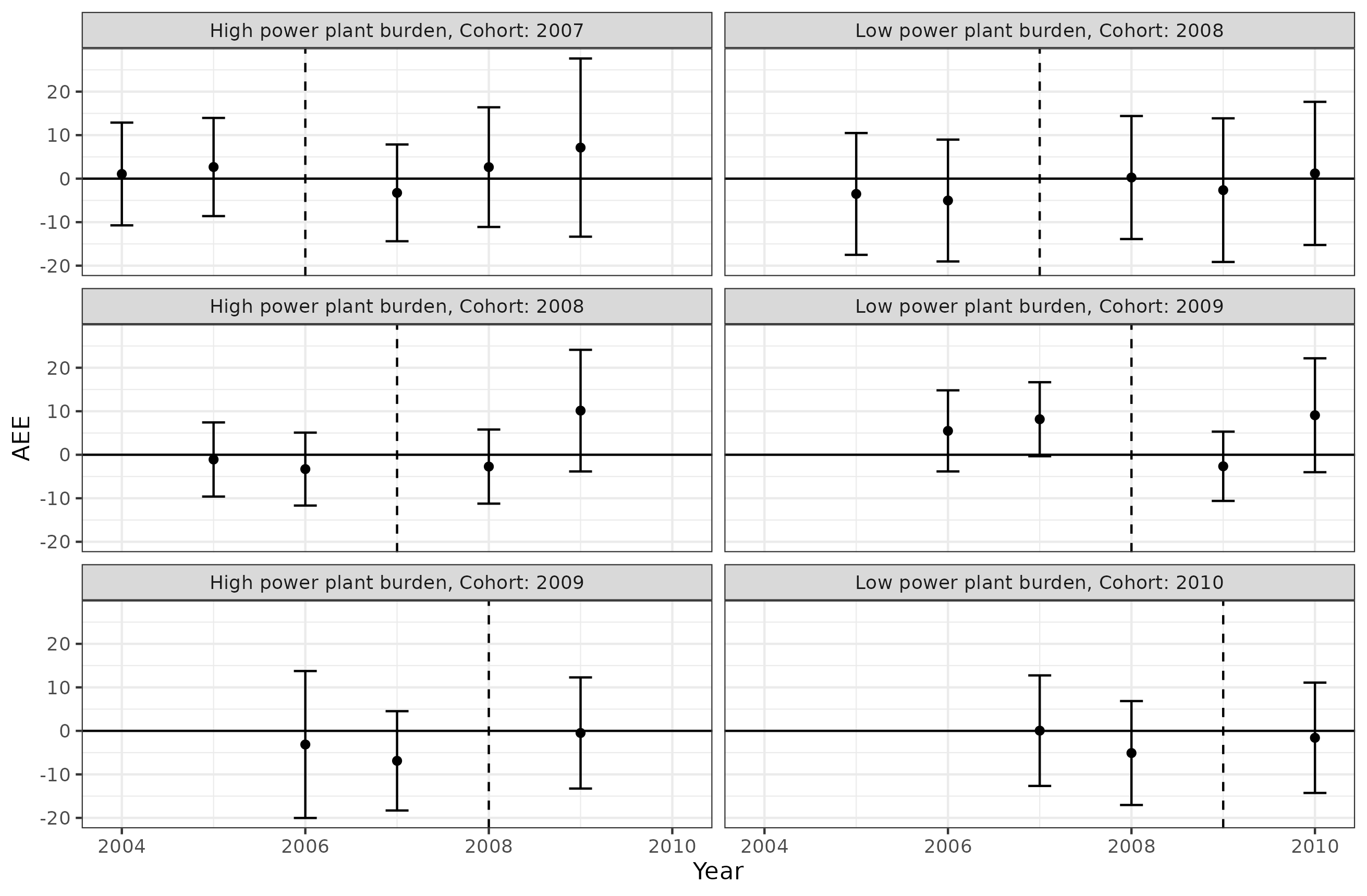}
    \caption{Estimated effect of coal power plant scrubber exposure on county-level deaths due to cardiovascular diseases, per 100,000 individuals per year, assuming unconditional parallel trends.}
    \label{fig:results_mean}
\end{figure}

An analysis was performed assuming unconditional parallel trends and found similar results as the main analysis. Figure \ref{fig:results_mean} displays the results. 

\newpage
\section*{References}

\printbibliography[heading=none]

\end{document}